\documentclass[12pt]{article}

\usepackage{theorem,amssymb,amsbsy,latexsym}

\newcommand{\mbf}[1]{{\boldsymbol {#1} }}

\newcommand{\eqq}[1]{\begin{equation} #1 \end{equation}}
\newcommand{\ar}[1]{\begin{eqnarray} #1 \end{eqnarray}}

\newcommand{\newsection}{\setcounter{equation}{0}\section}

\def\appendix#1{\addtocounter{section}{1}\setcounter{equation}{0}
\renewcommand{\thesection}{\Alph{section}}
\section*{Appendix \thesection\protect\indent \parbox[t]{11.715cm} {#1}}
\addcontentsline{toc}{section}{Appendix \thesection\ \ \ #1} }
\newcommand{\eq}{\begin{equation}}
\newcommand{\eqend}{\end{equation}}
\def\ddd{^{\dagger}}

\def\D{\delta}
\newcommand{\br}[1]{\left( #1 \right)}

\def\ep{\epsilon}

\newcommand{\vxi}{{\mbf \xi}}
\newcommand{\veff}{{\mbf f}}
\newcommand{\vbeta}{{\mbf \beta}}
\newcommand{\vell}{{\mbf \ell}}
\newcommand{\vem}{{\mbf m}}
\newcommand{\aosc}{{\sf a}}
\newcommand{\Aosc}{{\sf A}}
\newcommand{\bosc}{{\sf b}}

\newcommand{\mS}{{\Gamma}}
\newcommand{\complex}{{\mathbb C}} %% complex numbers
\newcommand{\zed}{{\mathbb Z}} %% integers
\newcommand{\nat}{{\mathbb N}} %% naturals
\newcommand{\real}{{\mathbb R}} %% real numbers
\newcommand{\reals}{{\mathbb R}} %% small real numbers
\newcommand{\zeds}{{\mathbb Z}} %% small integers
\newcommand{\id}{{1\!\!1}} %% identity operator
\def\nn{\nonumber}

\newcommand{\Tr}[1]{\:{\rm Tr}\,#1}
\def\e{{\,\rm e}\,}
\newcommand{\rf}[1]{(\ref{#1})}
\newcommand{\non}{\nonumber \\}

\def\be{\begin{equation}}
\def\ee{\end{equation}}
\def\bea{\begin{eqnarray}}
\def\eea{\end{eqnarray}}
\def\bd{\begin{displaymath}}
\def\ed{\end{displaymath}}

\def\DD{{\rm D}}
\def\dd{{\rm d}}

\def\ii{{\,{\rm i}\,}}

\def\K{{{\rm K}_0}}
\def\K1{{{\rm K}_1}}

\makeatletter
\newdimen\normalarrayskip              % skip between lines
\newdimen\minarrayskip                 % minimal skip between lines
\normalarrayskip\baselineskip
\minarrayskip\jot
\newif\ifold             \oldtrue            
\def\arraymode{\ifold\relax\else\displaystyle\fi} % mode of array entries
\def\@arrayskip{\ifold\baselineskip\z@\lineskip\z@
     \else
     \baselineskip\minarrayskip\lineskip2\minarrayskip\fi}
\def\@arrayclassz{\ifcase \@lastchclass \@acolampacol \or
\@ampacol \or \or \or \@addamp \or
   \@acolampacol \or \@firstampfalse \@acol \fi
\edef\@preamble{\@preamble
  \ifcase \@chnum
     \hfil$\relax\arraymode\@sharp$\hfil
     \or $\relax\arraymode\@sharp$\hfil
     \or \hfil$\relax\arraymode\@sharp$\fi}}
\def\@array[#1]#2{\setbox\@arstrutbox=\hbox{\vrule
     height\arraystretch \ht\strutbox
     depth\arraystretch \dp\strutbox
     width\z@}\@mkpream{#2}\edef\@preamble{\halign \noexpand\@halignto
\bgroup \tabskip\z@ \@arstrut \@preamble \tabskip\z@ \cr}%
\let\@startpbox\@@startpbox \let\@endpbox\@@endpbox
  \if #1t\vtop \else \if#1b\vbox \else \vcenter \fi\fi
  \bgroup \let\par\relax
  \let\@sharp##\let\protect\relax
  \@arrayskip\@preamble}
\makeatother

\newcommand{\beq}{\begin{eqnarray}}
\newcommand{\eeq}{\end{eqnarray}}

\newcommand{\g}{\gamma}

\newcommand{\nd}{{\phantom\dag}}

\begin{document}
\begin{titlepage}
\begin{flushright}

\baselineskip=12pt
HWM--03--9\\ EMPG--03--12\\ ITEP--TH--38/03\\ UUITP--14/03\\ hep-th/0308043\\
\hfill{ }\\ August 2003
\end{flushright}

\begin{center}

\baselineskip=24pt

{\Large\bf Exact Solution of Quantum Field Theory\\ on Noncommutative
  Phase Spaces}

\baselineskip=14pt

\vspace{1cm}

{\bf E. Langmann}
\\[2mm]
{\it Department of Physics -- Mathematical Physics, Royal Institute of
Technology\\ AlbaNova -- SCFAB,
SE-10691 Stockholm, Sweden}\\  {\tt langmann@theophys.kth.se}
\\[5mm]

{\bf R.J. Szabo}
\\[2mm]
{\it Department of Mathematics, Heriot-Watt University\\ Scott Russell
  Building, Riccarton, Edinburgh EH14 4AS, U.K.}\\  {\tt R.J.Szabo@ma.hw.ac.uk}
\\[5mm]

{\bf K. Zarembo}\footnote{\baselineskip=12pt Also at: Institute of Theoretical
and Experimental Physics, B. Cheremushkinskaya 25, 117 259 Moscow, Russia.}
\\[2mm]
{\it Department of Theoretical Physics, Uppsala University\\ Box 803, SE-75108
Uppsala, Sweden} \\  {\tt Konstantin.Zarembo@teorfys.uu.se}
\\[10mm]

{\sl In memory of Ian I. Kogan}

\vspace{5mm}

\end{center}

\begin{abstract}

\baselineskip=12pt

We present the exact solution of a scalar field theory defined with
noncommuting position and momentum variables. The model describes
charged particles in a uniform magnetic field and with an interaction
defined by the Groenewold-Moyal star-product. Explicit results are
presented for all Green's functions in arbitrary even spacetime
dimensionality. Various scaling limits of the field theory are
analysed non-perturbatively and the renormalizability of each limit
examined. A supersymmetric extension of the field theory is also
constructed in which the supersymmetry transformations are
parametrized by differential operators in an infinite-dimensional
noncommutative algebra.

\end{abstract}

\end{titlepage}

\newpage

\baselineskip=14pt

\tableofcontents

\newpage

\newsection{Introduction and Summary}

Despite the enormous amount of activity centered around the study of
noncommutative field theories (see~\cite{ks}--\cite{sz1} for reviews),
some of the most fundamental questions surrounding them have yet to be
answered to full satisfaction. Foremost among these is whether or
not these models can make sense as renormalizable, interacting quantum
field theories. Non-local effects such as the interlacing of
ultraviolet and infrared scales~\cite{MvRS} make conventional renormalization
schemes, such as the Wilsonian approach, seemingly hopeless. In fact,
this mixing is known to render some noncommutative field theories
non-renormalizable, even if their commutative counterparts are
renormalizable. A class of such models are asymptotically-free
noncommutative field theories~\cite{GGRdS,ADBS}, wherein the
combination of asymptotic freedom and UV/IR mixing forces the
couplings to run to their trivial Gaussian fixed points in the
ultraviolet scaling limit. Such theories possess non-trivial
interactions only when they are defined with a finite ultraviolet
cutoff. This result is obtained by expansion about a translationally
invariant vacuum of the quantum field theory.

The analysis of UV/IR mixing and the renormalization of noncommutative
scalar field theory to all orders of perturbation theory has been
carried out in~\cite{CR}--\cite{GW}. For at least certain classes of
noncommutative field theories, it may be that unusual effects such as
UV/IR mixing are merely perturbative artifacts which disappear when
resummed to all orders. This is of course a non-perturbative issue,
which motivates the search for examples of exactly solvable
noncommutative field theories. An example of such a model was
presented and analysed in~\cite{lsz}. It describes charged scalar
fields on two-dimensional Euclidean spacetime in a constant external
magnetic field and with a non-local quartic interaction. It may be
thought of as the natural modification of a quantum field theory on a
noncommutative space to one with noncommuting momentum space
coordinates as well, i.e. as a quantum field theory on a
noncommutative {\it phase} space. These models provide concrete
examples in which one can see genuine quantum field theoretic issues,
such as renormalization, at work in non-trivial exactly solvable
theories.

In~\cite{lsz} the exact two-point Green's function of this model was
obtained and the non-perturbative non-renormalizability of the field
theory established. In this paper we will elaborate on this analysis
and present various extensions and generalizations of this class of
noncommutative field theories. In particular, we will extend the
solution to arbitrary even dimensionality and higher order Green's
functions, as well as presenting a detailed analysis of the various
field theory limits which are possible in these models. In the course
of this analysis, we will encounter {\it new} types of complex matrix
models with external fields, for which we provide the necessary
technical tools for analysis. Other aspects of these types of
quantum field theories are discussed in~\cite{habara}--\cite{MM-KP},
while their one-particle sectors are analysed
in~\cite{DH}--\cite{Horvathy}. Related exact calculations in
nonrelativistic noncommutative field theory may be found
in~\cite{BKSY,BKSY1}.

\subsection{Description of the Model}

The model we shall study in this paper is a particular deformation of
scalar field theory in both position and momentum space. Consider
complex scalar fields $\Phi(x)$ on even dimensional Euclidean
spacetime $\real^{2n}$ in an external constant magnetic field $B_{ij}$,
and with an interaction defined by replacing the usual pointwise
multiplication of fields with the Groenewold-Moyal star-product,
defined such that the components of the spatial coordinates
$x=(x^1,\dots,x^{2n})$ satisfy the relations
\beq
x^i\star x^j-x^j\star x^i=2\ii\theta^{ij} \ .
\label{starcommrels}\eeq
{}From (\ref{starcommrels}) and the Baker-Campbell-Hausdorff formula it
follows that the star-product of two plane waves is given by
\beq
\e^{\ii k\cdot x}\star\e^{\ii q\cdot x}=\e^{-\ii k_i\,\theta^{ij}
  \,q_j}~\e^{\ii(k+q)\cdot x} \ .
\label{planewavestar}\eeq
Thus the interaction can be computed by expanding the fields in plane
waves, i.e. by Fourier transformation, and (\ref{planewavestar}) defines
the interaction part of the action for all Schwartz functions
$\Phi(x)$. The background magnetic field, on the other hand, may be
realized as the curvature of Landau momentum operators
$P_i=-\ii\,\frac\partial{\partial x^i}-B_{ij}\,x^j$ with
\beq
[P_i,P_j]=-2\ii B_{ij} \ .
\label{momcommrels}\eeq
The free part of the action is then defined by replacing the Laplacian
operator of conventional scalar field theory with the Landau
Hamiltonian $P_i^2$. The $2n\times2n$ constant skew-symmetric matrices
$\theta=(\theta^{ij})$ and $B=(B_{ij})$, characterizing respectively
the noncommutativity of the coordinates and momenta, are assumed to be
non-singular.

It was shown in~\cite{LS} that there is a natural regularization of
these models which allows, in a simple manner, to make precise
mathematical sense of the functional integrals defining the quantum
field theory. Rather than expanding the boson fields in plane waves,
it is more natural to expand them in the eigenfunctions of the Landau
Hamiltonian $P_i^2$ and thereby diagonalize the free part of the
action. The Landau eigenfunctions $\phi_{\vell,\vem}(x)$ are labelled by
two sets of quantum numbers
$\vell=(\ell_1,\dots,\ell_n),\vem=(m_1,\dots,m_n)\in\nat^{\,n}$, and
the corresponding energy eigenvalues depend on only half these quantum
numbers $\vell$. By expanding the fields as
$\Phi(x)=\sum_{\vell,\vem}M_{\vem,\vell}\,\phi_{\vell,\vem}(x)$, the
formal functional integration then becomes an integration over all
complex expansion coefficients $M_{\vem,\vell}$. This suggests a
regularization of the quantum field theory in which the collections of
Landau quantum numbers $\vell$ and $\vem$ are each restricted to a
finite set which is in one-to-one correspondence with the set
$\{1,2,\dots,N\}$, where $N<\infty$. With this regularization, the
functional integral becomes a finite dimensional integral over $N\times N$
complex matrices. Similar matrix approximation schemes have also been
studied in~\cite{GW},\cite{G-BV}--\cite{LVZ}. We remark that these
matrix models of noncommutative field theory are not the same as those
based on lattice regularization~\cite{AMNS,AMNS1}, and in particular
the scaling limits required in the two approaches are very different.

As discussed in~\cite{LS}, in the case of complex $\Phi^4$-theory deformed as
described above, a finite matrix dimension $N$ amounts to
having both an ultraviolet and infrared cutoff at the same time, as
there exist duality transformations which exchange infrared and
ultraviolet divergences, while the action and all regularized Green's
functions of the model are duality invariant. In this paper we will
carry this approach one step further and describe precisely how to
remove the regularization, i.e. pass to the field theory limit
$N\to\infty$. In general, the interaction part of the action has a
quite complicated form in the Landau basis (This is of course also
true in the commutative case~\cite{KLL}). However, the quantum duality
symmetry in this family of models suggests that they should be special
when $B=\pm\,\theta^{-1}$~\cite{LS}. The crucial point is that at the
special points $B=\pm\,\theta^{-1}$, the interactions acquire a simple form
as well owing to a remarkable closure property of the Landau
eigenfunctions under the star-product, and the field theory can be mapped
{\it exactly} onto a matrix model~\cite{L}. The significance of the
points $B=\pm\,\theta^{-1}$ has been previously noted within different
field theoretic contexts in~\cite{AMNS1}--\cite{Seiberg} and within the
context of noncommutative quantum mechanics in~\cite{DH,NP,BNS}.

In this paper we will concentrate for the most part on extracting Green's
functions of the noncommutative field theory in the limit where the
regulator is removed. The latter limit corresponds to the large $N$
limit of the matrix model, and we can make it non-trivial simply by
taking the 't~Hooft limit of the matrix model~\cite{lsz}. We thereby
determine, in a completely non-perturbative way, how to renormalize
the parameters of the model, i.e. give them a particular
dependence on the regulator $N$ such that the $N\to\infty$ limit exists
and is non-trivial. Even though this limit is essentially unique at
the level of the matrix model, we will see that there is still some
freedom left in defining the limit at the level of the field theoretic
Green's functions. In~\cite{lsz} the two-point Green's function was
derived for one such limit in which it is translationally
invariant. In the following we will use duality to argue that there is
another limit, which mathematically is simply obtained from that
of~\cite{lsz} by Fourier transformation, but which has a very
different physical interpretation. We will use this result to argue
that there may exist limits ``in between'' these other two scaling
limits.

\subsection{Summary of Main Results}

{}From the mapping sketched above onto a finite-dimensional matrix
model, we will derive a number of important consequences which lead to
the exact solution of the noncommutative quantum field
theory. Foremost among these features is the fact that the regularized
field theory, for any interaction potential, is exactly solvable in
the sense that there is a closed formula for the partition function at
finite $N$. In fact, we will establish the formal integrability of a
more general matrix model which in the field theory amounts to having,
in addition to the magnetic field, a confining electric field
background for the charged scalars. In~\cite{L} this was established
within the framework of the Hamiltonian quantization of the
nonrelativistic, $2n+1$-dimensional fermion version of this
field theory. It was shown that this model has a dynamical symmetry
group $GL(\infty)\times GL(\infty)$, with the free part of the model
given as a linear superposition of Cartan elements of the
representation and the interaction proportional to a tensor Casimir
operator. The dynamical symmetry allows the explicit and
straightforward construction of the eigenstates of the Hamiltonian,
and some remnant of it should survive in the high-temperature limit
of the corresponding partition function, which is essentially what we
deal with here. In the following we will explore the integrability of
this system from a different point of view. We will derive a
determinant formula for the matrix integral and hence show that it is
a tau-function of the integrable, two-dimensional Toda lattice
hierarchy.  The $GL(\infty)\times GL(\infty)$ dynamical symmetry then
presumably manifests itself as the rotational symmetry of the free
fermion representation of the vertex operator construction of the Toda
tau-function. The limit in which the electric potential is turned off
reduces the partition function {\it exactly} to a KP tau-function.

We will then study the large $N$ limit of the original matrix
model. The Schwinger-Dyson equations of the matrix
model succinctly sum up all divergences which one would encounter when
solving the quantum field theory in perturbation theory. We shall
compute the field theoretic Green's functions from the matrix model
correlators, which also requires us to perform sums
over Landau levels of certain combinations of the Landau
eigenfunctions. The latter quantities capture the spacetime dependence
of the Green's functions. As a warm-up and as a simple example where
we can give complete results, we will first consider the limiting case
in which the kinetic energy in the action is dropped. In this case we
compute {\it all} Green's functions explicitly.

We will then present the detailed computation of the two-point
function $G(x,y)=\langle\Phi^\dag(x)\,\Phi(y)\rangle$ of the full
model in various scaling limits of the quantum
field theory with a complex $\Phi^4$ interaction potential. In the limit
$N\to\infty$ with the quantity $\Lambda^{2n}=N\,\sqrt{\det(B/4\pi)}$
held fixed, we find that the exact propagator admits the Fourier
integral representation
\beq
G_{\rm ir}(x,y)=-\int\limits_{|p|\leq4\,\sqrt{\pi}\,\Lambda}
\frac{\dd^{2n}p}{(2\pi)^{2n}}~\Lambda^{-2}\,W\Bigl(\Lambda^{-2}\,
\left(p^2+\mu^2\right)\Bigr)~\e^{\ii p\cdot(x-y)} \ ,
\label{Guvxysumm}\eeq
where $\mu$ is the mass of the scalar field $\Phi(x)$, and $W(\lambda)$ is
the propagator of the underlying matrix model which may be determined
as the solution of the Schwinger-Dyson equations in the large $N$
limit. The quantity $\Lambda$ can be interpreted here as an
ultraviolet cutoff, and in the limit $\Lambda\to\infty$ the function
$W$ reduces to
\beq
W(\lambda)=\frac{\lambda-\sqrt{\lambda^2+4\Lambda^{2n-4}\,g}}
{2\Lambda^{2n-4}\,g}
\label{Wxisumm}\eeq
with $g$ the bare $\Phi^4$ coupling constant. This result was derived
in~\cite{lsz} in the two-dimensional case. We will then argue that all
higher Green's functions can be obtained from the two-point function
by using Wick's theorem, i.e. all connected Green's functions in this
particular scaling limit vanish, mainly as a consequence of the usual
large $N$ factorization in the matrix model. This type of scaling seems to be
generic to the matrix regularization of noncommutative field
theories~\cite{MRW,AMNS1}. In particular, the continuum limits of the
lattice derived matrix models of noncommutative field theory seem to
be most naturally associated with the trivial Gaussian fixed point of
the corresponding commutative theory~\cite{AC,BHN}. Notice, however,
that if we replace the bare coupling $g$ by the renormalized coupling
constant $\Lambda^{2n-4}\,g$, then the function
$\Lambda^{-2}\,W(\Lambda^{-2}\,\xi)$ has a more complicated form
reflecting the interacting nature of the quantum field theory,
provided we simply interpret $\Lambda$ as the finite mass scale set
by the magnetic field. Then there is no trouble going to the full
quantum field theory limit, and this suggests that an alternative
physical interpretation for the infrared limit may also be
given. This interpretation is consistent with the old folklore that
spacetime noncommutativity provides a means of regulating ultraviolet
divergences in quantum field theory.

We then exploit the duality discussed above to show that there exists
another scaling limit of the quantum field theory, leading to a
Green's function which is essentially the Fourier transform of
(\ref{Guvxysumm}) (up to rescalings). It corresponds to taking the
limit $N\to\infty$ with $\Lambda^{2n}=N^{-1}\,\sqrt{\det(B/4\pi)}$
fixed and it produces an ultra-local propagator
\beq
G_{\rm uv}(x,y)=-\delta^{(2n)}(x-y)~\Lambda^{-2}\,W\Bigl(
(4\pi\,\Lambda)^2+\Lambda^{-2}\,\mu^2\Bigr) \ , ~~
|x|\leq\Lambda^{-1}/\sqrt\pi \ ,
\label{Girxysumm}\eeq
where the parameter $\Lambda$ now has the interpretation of an
infrared cutoff. As remarked above, however, there is an alternative
interpretation of both of these limits in which $\Lambda$ may be
interpreted simply as the mass scale set by the background magnetic
field, i.e. $\Lambda^{2n}=\sqrt{\det(B/4\pi)}$. This follows from a
remarkable scale invariance of the regularized Green's functions (and
the underlying matrix model) which shows that, in the regularized
quantum field theory, the length scale is set by the magnetic length
$l_{\rm mag}=\det(B)^{-1/4n}$. The two scaling limits thus
described can be interpreted as zooming into short distances $\ll
l_{\rm mag}$ in space in the infrared limit leading to
(\ref{Guvxysumm}), and zooming out to large distances $\gg l_{\rm mag}$
in the ultraviolet limit leading to (\ref{Girxysumm}). With this
latter physical interpretation, one can proceed to analyse whether
there exist well-defined intermediate limits with no zooming in or out
of space at all. We thus find that the renormalization of the
interacting noncommutative field theory depends crucially on the
particular scaling limits that one takes, and also on their physical
interpretations.

We will also describe some other extensions of the present class of
models which have the potential of providing exactly solvable,
renormalizable, and interacting noncommutative field theories. In
particular, we present a detailed construction of a supersymmetric
extension of the noncommutative scalar field theory in a
background magnetic field. One of the most interesting aspects of
these models that we shall find is that even at the level of the
non-interacting theory, noncommutative geometry appears to play a
role, in that the supersymmetry transformations are parametrized by
elements of an infinite-dimensional, noncommutative algebra
(or equivalently by infinite matrices). We use this observation to
explicitly construct the most general naively renormalizable,
noncommutative supersymmetric field theory, while it does not seem
possible to construct local actions which possess this infinite parameter
supersymmetry. This model thereby represents a supersymmetric
extension of noncommutative field theory for which, like the scalar
models studied in this paper, no commutative counterpart exists and
noncommutativity is an intrinsic feature. We will not deal with the
problem of actually solving any of these generalizations in this
paper. Their exact solutions are left for future work which has the
potential of deepening our insights into the basic structures
underlying noncommutative quantum field theory.

\subsection{Outline}

In the ensuing sections we shall give detailed derivations of the
results reported above, and describe many other interesting features
of these noncommutative field theories. The structure of the remainder
of this paper is as follows. In Section~2 we give the precise
definition of the quantum field theory, describe its symmetries, and
briefly review the duality of~\cite{LS}. In Section~3 we derive in
detail, following~\cite{L}, the relationship between the quantum field
theory and a large $N$ matrix model, focusing our attention on the
two-dimensional case for simplicity and ease of notation. We then use
this mapping to describe the nonperturbative regularization that will be
employed for most of this paper. In Section~4 we prove that the
regularized field theory is exactly solvable by deriving a closed
formula for the finite $N$ partition function, and relate it to the
Toda lattice hierarchy. We also derive and explicitly solve the
equations of motion in the large $N$ limit, and then evaluate the
exact vacuum amplitude of the field theory. In Section~5 we study the
Green's functions of the field theory in two dimensions, starting with
those for which the free part of the action is dropped. All
connected Green's functions are obtained in this limit and
are invariant under Fourier transformation, as expected by duality. We
then give a detailed derivation of the two-point function obtained
in~\cite{lsz} in the infrared limit described above, and further argue that
the connected correlation functions of the field theory are all
trivial in this limit. In Section~6 we derive the generalizations of
the preceeding results to arbitrary even spacetime dimensionality. In
Section~7 the various other possible scaling limits of the quantum field theory
are explored, including the infrared limit, and the different physical
interpretations of the infrared and ultraviolet limits. We also
briefly describe how non-trivial interactions can be obtained by
(small) perturbations of the model. In Section~8 we present the
detailed construction of the relativistic fermion version of the
scalar noncommutative field theory, and combine it with the bosonic
model to construct a supersymmetric, interacting noncommutative field
theory in a background magnetic field. Appendix~A describes an
alternative interpretation of the model as a particular generalization
of mean field theory for conventional (commutative) complex
$\Phi^4$-theory, providing another motivation for the present work. In
Appendix~B we sketch a ``stringy'' interpretation of the duality of
the model. The remaining Appendices~C--G contain more technical
details of the calculations.

\newsection{Formulation of the Model}

In this section we will start by defining precisely the noncommutative
quantum field theory that we shall study in this paper. We shall also
present a detailed analysis of the symmetries possessed by
this model, and describe in what sense it may be regarded as an
exactly solvable quantum field theory.

\subsection{Definitions\label{BosDefs}}

We will consider the class of noncommutative scalar field theories
defined by classical actions of the form
\beq
\tilde S_{\rm b}=\int\dd^{2n}x~\Bigl[\Phi^\dag(x)
\left(-\sigma\,{\sf D}^2-\tilde\sigma\,\tilde{\sf D}^2+
\mu^2\right)\Phi(x)+V_\star\left(\Phi^\dag\star\Phi\right)(x)\Bigr] \ ,
\label{SVstartilde}\eeq
where $\sigma+\tilde\sigma\geq0$ and
\bea
{\sf D}_i&=&\partial_i-\ii B_{ij}\,x^j \ , \label{DiBfield}\\
\tilde{\sf D}_i&=&\partial_i+\ii B_{ij}\,x^j
\label{DiBfieldtilde}\eea
with $\partial_i=\partial/\partial x^i$, $i=1,\dots,2n$. Here $\Phi(x)$ is a
massive, complex scalar field of mass dimension $n-1$ coupled minimally to a
constant, background magnetic field proportional to the $2n\times2n$
real, non-degenerate antisymmetric matrix $B=(B_{ij})$, and in the
background of a confining electric potential proportional to
$(Bx)^2$. Throughout we work on even-dimensional
Euclidean spacetime $\real^{2n}$, and we use the notation
$\Phi^\dag(x)\equiv\overline{\Phi(x)}$ with the bar indicating complex
conjugation. The interaction term is defined by a
potential of the general form
\beq
V(w)=\sum_{k\geq2}\frac{g_k}k\,w^k \ ,
\label{Vpolyr}\eeq
and the subscript $\star$ means that the real scalar fields
$X(x)=\Phi^\dag\star\Phi(x)$ are multiplied together in
$V_\star(X)(x)$ using the usual Groenewold-Moyal star-product,
$(X)_\star^k\equiv X\star X\star\cdots\star X$ ($k$ times), where
\beq
f\star f'(x)=(2\pi)^{-4n}\,\int\dd^{2n}k~\int\dd^{2n}q~\tilde f(k)\,
\tilde f'(q)~\e^{\ii k\cdot\theta q}~\e^{\ii(k+q)\cdot x} \ .
\label{fstarg}\eeq
Here $\tilde f$ is the Fourier transform of the field $f$ with the
convention
\beq
\tilde f(k)=\int\dd^{2n}x~\e^{-\ii k\cdot x}\,f(x) \ ,
\label{Fouriertransfdef}\eeq
and for any two momentum space vectors $k$ and $q$ we have defined
their skew-product by
\beq
k\cdot\theta q=\theta^{ij}\,k_i\,q_j \ .
\label{wedgemom}\eeq
We will assume that the noncommutativity parameter matrix
$\theta=(\theta^{ij})$ is non-degenerate. We also define
\beq
x\cdot By=B_{ij}\,x^i\,y^j \ .
\label{wedgepos}\eeq

Most of the results we shall describe in the
following will be obtained from the noncommutative complex $\Phi^4$
field theory with interaction potential given by
\beq
V(w)=\frac g2\,w^2 \ ,
\label{Phi4pot}\eeq
and in the special instances when either $\sigma$ or $\tilde\sigma$
vanishes with $\sigma+\tilde\sigma=1$. These cases correspond to
charged particles in a background magnetic field alone. In
Section~\ref{BareGF} we shall consider the case where both
$\sigma=\tilde\sigma=0$. The cases $\sigma,\tilde\sigma>0$ will have
the effect of removing the degeneracies of the Landau levels in the
magnetic field problem. The special case $\sigma=\tilde\sigma=\frac12$
in (\ref{SVstartilde}) corresponds to scalar fields in a harmonic
oscillator potential alone and is closest to the conventional
noncommutative field theories with no background magnetic field. The
addition of such a harmonic oscillator potential to standard
noncommutative $\Phi^4$-theory is in fact necessary for its
renormalization to all orders in perturbation
theory~\cite{GW}. However, most of our analysis will focus on the
instance where $\sigma=1$, $\tilde\sigma=0$, and the interaction
potential is given by (\ref{Phi4pot}), i.e. the noncommutative field
theory with action
\beq
S_{\rm b}=\int\dd^{2n}x~\Bigl[\Phi^\dag(x)
\left(-{\sf D}^2+\mu^2\right)\Phi(x)+\frac g2\,\Phi^\dag\star\Phi
\star\Phi^\dag\star\Phi(x)\Bigr] \ .
\label{SVstar}\eeq

At the quantum level, the field theory is defined by the partition function
\beq
Z_{\rm b}=\int\DD\Phi~\DD\Phi^\dag~\e^{-\tilde S_{\rm b}}
\label{partfncont}\eeq
and the non-vanishing Green's functions
\bea
G_{\rm b}^{(2r)}(x_1,\dots,x_r;y_1,\dots,y_r)&=&\Bigl\langle\Phi^\dag(x_1)\,
\Phi(y_1)\cdots\Phi^\dag(x_r)\,\Phi(y_r)\Bigr\rangle\nn\\&\equiv&
\frac1{\tilde
  Z_{\rm b}}\,\int\DD\Phi~\DD\Phi^\dag~\prod_{I=1}^r\Phi^\dag(x_I)\,
\Phi(y_I)~\e^{-\tilde S_{\rm b}} \ ,
\label{Greensfnscont}\eea
with the usual, formal functional integration measure
$\DD\Phi~\DD\Phi^\dag=\prod_x\dd\,{\rm Re}\,\Phi(x)~\dd\,{\rm
  Im}\,\Phi(x)$. The connected parts of the Green's functions
(\ref{Greensfnscont}) may be extracted from the generating functional which is
defined by the functional integral with the coupling of the scalar
fields to external sources $J(x)$ and $\overline{J}(x)$,
\beq
Z_{\rm b}\left[J,\overline{J}\,\right]=
\int\DD\Phi~\DD\Phi^\dag~\e^{-\tilde S_{\rm b}+
\int\dd^{2n}x~\bigl[\Phi^\dag(x)\,J(x)+\overline{J}(x)\,\Phi(x)
\bigr]} \ .
\label{ZbJgen}\eeq
We then have
\beq
G_{\rm b\,,\, conn}^{(2r)}(x_1,\dots,x_r;y_1,\dots,y_r)=\left.
\prod_{I=1}^r\frac\delta{\delta J(x_I)}\frac\delta
{\delta\overline{J}(y_I)}\ln\frac{
  Z_{\rm b}\left[J,\overline{J}\,\right]}
{Z_{\rm b}}\right|_{J=\overline{J}=0} \ .
\label{Greensfnsconn}\eeq

To make the definition of the functional integral measure somewhat more
precise we will expand the fields in a convenient basis of
functions. Formally, one can choose any complete orthonormal set in
the Hilbert space of $L^2$-functions on $\real^{2n}$. However,
different choices of basis lead to different regularization
prescriptions for the quantum field theory. For example, if one uses
Fourier expansion of fields in a basis of plane waves $\e^{\ii k\cdot
  x}$, then the natural regularization imposed is the restriction of
momenta $k\in\real^{2n}$ to the annulus $\Lambda_0\leq
|k|\leq\Lambda$, where $\Lambda_0$ is interpreted as an infrared
cutoff and $\Lambda$ as an ultraviolet cutoff. Removing the cutoffs
then amounts to taking the limits $\Lambda_0\to0$ and
$\Lambda\to\infty$. This procedure is known to be rather difficult to
carry out in noncommutative quantum field theory, and has in fact been
done only at low orders in perturbation theory due to the
complications set in by UV/IR mixing. In what follows we will
circumvent such problems by using a different basis of functions that
enables us to make sense of the quantum field theory non-perturbatively.

\subsection{Spacetime Symmetries\label{Symms}}

Let us consider the spacetime symmetries of the field theory
defined by the action (\ref{SVstartilde}). If $n=1$, then the theory is
rotationally invariant, because a constant magnetic field is
rotationally invariant in two dimensions. However, parity is broken by
the magnetic field. A parity transformation amounts to changing the
sign of the magnetic field, $B\to-B$, or equivalently to interchanging
$\sigma\leftrightarrow\tilde\sigma$. What is much less obvious is that, in any
dimension, the model is also translationally invariant for
$\tilde\sigma=0$. For this, we accompany a translation of the scalar
field $\Phi(x)$ by a gauge transformation in the constant magnetic field,
\beq
\Phi_a(x)\equiv\e^{\ii a\cdot Bx}\,\Phi(x+a) \ .
\label{Phiaxgauge}\eeq
Note that the product on the right-hand side of (\ref{Phiaxgauge}) is just
{\it ordinary} pointwise multiplication of fields. Under
(\ref{Phiaxgauge}), the covariant derivative of $\Phi$ transforms
homogeneously,
\beq
{\sf D}_i\Phi_a=({\sf D}_i\Phi)_a \ .
\label{Dihomogeneous}\eeq
An elementary calculation using (\ref{fstarg}) shows that the
star-product also transforms in a simple way as
\beq
\Phi_a^\dag\star\Phi_a^\nd(x)=\Phi^\dag\star\Phi\Bigl(x+(\id-\theta B)\,a
\Bigr) \ .
\label{Phiastartransf}\eeq
It follows that the action (\ref{SVstartilde}) with $\tilde\sigma=0$
is invariant under the substitution of $\Phi$ by $\Phi_a$. We will
refer to this symmetry as ``magnetic translation invariance''.

If this symmetry persists at the quantum level, then the two-point
Green's function $G(x,y)=\langle\Phi^\dag(x)\,\Phi(y)\rangle$ assumes
the form
\beq
G(x,y)=\e^{-\ii x\cdot By}\,g(x-y)
\label{Gxytranslinv}\eeq
with $g(x)\equiv G(x,0)$ a function only of one variable. This implies
that the two-point function takes the same form in momentum
space. Computing the Fourier transform of (\ref{Gxytranslinv}) gives
\bea
\tilde G(p,q)&\equiv&\int\dd^{2n}x~\e^{\ii p\cdot x}\,\int\dd^{2n}y~
\e^{-\ii q\cdot y}\,G(x,y)\non&=&\det\left(2\pi B^{-1}\right)~
\e^{-\ii q\cdot B^{-1}p}\,g\Bigl(B^{-1}(q-p)\Bigr)~=~\det\left(
  2\pi B^{-1}\right)\,G\left(B^{-1}q,B^{-1}p\right) \ , \non&&
\label{Gpqtranslinv}\eea
which for $B=\theta^{-1}$ is a special case of the duality to be
discussed in Section~\ref{UVIR}. To interpret our later results it is
important to note that the limits $B\to0$ and $B\to\infty$ are
singular. At finite $B$ the Green's functions in position and momentum
space are both translationally invariant up to a phase. However, in
the limit $B\to0$ only the translational symmetry in position space is
maintained, with $G(x,y)\to g(x-y)$. In momentum space this limit is
ill-defined, but a direct computation yields $\tilde
G(p,q)=(2\pi)^{2n}~\delta^{(2n)}(p-q)~\tilde g(p)$ and the momentum
space Green's function becomes ultra-local in this limit. An analogous
remark applies to the limit $B\to\infty$ with the roles of $G$ and
$\tilde G$ interchanged.

One can now ask if the full symplectomorphism group of $\real^{2n}$ is
a symmetry of the field theory, as it is in the case of noncommutative gauge
theories~\cite{LSZ}. For this, we use the Weyl-Wigner correspondence
to associate, for each Schwartz function $f(x)$ on
$\real^{2n}\to\complex$, a corresponding compact Weyl operator
$\hat{f}$ by
\eq
\hat f\equiv f(\hat x)=
\int\frac{\dd^{2n}k}{(2\pi)^{2n}}~\tilde f(k)~\e^{\ii k\cdot\hat x} \ ,
\label{Weylopf}\eqend
where $\hat x^i$ are the Hermitian generators of the Heisenberg algebra
of $n$ degrees of freedom which is defined by the commutation
relations
\eq
\left[\hat{x}^i\,,\,\hat{x}^j\right]=2\ii\theta^{ij} \ .
\label{Heisencomm}\eqend
We will represent this algebra on a separable Hilbert space ${\cal
  H}$. With this correspondence, we can rewrite the action
  (\ref{SVstartilde}) by using the usual rules
\bea
\widehat{f\star g}&=&\hat f\,\hat g~=~f(\hat x)\,g(\hat x) \ ,
  \label{fstargrule}\\
\widehat{\overline{f}}&=&\hat f^\dag \ , \label{fdagrule}\\
\widehat{\partial_if}&=&\frac1{2\ii}\,\left(\theta^{-1}\right)_{ij}\,
\left[\hat x^j\,,\,\hat f\right] \ , \label{dfrule}\\
\int\dd^{2n}x~f(x)&=&\sqrt{\det(4\pi\theta)}~\Tr^{~}_{\cal H}\,(\hat f\,) \ ,
\label{intfrule}\eea
where the dagger in (\ref{fdagrule}) denotes the Hilbert space adjoint
and the trace of operators over $\cal H$ in (\ref{intfrule}) is
defined such that
\beq
\Tr^{~}_{\cal H}\left(\e^{\ii p\cdot\hat x}\right)=\frac{(2\pi)^{2n}}
{\sqrt{\det(4\pi\theta)}}~\delta^{(2n)}(p) \ .
\label{tracenorm}\eeq
By using (\ref{fstargrule}) and (\ref{dfrule}), along with the
star-product identity
\beq
x^i\star f(x)=x^i\,f(x)+\ii\theta^{ij}\,\partial_jf(x) \ ,
\label{xistarfxid}\eeq
we may also derive the correspondence
\beq
\widehat{x^i\,f}=\frac12\,\Bigl(\hat x^i\,\hat f+\hat f\,\hat
x^i\Bigr) \ .
\label{xifrule}\eeq

It follows that the actions of the covariant derivatives in
(\ref{SVstartilde}) on the Hilbert space $\cal H$ are given by
\bea
\widehat{{\sf D}_if}&=&-\frac\ii2\,\Bigl(\left(\theta^{-1}
  \right)_{ij}-B_{ij}\Bigr)\,\hat x^j\,\hat
  f+\frac\ii2\,\Bigl(\left(\theta^{-1}\right)_{ij}+B_{ij}\Bigr)\,\hat
  f\,\hat x^j \ , \non\widehat{\tilde{\sf D}_if}&=&
  -\frac\ii2\,\Bigl(\left(\theta^{-1}
  \right)_{ij}+B_{ij}\Bigr)\,\hat x^j\,\hat f+
  \frac\ii2\,\Bigl(\left(\theta^{-1}
  \right)_{ij}-B_{ij}\Bigr)\,\hat f\,\hat x^j \ .
\label{Difrules}\eea
We see that the points where
$B\,\theta=\pm\,\id$ have the special property that only one of the
terms on the right-hand sides of the equalities in (\ref{Difrules}) is
non-zero, and the thus the action assumes a particularly simple
form. In particular, at the special point in parameter space where
\beq
B\,\theta=\id \ ,
\label{specialpt}\eeq
the action (\ref{SVstartilde}) is mapped under the above
correspondence to
\bea
\tilde S_{\rm b}&=&\sqrt{\det(4\pi\theta)}~\Tr^{~}_{\cal H}\left[
  \Phi^\dag(\hat x)\Bigl(\tilde\sigma\,(B\hat x)^2+\mu^2\Bigr)
  \Phi(\hat x)+\sigma\,\Phi^\dag(\hat x)\Phi(\hat x)
  (B\hat x)^2\right.\non&&\left.+\,
  V\Bigl(\Phi^\dag(\hat x)\Phi(\hat x)\Bigr)
  \right] \ ,
\label{SVcalH}\eea
and an analogous formula in the case $B\,\theta=-\id$ with the
interchange $\sigma\leftrightarrow\tilde\sigma$. At these points, and
only at these points, the action depends solely on the combinations
$\Phi^\dag(\hat x)\Phi(\hat x)$ and $\Phi(\hat x)\Phi^\dag(\hat
x)$. As we will see, this property makes the field theory exactly
solvable at the full quantum level. Note that the representations of
the two covariant derivatives in (\ref{Difrules}) also commute at
these special points, so that the Hilbert space $\cal H$ provides a
Morita equivalence bimodule for the corresponding Heisenberg algebras
that they generate. This equivalence is described from another point
of view in Appendix~B.

The action (\ref{SVcalH}) for $\tilde\sigma=0$ possesses a large
symmetry
\beq
\Phi(\hat x)~\longmapsto~U(\hat x)\,\Phi(\hat x) \ , ~~
\Phi^\dag(\hat x)~\longmapsto~\Phi^\dag(\hat x)\,U(\hat x)^{-1} \ ,
\label{SVlargesym}\eeq
where $U(\hat x)$ is any invertible operator on $\cal H$. In the
spacetime picture, this $GL(\infty)$ symmetry corresponds to the
complex, fundamental star-gauge transformations
\beq
\Phi(x)~\longmapsto~U\star\Phi(x) \ , ~~
\Phi^\dag(x)~\longmapsto~\Phi(x)\star U^{\star-1}(x) \ ,
\label{SVlargesymspace}\eeq
where $U(x)$ is a star-invertible field on $\real^{2n}$,
\beq
U\star U^{\star-1}(x)=U^{\star-1}\star U(x)=1 \ .
\label{Ustarinv}\eeq
With both $\sigma=\tilde\sigma=0$ the action has an even larger
$GL(\infty)\times GL(\infty)$ symmetry
\beq
\Phi(\hat x)~\longmapsto~U(\hat x)\,\Phi(\hat x)\,V(\hat x)^{-1} \ , ~~
\Phi^\dag(\hat x)~\longmapsto~V(\hat x)\,\Phi^\dag(\hat x)\,U(\hat x)^{-1} \ .
\label{SVlargersym}\eeq
As we will see, this $GL(\infty)$ symmetry drastically simplifies the
calculation of the Green's functions of the model. The unitary
subgroup of this huge symmetry corresponds to invariance
of the model under canonical transformations of
$\real^{2n}$~\cite{LSZ}. The symplectomorphism $W_{1+\infty}$-symmetry
at the critical point (\ref{specialpt}) has also been observed within
a different context in the corresponding noncommutative quantum
mechanics~\cite{DH1,Horvathy}.

A somewhat different interpretation of the special point
(\ref{specialpt}) may also be given. The actions of the covariant
momentum operators (\ref{DiBfield}) and (\ref{DiBfieldtilde}) may be
written, by using (\ref{xistarfxid}), as
\bea
{\sf D}_i\Phi(x)&=&(\id+B\,\theta)_i^{~j}\,\Bigl(\partial_j\Phi(x)+
\ii B^\prime_{jk}\,x^k\star\Phi(x)\Bigr) \ , \non
\tilde{\sf D}_i\Phi(x)&=&(\id-B\,\theta)_i^{~j}\,\Bigl(\partial_j\Phi(x)+
\ii B^{\prime\prime}_{jk}\,x^k\star\Phi(x)\Bigr) \ ,
\label{Distaraction}\eea
where
\beq
B^\prime=\frac1{\id+B\,\theta}\,B \ , ~~ B^{\prime\prime}=
\frac1{\id-B\,\theta}\,B \ .
\label{Bprimedefs}\eeq
This relationship is a very simple version of the Seiberg-Witten map
between commutative and noncommutative descriptions of the same
theory~\cite{SW}. One can either work with the new background magnetic
fields $B^\prime$ and $B^{\prime\prime}$ using explicit noncommutative
star-products in the definition of the Landau momentum operators, or
else one can work with the original magnetic field $B$ using the
ordinary commutative product. The two descriptions are the same up to
an overall rescaling. However, one of the transformations in
(\ref{Bprimedefs}) is always undefined at the points
$B=\pm\,\theta^{-1}$, and in this sense they can be interpreted as
points of ``maximal'' noncommutativity where the quantum field theory
becomes exactly solvable.\footnote{\baselineskip=12pt Note that the
  transformations (\ref{Bprimedefs}) can be inverted to give
$B=B^\prime(\id-\theta\,B^\prime)^{-1}$ and
$B=B^{\prime\prime}(\id+\theta\,B^{\prime\prime})^{-1}$, in which the roles
of the two special points $B^\prime=+\theta^{-1}$ and
$B^{\prime\prime}=-\theta^{-1}$ are interchanged. The Seiberg-Witten
transformation then asserts that the two points are equivalent, in
that they lead to the same noncommutative quantum field theory.}

\subsection{UV/IR Duality\label{UVIR}}

As demonstrated in~\cite{LS}, the noncommutative field theory in the
  case of a quartic interaction also possesses a novel duality
  symmetry. As this duality will be important for the physical
  interpretations of the scaling limits of the regularized field
  theory that we will obtain, we shall now summarize the duality
  transformation rules, refering to~\cite{LS} for the technical
  details. The action (\ref{SVstar}) has a duality under Fourier
transformation, i.e. it retains the same form when written in position
or momentum space. Explicitly, it is invariant under the duality
transformation
\bea
\Phi(x)&\longmapsto&\sqrt{\Bigl|\det(B/2\pi)\Bigr|}~\tilde\Phi(B\,x) \ , \non
\theta&\longmapsto&-B^{-1}\,\theta^{-1}\,B^{-1} \ , \non
g&\longmapsto&\Bigl|\det(B\,\theta)\Bigr|^{-1}\,g \ ,
\label{dualitytransf}\eeq
with all other parameters of the model unchanged. This symmetry
extends to the quantum level as a symmetry of rewriting the Green's
functions (\ref{Greensfnscont}) in momentum space,
\beq
\tilde G_{\rm b}^{(r)}(p_1,\dots,p_r;q_1,\dots,q_r)=
\Bigl|\det\left(2\pi B^{-1}\right)
\Bigr|^r~G_{\rm b}^{(r)}(x_1,\dots,x_r;y_1,\dots,y_r)
\label{Greensfnsduality}\eea
with $x_I=B^{-1}\,p_I$, $y_I=B^{-1}\,q_I$, $I=1,\dots,r$, and the
transformations (\ref{dualitytransf}) implicitly understood on the
right-hand side of (\ref{Greensfnsduality}). The key to
the proof of this fact is the existence of a regularization of the
quantum field theory which respects the duality. This is essentially
the regularization that will be exploited in the subsequent sections.
Note that at the special points $B=\pm\,\theta^{-1}$, the field theory is
essentially ``self-dual'' in that it is completely invariant under
Fourier transformation, and the duality identifies the two special
points. This is consistent with what was found in the previous
subsection. In Appendix~B we point out an alternative ``stringy''
interpretation of this duality at the special point.

\newsection{Matrix Model Representation\label{Matrix}}

We are interested in expressing the action (\ref{SVstartilde}) with respect
to the expansion of functions on $\real^{2n}$ in a suitable basis. To
simplify the presentation of our results we will first focus on the
two-dimensional case $n=1$. The generalizations of our results to
arbitrary even dimensionality is remarkably simple and will be described
later on in Section~\ref{HigherD}. In this section, and until
Section~\ref{HigherD}, we define $\theta=\theta^{12}$ and
$B=B_{12}$, both of which are assumed to be positive parameters. We
will begin by deriving the matrix form of the action (\ref{SVstartilde})
for $n=1$, and then use this representation to define the formal
functional integral of the quantum field theory as an integral over
infinite-dimensional matrices. This will then enable us to describe a
natural non-perturbative regularization of the model.

\subsection{Mapping to a Matrix Model\label{MatrixMap}}

Let us begin by introducing the complex coordinates
\beq
z=x^1+\ii x^2 \ , ~~ \overline{z}=x^1-\ii x^2 \ ,
\label{complexcoords}\eeq
and the ladder operators
\bea
{\sf a}=\frac12\,\left(\sqrt\theta~\partial+\frac{\overline{z}}{\sqrt\theta}
\right) \ &,& ~~ {\sf a}^\dag=\frac12\,\left(-\sqrt\theta~\overline{\partial}+
\frac{z}{\sqrt\theta}\right) \ , \non{\sf b}=\frac12\,\left(\sqrt\theta~
\overline{\partial}+\frac{z}{\sqrt\theta}\right) \ &,& ~~
{\sf b}^\dag=\frac12\,\left(-\sqrt\theta~\partial+
\frac{\overline{z}}{\sqrt\theta}\right) \ .
\label{ladderops}\eea
Here $\partial=\partial_1-\ii\partial_2$ and
$\overline{\partial}=\partial_1+\ii\partial_2$. The operators
(\ref{ladderops}) generate two commuting copies of the usual harmonic
oscillator commutation relations
\bea
\left[{\sf a}\,,\,{\sf a}^\dag\right]&=&\left[{\sf b}
\,,\,{\sf b}^\dag\right]~=~1 \ , \non
\left[{\sf a}\,,\,{\sf b}\right]&=&\left[{\sf a}\,,\,{\sf b}^\dag
\right]~=~0 \ .
\label{harmoscrels}\eea

The Landau states are defined by
\beq
|\ell,m\rangle=\frac{(\hat{\sf a}^\dag)^{\ell-1}}{\sqrt{(\ell-1)!}}\,
\frac{(\hat{\sf b}^\dag)^{m-1}}{\sqrt{(m-1)!}}|{\rm vac}\rangle
\label{Landaustates}\eeq
where $|{\rm vac}\rangle$ is the two-particle Fock vacuum with
$\hat{\sf a}|{\rm vac}\rangle=\hat{\sf b}|{\rm vac}\rangle=0$, and $\ell,m$ are
positive integers. The normalized Landau wavefunctions in position
space are given by
\beq
\phi_{\ell,m}(x)=\langle x|\ell,m\rangle \ .
\label{Landaueigenfn}\eeq
The ground state wavefunction $\phi_{0}(x)\equiv\phi_{1,1}(x)$ can be computed
explicitly by solving the differential equations ${\sf a}\,\phi_0=0$ and
${\sf b}\,\phi_0=0$ using (\ref{ladderops}) to get
\beq
\phi_0(x)=\frac1{\sqrt{\pi\theta}}~\e^{-|z|^2/2\theta} \ ,
\label{groundstate}\eeq
while in general they can be computed by introducing the generating function
\beq
{\cal P}_{s,t}(x)=\sum_{\ell,m=1}^\infty\,\frac{s^{\ell-1}}{\sqrt{(\ell-1)!}}\,
\frac{t^{m-1}}{\sqrt{(m-1)!}}~\phi_{\ell,m}(x) \ .
\label{Landaugenfn}\eeq
Using (\ref{ladderops})--(\ref{groundstate}), an elementary
calculation gives (see for example~\cite{L}, Appendix~A)
\beq
{\cal P}_{s,t}(x)=\frac1{\sqrt{\pi\theta}}~\e^{-st}~\e^{(sz+t\,\overline{z})/
\sqrt\theta}~\e^{-|z|^2/2\theta} \ ,
\label{genfnexpl}\eeq
and from (\ref{genfnexpl}) the Landau wavefunctions may be extracted as
\beq
\phi_{\ell,m}(x)=\left.\frac1{\sqrt{(\ell-1)!\,(m-1)!}}\,
  \frac{\partial^{\ell-1}}
{\partial s^{\ell-1}}\,\frac{\partial^{m-1}}{\partial t^{m-1}}{\cal P}_{s,t}(x)
\right|_{s=t=0} \ .
\label{Landauextract}\eeq

The advantage of working with the functions $\phi_{\ell,m}(x)$ is two-fold.
First of all, using the generating function (\ref{genfnexpl}) it is easy to
check that they form a complete orthonormal basis of one-particle
wavefunctions, and
\beq
\int\dd^2x~\overline{\phi_{\ell,m}}\star\phi_{\ell',m'}(x)=
\delta_{\ell\ell'}\,\delta_{mm'}
\label{Landaucomplete}\eeq
with $\overline{\phi_{\ell,m}}=\phi_{m,\ell}$. Secondly, a straightforward
computation using (\ref{genfnexpl}) and the representation
(\ref{fstarg}) yields the identity (see for example~\cite{L},
Appendix~A)
\beq
{\cal P}_{s,t}\star {\cal P}_{s',t'}=\frac1{\sqrt{4\pi\theta}}~
\e^{s't}~{\cal P}_{s,t'} \ ,
\label{Fstprojrel}\eeq
which is equivalent to the star-product projector relation
\beq
\phi_{\ell,m}\star\phi_{\ell',m'}=\frac1{\sqrt{4\pi\theta}}\,
\delta_{m\ell'}~\phi_{\ell,m'} \ .
\label{Landauprojrel}\eeq
This relation of course just reflects the well-known fact that the functions
$\sqrt{4\pi\theta}\,\phi_{\ell,m}$ provide the Wigner representation of the
rank~1 Fock space operators $\hat\phi_{\ell,m}=|\ell\rangle\langle m|$.

These facts suggest an expansion of the scalar fields in the action
(\ref{SVstartilde}) with respect to the Landau basis as
\beq
\Phi(x)=\sqrt{4\pi\theta}~\sum_{\ell,m=1}^\infty M_{\ell
  m}~\overline{\phi_{\ell,m}}(x) \ , ~~ \Phi^\dag(x)=\sqrt{4\pi\theta}~
\sum_{\ell,m=1}^\infty \overline{M_{\ell m}}~\phi_{\ell,m}(x) \ .
\label{PhiLandauexp}\eeq
It is natural to assemble the dimensionless complex numbers $M_{\ell
  m}$ into an infinite, square complex matrix $M=(M_{\ell
  m})_{\ell,m\geq1}$. To ensure convergence of the expansion
  (\ref{PhiLandauexp}), we will regard $M$ as a compact operator
  acting on the separable Hilbert space ${\cal H}={\cal S}(\nat)$ of
  Schwartz sequences $(a_m)_{m\geq1}$ with sufficiently rapid decrease as
  $m\to\infty$. Then, the projector relation (\ref{Landauprojrel})
  implies that the star-product of fields $\Phi(x)$ and $\Phi'(x)$
  corresponds to ordinary matrix multiplication~\cite{G-BV},
\beq
\Phi\star\Phi'(x)=\sqrt{4\pi\theta}~\sum_{\ell,m=1}^\infty
(M\cdot M')_{\ell m}~\overline{\phi_{\ell,m}}(x) \ ,
\label{PhiMatrixprod}\eeq
while (\ref{Landauprojrel}) along with the orthogonality relation
(\ref{Landaucomplete}) implies that spacetime averages of fields
correspond to traces over the Hilbert space $\cal H$,
\beq
\int\dd^2x~\Phi(x)=4\pi\theta\,\sum_{m=1}^\infty M_{mm}\equiv4\pi\theta\,
\Tr^{~}_{\cal H}(M) \ .
\label{PhiTrace}\eeq
The star-product powers $(\Phi^\dag\star\Phi)_\star^k$ in the interaction
potential $V_\star(\Phi^\dag\star\Phi)(x)$ may thereby be succinctly
written as traces of the matrices $(M^\dag M)^k$. For the kinetic energy
terms in (\ref{SVstartilde}), we use the definitions (\ref{DiBfield}) and
(\ref{ladderops}) to write
\bea
{\sf D}^2&=&-\frac1\theta\,\left[\left(1+B\theta\right)^2\left({\sf a}^\dag
{\sf a}+\frac12\right)+\left(1-B\theta\right)^2\left({\sf b}^\dag
{\sf b}+\frac12\right)\right.\non&&+\Biggl.\left(B^2\theta^2-1\right)
\left({\sf a}^\dag{\sf b}^\dag+{\sf a}{\sf b}\right)\Biggr] \ ,
\label{Boxab}\eea
along with the usual harmonic oscillator creation and annihilation relations
applied to the Landau wavefunctions (\ref{Landaueigenfn}) to get
\bea
{\sf a}\,\phi_{\ell,m}=\sqrt{\ell-1}~\phi_{\ell-1,m} \ &,& ~~
{\sf a}^\dag\,\phi_{\ell,m}=\sqrt{\ell}~\phi_{\ell+1,m} \ , \non
{\sf b}\,\phi_{\ell,m}=\sqrt{m-1}~\phi_{\ell,m-1} \ &,& ~~
{\sf b}^\dag\,\phi_{\ell,m}=\sqrt{m}~\phi_{\ell,m+1} \ .
\label{laddersonLandau}\eea
The operator $\tilde{\sf D}^2$ is obtained from (\ref{Boxab}) by a
reflection of the magnetic field $B\to-B$.

We thereby find that the total action (\ref{SVstartilde}) may be expressed
as the infinite-dimensional complex matrix model action
\bea
\tilde S_{\rm b}&=&\left(\sigma+\tilde\sigma\right)
\left(B^2\theta^2-1\right)\,\Tr^{~}_{\cal
    H}\left(\Gamma_\infty^\dag\,M^\dag\,\Gamma^\nd_\infty\,M+
    M^\dag\,\Gamma_\infty^\dag\,M\,\Gamma^\nd_\infty\right)
\nn\\&&+\,\Bigl(\sigma+\tilde\sigma+(\sigma-\tilde\sigma)
B^2\theta^2\Bigr)\,\Tr^{~}_{\cal H}\left(M\,{\cal
    E}\,M^\dag\right)\non&&+\,\Bigl(\sigma+\tilde\sigma-(\sigma-\tilde\sigma)
B^2\theta^2\Bigr)\,\Tr^{~}_{\cal H}\left(M^\dag\,{\cal
    E}\,M\right)\non&&+\,4\pi\theta\,\mu^2\,\Tr^{~}_{\cal H}
\left(M^\dag M\right)+4\pi\theta\,\Tr_{\cal H}^{~}\,V\left(M^\dag M\right) \ ,
\label{SVinfmatrix}\eea
where we have introduced the diagonal matrix ${\cal E}=({\cal E}_{\ell
  m})$ with
\beq
{\cal E}_{\ell m}=4\pi\left(\ell-\frac12\right)\,\delta_{\ell m}
\label{Lambdadef}\eeq
and the infinite shift matrix $\Gamma_\infty=(\Gamma_{\infty,\ell m})$ with
\beq
\Gamma_{\infty,\ell m}=\sqrt{4\pi(m-1)}~\delta_{\ell,m-1} \ .
\label{Gammadef}\eeq
Note that the noncommutativity parameter $\theta$ now only appears as
an explicit coupling parameter in the action
(\ref{SVinfmatrix}).  The noncommutativity of spacetime has simply
become the noncommutativity of matrix multiplication. The first three
traces in the action (\ref{SVinfmatrix}) are similar to
the clock and hopping terms that appear as the kinetic parts in the
lattice derived matrix models of noncommutative scalar field
theory~\cite{AMNS1}.

The maximal symmetry group of area-preserving
diffeomorphisms, which appears at the special points in
parameter space where either $\sigma=0$ or $\tilde\sigma=0$, and
$B\theta=\pm\,1$, is particularly transparent within
this formalism. At these points, the first term in (\ref{SVinfmatrix})
vanishes and the resulting action depends only on the combinations
$M^\dag M$ and $MM^\dag$. In particular, for $\tilde\sigma=0$ it
possesses the $GL(\infty)$ symmetry
\beq
M~\longmapsto~U\cdot M \ , ~~ M^\dag~\longmapsto~M^\dag\cdot U^{-1} \ .
\label{MUinftysym}\eeq
As we will discuss below, in this case the
noncommutative quantum field theory has the same fixed
point as the complex one-matrix model in the 't~Hooft limit. Note also
that in this case the Landau states diagonalize the Landau Hamiltonian
operator ${\sf D}^2$, with the standard harmonic oscillator spectrum
\beq
-{\sf D}^2\phi_{\ell,m}=4B\left(\ell-\frac12\right)\,\phi_{\ell,m} \ .
\label{Didiag}\eeq
The $GL(\infty)$ symmetry can thereby be physically understood as a
consequence of the infinite degeneracies of the Landau levels $\ell$, and it
acts through rotations of the magnetic quantum numbers $m$. The phase
space now becomes degenerate and the Landau wavefunctions depend
on only half of the position space coordinates, leading to a reduction
of the quantum Hilbert space at $B=\theta^{-1}$. Henceforth we shall
deal only with the quantum field theory defined at this special point.
Note that from (\ref{Boxab}) it follows that while the
${\sf D}^2$ operator at $B\theta=1$ is given by the ${\sf a},{\sf
  a}^\dag$ oscillators, the Hamiltonian $\tilde{\sf D}^2$ is given in
terms of the commutant ${\sf b},{\sf b}^\dag$ oscillators and the
eigenvalue equation (\ref{Didiag}) is modified to
\beq
-\tilde{\sf D}^2\phi_{\ell,m}=4B\left(m-\frac12\right)\,\phi_{\ell,m} \ .
\label{tildeDidiag}\eeq

\subsection{Regularization\label{Reg}}

Let us consider now the quantum field theory
in the matrix model representation (\ref{SVinfmatrix}), whereby the
scalar field variables can be identified with infinite matrices. The
Jacobian of this transformation is formally $1$, and the partition
function (\ref{partfncont}) is then given by the functional
integral for the $N=\infty$ complex matrix model. With this rewriting
we can now properly define the functional integration measure
appearing in (\ref{partfncont}). Namely, we replace the infinite
matrix variables by finite $N\times N$ matrices, and then take the
large $N$ limit of the matrix model. In other words, we restrict the
quantum numbers of the Landau wavefunctions to
$\ell,m=1,\dots,N$ with $N<\infty$. There are several advantages to
this matrix regularization. For example, as discussed in~\cite{LS}, a finite
matrix rank $N$ provides both a short distance and low momentum cutoff
simultaneously, and thereby avoids the mixing of ultraviolet and
infrared divergences which seem to make the analysis of noncommutative
perturbation theory to all orders hopeless. Introducing an ultraviolet
cutoff alone in the matrix form is not enough to regulate all possible
divergences. Ultraviolet and infrared divergences cannot be clearly
separated and one needs to regulate them both at the same time. This
point will be further elucidated in Section~\ref{Scaling}. In
this respect, the Landau basis for the expansion of noncommutative
fields is far superior to the conventional Fourier basis, in that it
naturally leads to a non-perturbative regularization which avoids
dealing with the usual difficulties of noncommutative quantum field
theory. It is important to note though that this matrix regularization
does not impose any {\it a priori} restrictions on the allowed noncommutativity
parameters $\theta$, in contrast to what occurs in lattice
regularization~\cite{AMNS1}.

However, while the matrix regularization is {\it close} to a cutoff in the
kinetic energy term of the action (\ref{SVstartilde}), $N$ cannot itself be
an ultraviolet cutoff, because it is not determined by any mass scales
of the theory. Without a dimensionful scale there can be no
dimensional transmutation of the coupling constants of the model,
which is the essence of renormalization in quantum field theory. The
renormalization ensures finiteness of physical quantities and requires
some set of observables to define renormalization conditions. The way
to define the theory is explained in~\cite{MRW} and consists of first
introducing the kinetic energy cutoff, then taking the limit
$\theta\to\infty$ with the cutoff scaled such that it remains finite,
and finally removing the cutoff. This limit is equivalent to the
conventional large $N$ limit of the complex matrix
model~\cite{tHooft,BIPZ}. In other words, we should take the limit
$N\to\infty$ while keeping fixed the dimensionful ratio
\beq
\Lambda^2=\frac N{4\pi\theta} \ ,
\label{LambdaNtheta}\eeq
which simply defines the large $\theta$ limit of the model
(\ref{SVstartilde}). The true ultraviolet cutoff (\ref{LambdaNtheta}) has a
very natural physical interpretation. At $B=\theta^{-1}$, it is
associated with the energy $2B(2N-1)=16\pi\Lambda^2-2B$ of the $N^{\rm
  th}$ Landau level, which remains finite in the large $N$ limit just
described. Since $B\to0$ as $N\to\infty$, the spacing between Landau
levels also vanishes. Thus taking the limit described above is
equivalent to filling the finite energy interval $[0,16\pi\Lambda^2]$
with infinitely many Landau levels and an infinite density of
states. This limit can be thereby regarded as a combination of those
found in the corresponding noncommutative quantum mechanics
in~\cite{DH,DH1,Horvathy}, whereby the Laughlin theory of the lowest
Landau level is recovered, and in~\cite{NP}, where all Landau levels
survive.

The large $\theta$ limit of the model (\ref{SVstartilde}) thereby defines a
field theory with a finite cutoff. The quantum field theory in this
limit coincides with the 't~Hooft limit of the matrix model and, as we
will see below, is exactly solvable (at the special point where
$B\theta=1$). At this stage, we should renormalize the regulated
quantum field theory, i.e. fine-tune the parameters of the model in
such a way that the limit $\Lambda\to\infty$ is well-defined and
non-trivial. For this, we rescale the massive coupling constants
$\mu^2$ and $g_k$ to the dimensionless ones
\beq
\tilde\mu^2=\frac{\mu^2}{\Lambda^2} \ , ~~ \tilde g_k=\frac{g_k}
{\Lambda^2} \ .
\label{dimlesscouplings}\eeq

Let us now specialize to the case $\sigma=1$, $\tilde\sigma=0$ in
(\ref{SVstar}). The functional integral (\ref{partfncont}) may then be
properly defined through the large $N$ limit of the matrix integral
\bea
Z_N(E)&=&\int\DD M~\DD M^\dag~
\e^{-N\,\Tr\bigl(M\,E\,M^\dag+\tilde V(M^\dag M)\bigr)} \ , \non
\DD M~\DD M^\dag&\equiv&\prod_{\ell,m=1}^N\frac{\dd M_{\ell m}~
\dd\overline{M_{\ell m}}}{\ii\pi} \ ,
\label{ZNEdef}\eeq
with the integration extending over the finite-dimensional linear
space of $N\times N$ complex matrices $M=(M_{\ell m})_{1\leq\ell,
  m\leq N}$, and
\beq
\tilde V(w)=\sum_{k\geq2}\frac{\tilde g_k}k\,w^k
\label{tildeVdef}\eeq
the renormalized interaction potential. Expectation values in the matrix model
(\ref{ZNEdef}) are defined in the usual way as
\beq
\Bigl\langle{\cal O}\left(M,M^\dag\right)\Bigr\rangle_E=
\frac1{Z_N(E)}\,\int\DD M~\DD M^\dag~
\e^{-N\,\Tr\bigl(M\,E\,M^\dag+\tilde V(M^\dag M)\bigr)}
{}~{\cal O}\left(M,M^\dag\right) \ ,
\label{ZNEavg}\eeq
such that they are well-defined and finite in the limit
$N\to\infty$. For later convenience, we have introduced in
(\ref{ZNEdef}) a more general complex matrix model depending on a
generic $N\times N$, Hermitian external field $E$, which after
calculation should be set equal to the diagonal matrix
\beq
\tilde{\cal E}_{\ell m}=\left(\frac{16\pi\left(\ell-\frac12\right)}
  N+\tilde\mu^2\right)\,\delta_{\ell m}
\label{tildecalE}\eeq
appropriate to the original field theory, with
$1\leq\ell,m\leq N$. This provides a concrete, non-perturbative
definition of the noncommutative quantum field theory. Of course
afterwards the limit in which the ultraviolet cutoff
(\ref{LambdaNtheta}) is removed should also be taken.

In what follows we will not be very precise about the convergence of
the functional integrals (\ref{ZNEdef}) and (\ref{ZNEavg}). In the regularized
theory, the convergence of the traces is obvious. This is very similar
to the situation in conventional quantum field theory, whereby one
usually has an action functional which is only well-defined on fields
which are sufficiently smooth or of Schwartz-type, while at the quantum
level the functional integral also takes into account less well-behaved
functions. At the algebraic level, the large $N$ limit should be
rigorously defined by using the appropriate analog of that described for
the noncommutative torus in~\cite{LLS}.

While the above prescription is not the only possible definition of
the quantum theory, we will see that it leads to a physically sensible
and well-defined formulation of the quantum field theory, which is
moreover exactly solvable. For this, it is crucial that an
overall factor of $N$ has scaled out in front of the action in
(\ref{ZNEdef}). With this convention, the correct scaling of the free
energy which is finite in the limit $N\to\infty$ is given by
\beq
\Xi_N(E)=-\frac1{N^2}\,\ln Z_N(E) \ .
\label{freeenergydef}\eeq
{}From a field theoretical point of view, this follows from the fact
that the free energy should be proportional to the number of
one-particle degrees of freedom. In a free scalar quantum field
theory, this would be given by
$\Tr\ln(-\partial^2+\mu^2)\propto\Lambda^2\,{\cal V}$, where $\cal V$
is the spatial area and $\Lambda^2$ the area in momentum space. By
using (\ref{LambdaNtheta}), this quantity is proportional to $N^2$ in
the large $N$ limit. We will examine the issue of other possible
scaling definitions of the matrix model representation in
Section~\ref{Scaling}.

It is important to note here that, unlike the conventional large
$\theta$ scalar field theories studied in~\cite{MRW} where derivative
terms in the action contribute only at leading order in $\frac1N$, the
model we solve here retains its kinetic energy terms exactly at
$N=\infty$. The exact solvability of the model is not ruined in this
case because of the persistent $GL(\infty)$ symmetry, contrary to the
cases of~\cite{MRW} where the kinetic energy breaks this symmetry.
In fact, already at the classical level one can see some important
differences. The classical solutions of the $\theta=\infty$ models
studied in~\cite{MRW} are given by Hermitian matrices $X$ obeying the
matrix equation $\tilde V'(X)=0$. They are of the form
\beq
X=\sum_\ell\mu_\ell~\hat\phi_{\ell,\ell} \ ,
\label{solitonX}\eeq
where $\mu_\ell$ are the distinct real critical points of the
potential $\tilde V(w)$. The corresponding field configurations
correspond to the standard $\theta=\infty$ GMS solitons~\cite{GMS} and
possess an infinite degeneracy under $U(\infty)$ rotations of the matrices $X$,
leading to a non-trivial soliton moduli space which is an
infinite-dimensional Grassmannian~\cite{GHS}. On the other hand, in the present
case the classical equations of motion are
\beq
\tilde{\cal E}+\tilde V'(X)=0 \ ,
\label{solitonM}\eeq
where $X=M^\dag M$. The degeneracy of the soliton solutions (and the
existence of the ensuing moduli space) is now lifted. The solutions of
(\ref{solitonM}) are still of the form (\ref{solitonX}), but now the
real numbers $\mu_\ell$ satisfy the equation $\tilde{\cal E}_{\ell\ell}+\tilde
V'(\mu_\ell)=0$. For any polynomial potential $V$, this latter
equation at large $\ell$ generically has only a single solution, since
$\tilde{\cal E}_{\ell\ell}\to\infty$ as $\ell\to\infty$ and $\tilde
V'(w)\simeq\tilde g_{{\rm deg}\,V}\,w^{{\rm deg}\,V-1}$ for
$|w|\to\infty$. Because of the kinetic energy term there exist only
finitely many real soliton solutions. As we will see in the following,
this fact also makes a big difference at the quantum level, wherein
the saddle-point solutions may be regarded as perturbations around the
GMS soliton solutions~\cite{MRW}, and the functional integral as an
integral over the soliton moduli space. Indeed, the expansion of
functions on $\real^2$ with respect to the Landau basis may be thought
of as a ``soliton'' expansion of fields on the noncommutative plane in
terms of projection operators and partial isometries of the Heisenberg
algebra.

As we will demonstrate in the next section, the crucial simplification
which occurs at the special point $B=\theta^{-1}$ is that the partition
function and observables of the matrix model depend only on the
eigenvalues of the external field $E$. This enables an explicit
solution of the matrix model. In Section~\ref{CorrFns} we will then
see that the computations of the Green's functions
(\ref{Greensfnscont}) boil down to extracting the matrix model
solution, and sums over Landau levels of certain combinations of the
Landau wavefunctions which give the spacetime dependence of the
correlation functions. This will then enable a non-perturbative
analysis of the possible scaling regimes in which the cutoff
(\ref{LambdaNtheta}) can be removed.

\newsection{Exact Solution\label{Exact}}

We will now show how to solve the quantum field theory  at the special
point in parameter space  where $B\theta=1$. At this point we can use
the correspondence with the large $N$ limit of the complex one-matrix
model to obtain an exact, non-perturbative solution of the noncommutative field
theory. We will begin by showing how to solve the general external
field problem defined by (\ref{ZNEdef}) at finite $N$, thereby
establishing the formal integrability of the model. Then we shall
demonstrate how to obtain the solution of (\ref{ZNEdef}) in the large
$N$ limit, and from this extract the solution of the noncommutative
field theory defined by the action (\ref{SVstar}).

\subsection{Integrability\label{Toda}}

We will first derive a very simple determinant formula for the
partition function (\ref{ZNEdef}) at finite $N$. While this expression
will not be particularly useful for extracting the solution of the
corresponding continuum noncommutative field theory, it will formally
prove that the model is exactly solvable and provide insights into the
potential extensions to more general exactly solvable noncommutative
quantum field theories. For this, we will consider the more general
class of noncommutative field theories with actions (\ref{SVstartilde}).
The regulated field theory is obtained through a modification of the
matrix model (\ref{ZNEdef}) defined by the partition function
\beq
Z_N\left(E\,,\,\tilde E\right)=\int\DD M~\DD M^\dag~
\e^{-N\,\Tr\bigl(M\,E\,M^\dag+M^\dag\,\tilde E\,M+
\tilde V(M^\dag M)\bigr)} \ ,
\label{ZNEtildeE}\eeq
where the external fields $E$ and $\tilde E$ have respective
eigenvalues $\lambda_\ell$ and $\tilde\lambda_m$,
$\ell,m=1,\dots,N$. The additional external field $\tilde E$
explicitly breaks the $U(N)$ symmetry of the original
model. Remarkably, this extended matrix model is nevertheless still
exactly solvable. For this, we use the polar coordinate decomposition
of a generic $N\times N$ complex matrix $M$ which enables us to write
\beq
M=U^\dag~{\rm diag}(r_1,\dots,r_N)~V \ ,
\label{polardecomp}\eeq
where $U$ and $V$ are $N\times N$ unitary matrices, and
$r_\ell\in[0,\infty)$. This transformation introduces a Jacobian
\beq
\DD M~\DD M^\dag=[\dd U]~[\dd V]~\prod_{\ell=1}^N\dd h_\ell~
\Delta_N[h]^2 \ ,
\label{DMJacobian}\eeq
where $[\dd U]$ denotes the invariant Haar measure on the unitary
group $U(N)$, $h_\ell=r_\ell^2$, and
\beq
\Delta_N[h]=\prod_{1\leq\ell<m\leq N}(h_\ell-h_m)
\label{Vandermonde}\eeq
is the Vandermonde determinant. Here and in the following we will not
write irrelevant numerical constants (depending only on $N$)
explicitly. Upon substituting (\ref{polardecomp}) and
(\ref{DMJacobian}) into (\ref{ZNEtildeE}), we observe
that the integrations over $U,V\in U(N)$ decouple over the two types
of external field terms in the integrand. With $H={\rm
  diag}(h_1,\dots,h_N)$, the integration over the angular
degree of freedom $U$ may be done by using the
Harish-Chandra-Itzykson-Zuber formula~\cite{H-C,IZ}
\beq
\int\limits_{U(N)}[\dd U]~\e^{-N\,\Tr(E\,U^\dag\,H\,U)}=
\frac1{\Delta_N[\lambda]\,\Delta_N[h]}~\det_{1\leq\ell,m\leq N}
\left(\e^{-N\lambda_\ell h_m}\right) \ .
\label{IZformula}\eeq
An analogous formula holds for the integration over
$V$ with $\tilde E$ and $\tilde\lambda_m$. By expanding the
determinants into sums over permutations of $N$ objects and using the
antisymmetry of the Vandermonde determinants (\ref{Vandermonde}), the
partition function (\ref{ZNEtildeE}) may thereby be written as the
eigenvalue model
\beq
Z_N\left(E\,,\,\tilde E\right)=\sum_{\pi\in S_N}\frac{(-1)^{|\pi|}}
{\Delta_N[\lambda]\,\Delta_N[\tilde\lambda]}
{}~\prod_{\ell=1}^N\,\int\limits_0^\infty\dd h_\ell~
\e^{-N\bigl[\tilde V(h_\ell)+h_\ell(\lambda_\ell+\tilde
\lambda_{\pi(\ell)})\bigr]} \ .
\label{ZNEafterIZ}\eeq

The eigenvalue integral (\ref{ZNEafterIZ}) illustrates the necessity of
the scaling $\theta\sim N$ in (\ref{LambdaNtheta}) in the large $N$
limit. It is the only scaling that correctly takes into account the
classical and quantum fluctuation terms in (\ref{ZNEafterIZ}), such that
the effective action for the $N$ eigenvalues $h_\ell$ is of order
$N^2$. It may be expressed in the compact form
\beq
Z_N\left(E\,,\,\tilde E\right)=\frac1{\Delta_N[\lambda]\,
\Delta_N[\tilde\lambda]}~\det_{1\leq\ell,m\leq N}\,
\Biggl(f\left(\lambda_\ell+\tilde\lambda_m\right)\Biggr) \ ,
\label{ZNEtildeEexact}\eeq
where we have introduced the special function
\beq
f(\lambda)=\int\limits_0^\infty\dd h~\e^{-N\bigl(\tilde V(h)+\lambda\,h
\bigr)} \ .
\label{specialfn}\eeq
Note that here $V$ can be any function such that the integral in
(\ref{specialfn}) makes sense.

One way to understand the exact solvability of this model is to note
that the partition function (\ref{ZNEtildeEexact}) provides a
non-trivial solution of the two-dimensional Toda lattice
hierarchy~\cite{Toda}.\footnote{\baselineskip=12pt The analysis in the
  remainder of this subsection is geared at integrability features. It is
  not required in the remainder of this paper and may be skipped. The
  important result which we will need is the derivation of
  (\ref{largeNWxeqn}). An alternative derivation of this equation is
  given in Appendix~C.} For any polynomial potential $V$, we may write
the determinant formula (\ref{ZNEtildeEexact}) formally as
\bea
Z_N\left(E\,,\,\tilde E\right)&=&\det_{0\leq\ell,m\leq N-1}\,
\left(~\oint\limits_{z=0}\frac{\dd z}{2\pi\ii z}~z^m\,
\oint\limits_{\tilde z=0}\frac{\dd\tilde z}{2\pi\ii\tilde z}~
\tilde z^\ell\,f\left(z^{-1}+\tilde z^{-1}\right)\right.\nn\\&&
\times\left.\prod_{k=1}^\infty
\e^{N(t_k\,z^k+\tilde t_k\,\tilde z^k)}\right) \ ,
\label{ZNEtildeEtau}\eea
where we have defined the time variables
\beq
t_k=\frac1{Nk}\,\sum_{\ell=1}^N\lambda_\ell^k \ , ~~
\tilde t_k=\frac1{Nk}\,\sum_{m=1}^N\tilde\lambda_m^k
\label{Todatimes}\eeq
with $k\geq1$. This formula is derived by using the Cauchy identity to
get
\beq
\prod_{k=1}^\infty\e^{N(t_k\,z^k+\tilde t_k\,\tilde z^k)}
=\prod_{\ell=1}^N\frac1{(1-z\,\lambda_\ell)(1-\tilde z\,\tilde
\lambda_\ell)}
\label{Cauchyid}\eeq
and expanding the integration contours in (\ref{ZNEtildeEtau}) to
catch the residues of the poles at $z=1/\lambda_\ell$ and $\tilde
z=1/\tilde\lambda_m$. This makes $Z_N(E,\tilde E)$ a tau-function
$\tau[t,\tilde t\,]$ of the integrable two-dimensional Toda lattice
hierarchy. Its large $N$ limit may thereby be determined from the
equations of the dispersionless Toda hierarchy~\cite{Todadisp},
although we shall not pursue this interesting line of approach to
solving the noncommutative field theory (\ref{SVstartilde}) in this
paper. A more transparent indication of the exact solvability of the
matrix model (\ref{ZNEtildeE}) is described in Appendix~C.

Let us now go back to the original model (\ref{ZNEdef}), which is the
$\tilde E=0$ limit of (\ref{ZNEtildeE}) and hence should correspond to
a particular reduction of the Toda lattice hierarchy above. In
this limit, the expression (\ref{ZNEtildeEexact}) for the matrix
integral is indeterminate. One can take care of this by regularizing
the $\tilde E\to0$ limit in any way that removes the degeneracy, and
applying l'H\^{o}pital's rule to (\ref{ZNEtildeEexact}). In this way
we get
\beq
Z_N(E)=\frac1{\Delta_N[\lambda]}~\det_{1\leq\ell,m\leq N}\,
\Bigl(f^{(\ell-1)}(\lambda_m)\Bigr) \ ,
\label{ZNEexact}\eeq
where
\beq
f^{(m)}(\lambda)=\frac1{(-N)^m}\,\frac{\partial^mf(\lambda)}
{\partial\lambda^m}=\int\limits_0^\infty\dd h~h^m~
\e^{-N\bigl(\tilde V(h)+\lambda\,h\bigr)} \ .
\label{specialfnderivs}\eeq
Thus once the function (\ref{specialfn}) is known explicitly, then so
is the partition function (\ref{ZNEdef}) for all $N$. The determinant
formula (\ref{ZNEexact}) implies that the partition function $Z_N(E)$
is a tau-function $\tau[t]$ for the integrable KP
hierarchy~\cite{KMMMZ} in the times $t_k$ of (\ref{Todatimes}) with
one-particle wavefunctions $f^{(0)}(\lambda^{-1})$,
$f^{(1)}(\lambda^{-1})$, $f^{(2)}(\lambda^{-1})$, $\dots$. This
follows from the fact that, for any polynomial potential $V$, the
functions (\ref{specialfnderivs}) have the correct asymptotics
\beq
N^{\ell-1}\,f^{(\ell-1)}\left(\lambda^{-1}\right)=\lambda^{\ell-1}\,
\Bigl(1+O\left(\lambda^{-1}\right)\Bigr)~~~~{\rm as}~~\lambda\to\infty
\label{fellasympt}\eeq
such that (\ref{ZNEexact}) is a KP tau-function. Note while generic
external field matrix models have other properties and correspond to
particular reductions of the KP hierarchy~\cite{KMMMZ}, the partition
function (\ref{ZNEdef}) is {\it exactly} a KP tau-function. Its
$W_{1+\infty}$ symmetry then manifests itself as the symmetry of the
noncommutative quantum field theory under area-preserving
diffeomorphisms which was described in Section~\ref{Symms}.

We can also see this connection at the level of the equations of
motion. The Schwinger-Dyson equations for the matrix model
(\ref{ZNEdef}) follow from the identities
\beq
\int\DD M~\DD M^\dag~\sum_{\ell=1}^N\frac\partial
{\partial M_{\ell m}}\left(M_{\ell m'}~\e^{-N\,\Tr\bigl(M\,E\,M^\dag+
\tilde V(M^\dag M)\bigr)}\right)=0 \ ,
\label{SDeqns}\eeq
which give rise to a system of differential equations for the
partition function (\ref{ZNEexact}),
\beq
\frac1N\,\sum_{\ell=1}^N\frac\partial{\partial E_{m\ell}}\left[
\tilde V'\left(-\frac1N\,\frac\partial{\partial E}\right)_{\ell m'}+
E_{m'\ell}\right]Z_N(E)=0 \ .
\label{DEZLambda}\eeq
Both the integration measure and action
in (\ref{ZNEdef}) are invariant under arbitrary unitary
transformations of $E$. The partition function $Z_N(E)=Z[\lambda]$ thereby
depends only on the $N$ eigenvalues $\lambda_\ell$ of the external
field $E$ and is symmetric under permutation of them (This is manifest
in the determinant form (\ref{ZNEexact})). Thus only $N$ of
the $N^2$ Schwinger-Dyson equations (\ref{DEZLambda}) are
independent. Writing the diagonalization explicitly as
\beq
E={\cal U}^\dag~{\rm diag}(\lambda_1,\dots,\lambda_N)~{\cal U}
\label{diagE}\eeq
with $\cal U$ unitary, by using the chain rule the derivatives appearing
in (\ref{DEZLambda}) may be rewritten via second order perturbation
theory according to
\beq
\frac\partial{\partial E_{mm'}}=\sum_{\ell=1}^N{\cal U}^\dag_{m'\ell}\,
{\cal U}_{\ell m}\,\frac\partial{\partial\lambda_\ell}+
\sum_{\ell'\neq\ell}\frac{{\cal U}_{\ell'm}\,{\cal U}_{m'\ell}^\dag}
{\lambda_{\ell'}-\lambda_\ell}\,\sum_{k=1}^N{\cal U}_{\ell k}^\dag\,
\frac\partial{\partial{\cal U}_{\ell'k}} \ .
\label{dEchainrule}\eeq
Let us now specialize to our main model of interest, the
noncommutative complex $\Phi^4$ field theory defined by the potential
(\ref{Phi4pot}). In this case (\ref{specialfn}) is a transcendental
error function that can be written in the form~\cite{ASHandbook}
\beq
f(\lambda)=\frac1{\zeta+\sqrt{\zeta^2+\alpha(\zeta)}} \ , ~~ \zeta=
\lambda~\sqrt{\frac N{2\tilde g}} \ ,
\label{flambdaPhi4}\eeq
where $\alpha(\zeta)$ is a monotonic function that increases from
$\alpha(0)=\frac4\pi$ to $\alpha(\infty)=2$. It is important to note
though that the integrability property holds for any potential $V$. The
exact solvability of the noncommutative field theory is a consequence
of its underlying matrix representation, rather than of a special choice of
interaction.

In this case, written in terms of the eigenvalues, the $N$ independent
equations of motion read
\beq
\left[-\frac{\tilde g}{N^2}\,\frac{\partial^2}{\partial\lambda_m^2}-
\frac{\tilde g}{N^2}\,\sum_{\ell\neq m}\frac1{\lambda_\ell-
\lambda_m}\,\left(\frac\partial{\partial\lambda_\ell}-\frac\partial
{\partial\lambda_m}\right)+\frac1N\,\lambda_m\,\frac\partial
{\partial\lambda_m}+1\right]Z[\lambda]=0 \ .
\label{SDNlambda}\eeq
Contracting (\ref{SDNlambda}) with $\lambda_m^{-k'}$ and collecting
coefficients of the terms $\lambda_m^{-k}/N^2$ leads after some algebra
to a set of differential constraint equations in terms of the time
variables $t_k$ of (\ref{Todatimes}) of the form
\beq
\sum_{k=2}^\infty t_{k+k'-2}~{\sf L}_k[t]Z[\lambda]=0 \ ,
\label{Virconstrs}\eeq
where
\beq
{\sf L}_k[t]=\tilde g\,\sum_{n=1}^{k-1}\frac\partial{\partial t_n}
\,\frac\partial{\partial t_{k-n}}+\sum_{n=1}^\infty n\,t_n\,
\frac\partial{\partial t_n}+\frac\partial{\partial t_k}+
\frac\partial{\partial t_{k-2}}-2\,\delta_{k,2} \ .
\label{Virgens}\eeq
These equations coincide with the Virasoro constraints of the KP
hierarchy~\cite{KMMMZ}. However, while all of this provides a nice
characterization of the exact solvability of the underlying matrix
model, it is not very convenient for dealing explicitly with its large
$N$ limit, which is what is needed for the precise connection with the original
noncommutative field theory. The way to tackle the $N\to\infty$ limit
is the subject of the next subsection.

\subsection{The Master Equation\label{Master}}

To deal with the solution of (\ref{SDNlambda}) in the
large $N$ limit, we define the function
\beq
W(\lambda_m)=\frac1N\,\frac{\partial\ln Z[\lambda]}
{\partial\lambda_m} \ ,
\label{WxlnZdef}\eeq
and introduce the spectral density of eigenvalues of the external
matrix $E$,
\beq
\rho(\lambda)=\frac1N\,\Tr\,\delta(\lambda-E)=
\frac1N\,\sum_{\ell=1}^N\delta(\lambda-\lambda_\ell) \ .
\label{rhoE}\eeq
At $N=\infty$, the eigenvalues $\lambda_\ell$ are assumed to be of
order~1, and (\ref{rhoE}) becomes a continuous function of $\lambda$ with
support on some finite interval $[a_1,a_2]$ on the real line. Setting
$\lambda_m=\lambda$ in (\ref{WxlnZdef}), a simple power counting argument
shows that the derivative $\partial W/\partial\lambda$ which arises in the
first term of (\ref{SDNlambda}) is of order $\frac1N$ and so can be dropped at
$N=\infty$. Then, the differential equations (\ref{SDNlambda}) become
\beq
-\tilde g\,W^2(\lambda)-\tilde g\,\int\limits_{a_1}^{a_2}\dd\lambda'~
\rho(\lambda')\,\frac{W(\lambda')-W(\lambda)}{\lambda'-\lambda}+
\lambda\,W(\lambda)+1=0 \ , ~~ \lambda\in[a_1,a_2] \ .
\label{largeNWxeqn}\eeq

It is straightforward to obtain a perturbative solution to the
non-linear integral equation (\ref{largeNWxeqn}) by writing a power
series expansion for the function (\ref{WxlnZdef}),
\beq
W(\lambda;\tilde g)=\sum_{k=0}^\infty\tilde g^k~W^{(k)}(\lambda) \ .
\label{Wpowerseries}\eeq
By substituting this expansion into (\ref{largeNWxeqn}), we can then compute
the iterative solution
\bea
W^{(0)}(\lambda)&=&-\frac1\lambda \ , \nn\\W^{(k)}(\lambda)&=&
\sum_{l=0}^{k-1}\,\frac{W^{(l)}(\lambda)\,W^{(k-l-1)}(\lambda)}
\lambda\nn\\&&+\,
\int\limits_{a_1}^{a_2}\frac{\dd\lambda'}\lambda~\rho(\lambda')~
\frac{W^{(k-1)}(\lambda')-W^{(k-1)}(\lambda)}{\lambda'-\lambda}
\label{loopiterative}\eea
with $\lambda\in[a_1,a_2]$ and $k\geq1$. The free solution
$W^{(0)}(\lambda)$ of course follows directly from the Gaussian,
$\tilde g=0$ partition function
\beq
Z_N^{(0)}(E)=\e^{-N\,\Tr\ln E} \ .
\label{freepartfn}\eeq
An alternative interpretation of the function (\ref{WxlnZdef}) as the resolvent
of a related
Hermitian one-matrix model is given in Appendix~C.

The exact, non-perturbative solution of (\ref{largeNWxeqn}) is worked
out in Appendix~D. For the logarithmic derivative of the partition
function, we find
\bea
W(\lambda_m)&=&\frac{\lambda_m}{2\tilde g}-
\frac{\sqrt{(\lambda_m-b_1)(\lambda_m-b_2)}}{2\tilde g}\nn\\&&-\,
\frac1{2N}\,\sum_{\ell=1}^N
\frac{\sqrt{(\lambda_\ell-b_1)(\lambda_\ell-b_2)}-
\sqrt{(\lambda_m-b_1)(\lambda_m-b_2)}}{(\lambda_m-\lambda_\ell)\,
\sqrt{(\lambda_\ell-b_1)(\lambda_\ell-b_2)}} \ .
\label{Wxfinal}\eea
The parameters $b_1$ and $b_2$ of this rational solution are
determined by the non-linear constraints
\bea
b_1+b_2&=&-\frac{2\tilde g}N\,\sum_{\ell=1}^N\frac1
{\sqrt{(\lambda_\ell-b_1)(\lambda_\ell-b_2)}} \ , \nn\\
3\left(b_1^2+b_2^2\right)+2\,b_1b_2+8\tilde g&=&-\frac{8\tilde g}N
\,\sum_{\ell=1}^N\frac{\lambda_\ell}{\sqrt{(\lambda_\ell-b_1)
(\lambda_\ell-b_2)}} \ .
\label{bcnonlinconstr}\eea

\subsection{The Exact Vacuum Amplitude\label{ExactVac}}

The solution of the previous subsection represents the large $N$ limit
of the general external field matrix model (\ref{ZNEdef}). For the
noncommutative field theory of interest, we should now set the
external field $E$ equal to (\ref{tildecalE}). For this, we shift the
eigenvalues $\lambda_m\mapsto\lambda_m-\tilde\mu^2$ in
(\ref{Wxfinal},\ref{bcnonlinconstr}), and set
\beq
\lambda_m=16\pi\,\frac mN \ ,
\label{NClambda}\eeq
where we ignore the zero-point energy shift which vanishes in the
large $N$ limit. We then need the function $W(\lambda)$ in the limit
$N\to\infty$, $m\to\infty$ with $\frac mN\in[0,1]$ fixed. The spectral
density
\beq
\rho(\lambda)=\frac1{16\pi}
\label{rho2D}\eeq
is flat, and it is supported on
$\lambda\in[0,16\pi]$. The resulting integrals appearing in
(\ref{Wxfinal},\ref{bcnonlinconstr}) can be done explicitly, and
after some algebra we arrive finally at
\bea
W_{\rm b}(\lambda)&=&\frac{\lambda+\tilde\mu^2}{2\tilde g}-
\frac{\sqrt{(\lambda-b_1)(\lambda-b_2)}}{2\tilde g}\non&&+\,\frac1{16\pi}\,
\ln\left|\frac{\sqrt{b_1(\lambda-b_2)}-\sqrt{b_2(\lambda-b_1)}}
{\sqrt{(b_1-16\pi)(\lambda-b_2)}-\sqrt{(b_2-16\pi)(\lambda-b_1)}}\right|
\ ,
\label{WxfinalNC}\eea
where the parameters $b_1$ and $b_2$ are determined through
\bea
b_1+b_2+2\tilde\mu^2&=&\frac{\tilde g}{4\pi}\,\ln\left|
\frac{\sqrt{b_1}-\sqrt{b_2}}{\sqrt{b_1-16\pi}-\sqrt{b_2-16\pi}}
\right| \ , \nn\\
(b_1+b_2)^2&=&\frac{\tilde g}{2\pi}\,\Bigl(\sqrt{b_1b_2}-
\sqrt{(b_1-16\pi)(b_2-16\pi)}\,\Bigr)-3\tilde\mu^4-8\tilde g \ .
\label{bcnonlinconstrNC}\eea

The free energy $-\ln Z_{\rm b}$ of the noncommutative quantum field theory,
representing the exact (connected) vacuum amplitude,
can now be obtained by integrating the function (\ref{WxfinalNC}) over
$\lambda\in[0,16\pi]$. Using the permutation symmetry of the finite
$N$ partition function $Z[\lambda]$ in the eigenvalues
$\lambda_\ell$ and the boundary conditions (\ref{bcnonlinconstrNC}),
we arrive after some algebra at
\bea
\ln Z_{\rm b}&=&\lim_{N\to\infty}\frac{\ln
  Z_N}{N^2}\non&&{~~~~}^{~~}_{~~}\non
&=&\frac1{32\pi}\,\left[\left(\sqrt{b_1b_2}+
\sqrt{(b_1-16\pi)(b_2-16\pi)}\,\right)\right.\nn\\&&\times
\ln4\left(\sqrt{b_1}-
\sqrt{b_2}\,\right)\left(\sqrt{b_1-16\pi}-\sqrt{b_2-16\pi}\,\right)
\nn\\&&-\left.\sqrt{b_1b_2}\,\ln b_1b_2-
\sqrt{(b_1-16\pi)(b_2-16\pi)}\,\ln(b_1-16\pi)(b_2-16\pi)\right]\nn\\&&
+\,\frac{2\pi}{\tilde g}\,\left(4\tilde\mu^2+64+b_1+b_2-2\,
\sqrt{(b_1-16\pi)(b_2-16\pi)}\,\right)\nn\\&&-\frac\pi{4\tilde g^2}
\,\left[(b_1+b_2)\Bigl((b_1+b_2)^2-3\tilde\mu^4\Bigr)+(b_1-b_2)^2
\Bigl(2(b_1+b_2)+4\tilde\mu^2\Bigr)\right] \ . \nn\\&&
\label{freeenergyNC}\eea
The loop amplitude (\ref{WxfinalNC}) is a multi-valued analytic
function on the complex $\lambda$-plane with a square root branch cut
along the interval $[b_1,b_2]$. From the constraint equations
(\ref{bcnonlinconstrNC}), it is straightforward to see that the two
branch points $b_1$ and $b_2$ are always complex-valued. Moreover, the
one-cut solution (\ref{WxfinalNC}) is an analytic function of the
parameters $\tilde\mu^2$ and $\tilde g$. Thus the matrix model
exhibits {\it no} non-trivial critical behaviour as the coupling
constant $\tilde g$ is varied, and there is no scaling limit at
large $N$ with an approach to a phase transition. This is completely
consistent with the fact that the regulated noncommutative field
theory already describes a {\it continuum} model. The problem of
removing the ultraviolet regularization (\ref{LambdaNtheta}) will be
analysed in the next section.

\newsection{Correlation Functions\label{CorrFns}}

In this section we shall investigate the spacetime dependence of
the Green's functions (\ref{Greensfnscont}) using the exact,
non-perturbative solution of the matrix model that we obtained in the
previous section. As an illustration of the power of working in the
Landau basis for the expansion of noncommutative fields, and also as a
warmup to some of the general effects that we will see, we shall first
investigate a novel property of these correlators in a particular
truncation of the field theory which has an extended
$GL(\infty)\times GL(\infty)$ symmetry. Then we will proceed to obtain
{\it exact} expressions, valid to all loop orders, for the connected
Green's functions of the noncommutative quantum field
theory. This will enable us to describe
explicitly the non-perturbative scaling limits of the model.

\subsection{Green's Functions in Static and Free Limits\label{BareGF}}

An important feature in the analysis of the Green's functions
(\ref{Greensfnscont}) is that we use {\it ordinary} products of the
fields. A natural question which then arises is as to what the difference
is between these functions and those defined using star-products, as
may seem more appropriate to a noncommutative field theory with a
star-unitary symmetry. To get a flavour for the answer to this
question, we will find an $L^2$-integration kernel ${\cal
  G}_L(x_1,\dots,x_L)$ with the property that
\beq
\int\dd^2x~f_1\star\cdots\star f_L(x)=\int\dd^2x_1~\cdots
\int\dd^2x_L~{\cal G}_L(x_1,\dots,x_L)~f_1(x_1)\cdots f_L(x_L)
\label{calGdef}\eeq
for any collection of $L$ fields $f_1,\dots,f_L$. For $L=2$ this
kernel was computed in~\cite{LS}. The identity (\ref{calGdef})
thereby produces a convolution formula
\bea
&&\prod_{I=1}^r\,\int\dd^2x_I~\int\dd^2y_I~{\cal G}_{2r}(x_1,y_1,
\dots,x_r,y_r)~G_{\rm b}^{(r)}(x_1,\dots,x_r;y_1,\dots,y_r)
\nn\\&&~~~~~~~~~~=~~\int\dd^2x~\Bigl\langle\left(\Phi^\dag\star\Phi
\right)^r_\star(x)\Bigr\rangle
\label{convolution}\eea
for the integrated Green's functions of the noncommutative quantum field
theory in terms of correlators of composite operators.

The kernel ${\cal G}_L$ bears a remarkable relation to the
matrix model constructed in this paper. Defining
$\delta_y(x)=\delta^{(2)}(x-y)$, from the definition
(\ref{calGdef}) it follows that
\beq
{\cal G}_L(x_1,\dots,x_L)=\int\dd^2x~\delta_{x_1}\star\cdots\star
\delta_{x_L}(x) \ .
\label{calGdeltas}\eeq
Using completeness of the Landau wavefunctions on $\real^2$ to write
\beq
\delta_y(x)=\sum_{\ell,m=1}^\infty\phi_{\ell,m}(x)\,\phi_{m,\ell}(y)
\label{Landaudelta}\eeq
and the projector relations (\ref{Landauprojrel}), we may compute the
star-products appearing in (\ref{calGdeltas}) according to
\beq
\delta_{x_1}\star\delta_{x_2}(x)=\frac1{\sqrt{4\pi\theta}}~
\sum_{\ell,m=1}^\infty\phi_{\ell,m}(x)\,\sum_{k=1}^\infty
\phi_{\ell,k}(x_2)\,\phi_{k,m}(x_1) \ .
\label{deltaxstars}\eeq
By doing this an additional $L-2$ times in (\ref{calGdeltas}), and then
integrating over $x$ using the orthonormality relation
(\ref{Landaucomplete}), we arrive at a trace formula for
products of the Landau wavefunctions,
\bea
{\cal G}_L(x_1,\dots,x_L)&=&\left(\frac1{\sqrt{4\pi\theta}}
\right)^{L-2}~\sum_{m_1,\dots,m_L=1}^\infty
\phi_{m_1,m_2}(x_L)\,\phi_{m_2,m_3}(x_{L-1})\cdots\phi_{m_L,m_1}(x_1)
\nn\\&\equiv&\left(\frac1{\sqrt{4\pi\theta}}
\right)^{L-2}~\Tr^{~}_{\cal H}\Bigl(\phi(x_L)\,\phi(x_{L-1})\cdots
\phi(x_1)\Bigr) \ .
\label{Landautraceformula}\eeq
The physical significance of this formula is clear. Before
regularization and at $B=\theta^{-1}$, the $\theta\to\infty$ limit of
the matrix model action (\ref{SVinfmatrix}) is an invariant function of the
combination $M^\dag M$ and thereby possesses an extended
$GL(\infty)\times GL(\infty)$ symmetry $M\mapsto U\cdot M\cdot V^{-1}$,
$M^\dag\mapsto V\cdot M^\dag\cdot U^{-1}$. In this limit, the kinetic
energy completely drops out and the field theory possesses only static
configurations. Because of this huge symmetry, the only non-vanishing
correlators of the large $N$ complex matrix model are those which
depend solely on traces of powers $(M^\dag M)^r$. Using the field
expansions (\ref{PhiLandauexp}), it follows then that the right-hand
side of (\ref{Landautraceformula}) gives the exact spacetime
dependence of an $L$-point Green's function (\ref{Greensfnscont}) in
this limit, with $L=2r$. In other words, the solution to the
star-product relation (\ref{calGdef}) is determined by the {\it bare}
Green's functions in the limits of interest here, obtained essentially
by dropping the kinetic energy terms in (\ref{SVstartilde}),
i.e. $\sigma=\tilde\sigma=0$.

The integration kernel (\ref{Landautraceformula}) is computed explicitly for
all $L$ in Appendix~E. For the cases $L=2r$ which are
relevant for the complex scalar field theory, we have
\bea
{\cal G}_{2r}(x_1,\dots,x_{2r})&=&\left(\frac1{2\pi\theta}
\right)^{2(r-1)}~\delta^{(2)}\left(\,\sum_{I=1}^{2r}(-1)^I\,x_I\right)
\nn\\&&\times\,\exp\left(\frac\ii r\,\sum_{I<J}(-1)^{J-I}\,
(J-I-r)\,x_I\cdot Bx_J\right) \ .
\label{calGeven}\eea
Note that this result holds for any interaction potential $V$. The
full Green's functions (\ref{Greensfnscont}) in
the given limit are obtained by taking the sum of (\ref{calGeven})
over all permutations of the coordinates
$x_1,\dots,x_{2r}$,\footnote{\baselineskip=12pt Note that
  (\ref{calGeven}) is not real-valued, because even for real functions
  $f$ and $g$, $\overline{f\star g}=g\star f\neq f\star g$ in
  general. It becomes real after symmetrization over the spacetime
  coordinates.} and then multiplying it by matrix model correlators of
$\Tr(M^\dag M)^r$ which carry the detailed information about the
potential (\ref{Vpolyr}) of the quantum field theory. These matrix
averages are well-known to be calculable through a generating
function which is defined by the contour integration~\cite{AJM}
\beq
\frac12~\oint\limits_{\stackrel{\scriptstyle[-\alpha,\alpha]}
{\scriptstyle w\neq z}}\frac{\dd w}{4\pi\ii}~
\frac{w\,\tilde V'(w)}{z^2-w^2}~\sqrt{\frac{z^2-\alpha^2}{w^2-\alpha^2}}=
\frac1N\,\sum_{r=0}^\infty\frac1{z^{2r+1}}~\Bigl\langle\Tr\left(
M^\dag M\right)^r\Bigr\rangle_{E=0} \ ,
\label{matrixavggenfn}\eeq
where the normalized expectation values on the right-hand side of
(\ref{matrixavggenfn}) are defined by (\ref{ZNEavg}) in the large $N$
limit, and the parameter $\alpha$ is determined in terms of the rescaled
coupling constants of the potential (\ref{tildeVdef}) as the real
positive solution of the equation\footnote{\baselineskip=12pt This
  result is derived by introducing the resolvent function
  $R(z)=\frac1N\,\bigl\langle\Tr\,\frac z{z^2-M^\dag M}\bigr\rangle_{E=0}$,
  whose large~$z$ expansion coincides with the right-hand
  side of (\ref{matrixavggenfn}). One can then write an appropriate
  Schwinger-Dyson equation analogous to (\ref{SDeqns}) for this
  correlator, which at $N=\infty$ takes the form~\cite{AJM}
$$
\oint\limits_{\cal C}\frac{\dd w}{4\pi\ii}~\frac{w\,\tilde V'(w)}
{z^2-w^2}~R(w)=R^2(z)+O\left(\frac1{N^2}\right)
$$
where the contour $\cal C$ encloses the singularities of $R(z)$, but
not the point $w=z$, with counterclockwise orientation in the complex
$z$-plane. Assuming that the only singularity of $R(z)$ is a branch
cut across a single connected interval
$[-\alpha,\alpha]\subset\reals$, the solution of this equation
coincides with the left-hand side of (\ref{matrixavggenfn}). Then
(\ref{alpha2eqn}) follows by imposing the asymptotic boundary
condition $R(z)\simeq\frac1z+O(z^{-2})$ as $z\to\infty$ on this solution.}

\beq
\frac{\tilde\mu^2}2\,\alpha^2+\sum_{k\geq2}\,\frac{(2k-1)!}{k\,
\Bigl[(k-1)!\Bigr]^2\,4^k}~\tilde g_k~\alpha^{2k}=1 \ .
\label{alpha2eqn}\eeq
The connected part of the $2r$-point Green's function is of order
$1/N^{r-1}$ in the large $N$ limit, due to the usual factorization of
correlators in the matrix model at $N=\infty$. We will elucidate this
point further in Section~\ref{Connected}. Notice also that the large
$\theta$ bare propagator
\beq
{\cal G}_2(x,y)=\delta^{(2)}(x-y)
\label{calG2}\eeq
is ultra-local due to the persistent $GL(\infty)\times GL(\infty)$
symmetry.\footnote{\baselineskip=12pt The form (\ref{calG2}) follows
  immediately from (\ref{Landautraceformula}) by completeness of the
  Landau wavefunctions, or alternatively from the definition
  (\ref{calGdef}) by the well-known star-product identity
  $\int\dd^2x~f\star f'(x)=\int\dd^2x~f(x)\,f'(x)$.} This is no
longer true, however, for the higher order Green's functions, which
respect the magnetic translation symmetry described in
Section~\ref{Symms}.

Now let us examine the opposite extreme where the interaction
potential is turned off, and compare these exact results
with the {\it free} propagator
\beq
C_\mu(x,y)=\langle x|\frac1{\hat{\sf D}^2+\mu^2}|y\rangle
\label{freepropdef}\eeq
of a charged scalar particle of mass $\mu$ in the constant
electromagnetic background. The computation of (\ref{freepropdef}) is
presented in Appendix~F and one finds
\beq
C_\mu(x,y)=\frac1{4\pi}~\e^{-|x-y|^2/2\theta-\ii x\cdot By}~
\int\limits_0^\infty\dd u~\frac{\e^{-u}}{\sqrt{u^2+|x-y|^2\,u/
\theta}}~\left(\frac u{u+|x-y|^2/\theta}\right)^{\mu^2\,\theta/4} \
. \nn\\
\label{freepropmassive}\eeq
In particular, by setting $\mu^2=0$ in (\ref{freepropmassive}), the
integral can be done explicitly and yields the free massless
propagator
\beq
C_0(x,y)=\frac1{4\pi}~\e^{-\ii x\cdot By}~K_0\Bigl(|x-y|^2/2\theta
\Bigr) \ ,
\label{freepropmassless}\eeq
where $K_0$ is the modified Bessel function of the second kind of
order~0. From (\ref{freepropmassive}) we see that the ultraviolet
problem is unchanged by the presence of the electromagnetic field
$B=1/\theta$, i.e. the singularity at $x=y$ is the same as that for
$B=0$, while from (\ref{freepropmassless}) we see that the magnetic
field acts like a mass and cures the infrared problem in the massless
case. In particular, as shown in Appendix~F, the $\theta\to\infty$
limit of the free propagator yields an ultra-local spacetime
dependence {\it only} for massive fields $\mu^2>0$, while the result
(\ref{calG2}) is valid for all values of the coupling constants of the
field theory. In the massless case (\ref{freepropmassless}), the
standard logarithmic ultraviolet behaviour appears only at very short
length scales $|x-y|\ll\sqrt\theta$. As we will see more of in the
following, these properties are the remnants of the UV/IR mixing
phenomenon within the present class of noncommutative quantum field
theories.

\subsection{The Exact Propagator\label{ExactProp}}

In this subsection we will compute the exact two-point function of the quantum
field theory, which by using the field expansions (\ref{PhiLandauexp}) can be
written formally as
\beq
G(x,y)=\Bigl\langle\Phi^\dag(x)\,\Phi(y)\Bigr\rangle=
4\pi\theta~\sum_{\ell,m,\ell',m'=1}^\infty\,\left\langle
M^\dag_{m\ell}\,M^\nd_{\ell'm'}\right\rangle_{E=\tilde{\cal E}}~
\phi_{\ell,m}(x)\,\phi_{m',\ell'}(y) \ .
\label{2ptfndef}\eeq
To make sense of (\ref{2ptfndef}), we restrict the sums to
$\ell,\ell'<N$ (the sums over $m$ and $m'$ need no regularization) and
compute the matrix correlator in the 't~Hooft limit of the matrix model.
The action and measure in the functional integral (\ref{ZNEavg}) are
invariant under unitary transformations (\ref{MUinftysym}) with
$U^\dag=U^{-1}$. We can thereby explicitly make this transformation in
the matrix integral and then integrate over the unitary group. At
finite $N$ the integral we need is~\cite{DZ}
\beq
\int\limits_{U(N)}[\dd U]~U_{k\ell}^\dag\,U^\nd_{\ell'k'}=\frac1N~
\delta_{kk'}~\delta_{\ell\ell'} \ ,
\label{UN2ptint}\eeq
which can be derived by using the invariance properties of the Haar measure. As
a consequence, the matrix expectation values appearing in (\ref{2ptfndef}) are
given by
\beq
\left\langle M^\dag_{m\ell}\,M^\nd_{\ell'm'}\right
\rangle_{E=\tilde{\cal E}}=\frac1N\,\delta_{\ell\ell'}\,
\Bigl\langle\left(M^\dag M\right)_{mm'}\Bigr\rangle_{E=\tilde{\cal E}}
=\left.-\frac1{N^2}\,\delta_{\ell\ell'}\,\frac{
\partial\ln Z_N(E)}{\partial E_{m'm}}\,\right|_{E=\tilde{\cal E}} \ .
\label{matrixavg2ptWb}\eeq
Using (\ref{WxlnZdef}) and (\ref{NClambda}), the Green's function
(\ref{2ptfndef}) is therefore given by the large $N$ limit of
\beq
G_N(x,y)=-\frac1N\,\sum_{\ell=1}^NW(\lambda_\ell)\,
\gamma^{~}_\ell(x,y) \ ,
\label{2ptfnWgammaell}\eeq
where we have introduced the sum over Landau levels
\beq
\gamma^{~}_\ell(x,y)=4\pi\theta\,\sum_{m=1}^\infty\phi_{m,\ell}(x)\,
\phi_{\ell,m}(y) \ .
\label{gammaellxy}\eeq

The function (\ref{gammaellxy}) is computed in Appendix~G. It follows
from (\ref{gammaellxyfinite}) that, at finite $N$, the propagator
(\ref{2ptfnWgammaell}) respects the magnetic translation invariance
described in Section~\ref{Symms}. As discussed there, however, this
will no longer be true in the desired limit $N\to\infty$ with
$\theta/N$ finite. The two-point function (\ref{2ptfnWgammaell}) is to
be evaluated in the limit $N\to\infty$, $\ell\to\infty$ with
$\frac\ell N\in[0,1]$ fixed. The function $W(\lambda_\ell)$ in this
scaling limit is given by (\ref{WxfinalNC}), while the sum over Landau
wavefunctions (\ref{gammaellxy}) is computed in Appendix~G with the
result
\beq
\gamma^{~}_\ell(x,y)=4\,J_0\Bigl(\Lambda\,\sqrt{\lambda_\ell}\,|x-y|
\Bigr) \ ,
\label{gammaellJ0lambda}\eeq
where $J_0$ is the Bessel function of the first kind of order~0. In the
large $N$ limit, the function (\ref{2ptfnWgammaell}) thereby becomes
\bea
G(x,y)&=&-\int\limits_0^{16\pi}\frac{\dd\lambda}{4\pi}~
W_{\rm b}(\lambda)\,J_0\Bigl(\Lambda\,\sqrt\lambda\,|x-y|\Bigr)
\non&=&-\int\limits_0^{4\,\sqrt\pi\,\Lambda}\frac{\dd p~p}{2\pi\Lambda^2}~
W_{\rm b}\left(p^2/\Lambda^2\right)\,J_0\Bigl(p\,|x-y|\Bigr) \ ,
\label{2ptfnlargeN}\eeq
which by using the angular integral representation of the Bessel
function can be written in the form
\beq
G(x,y)=-\frac1{\Lambda^2}~\int\limits_{|p|\leq4\,\sqrt\pi\,\Lambda}~
\frac{\dd^2p}{(2\pi)^2}~W_{\rm b}\left(p^2/\Lambda^2\right)~
\e^{\ii p\cdot(x-y)} \ .
\label{2ptfnniceform}\eeq
The result (\ref{2ptfnniceform}) illustrates, along with the Green's
functions of the previous subsection, a remarkable property of the large
$\theta$ limit of the quantum field theory described in
Section~\ref{Reg}. In this limit, the underlying spacetime is expected
to disappear or to degenerate and all spacetime symmetries to be
maximally violated. Instead, the Green's functions are both rotationally
and translationally invariant. This is characteristic of a ``fuzzy''
regularization scheme, which typically preserves all spacetime
symmetries of the original continuum field theory (in
contrast to lattice regularization). The remnants of UV/IR mixing here
appear in the far infrared at $|x-y|\sim\sqrt\theta$. In deriving
(\ref{2ptfnniceform}), these distances have been effectively scaled
out, so that all results here are valid at $|x-y|\ll\sqrt\theta$. Thus
the spacetime picture which emerges in the limit described in
Section~\ref{Reg} is not pathological.

{}From (\ref{2ptfnniceform}) it also follows that $\Lambda$ is clearly
the ultraviolet cutoff in the quantum field theory. More precisely,
the quantity $4\,\sqrt\pi\,\Lambda$ is a sharp cutoff in momentum
space. Thus the large $\theta$ limit defines a rotationally and
translationally invariant field theory with a finite cutoff. Moreover,
the exact solution of the matrix model Schwinger-Dyson equations has a
direct physical meaning as the exact propagator of the quantum field
theory. From (\ref{2ptfnniceform}) we see that the function
(\ref{WxfinalNC}) is essentially the Fourier transform of the exact
two-point function,
\beq
\tilde G(p)=-\frac1{\Lambda^2}\,W_{\rm b}\left(p^2/\Lambda^2\right)
\label{2ptfnFourierWb}\eeq
with $|p|\leq4\,\sqrt\pi\,\Lambda$. The crucial issue now is whether or
not the momentum cutoff can be removed, i.e. if the limit
$\Lambda\to\infty$ can be taken in the interacting quantum field
theory. It is straightforward to see that the two-point function is
not finite in the naive scaling limit $\tilde\mu=\mu/\Lambda$, $\tilde
g=g/\Lambda^2$. This can be seen directly in perturbation
theory. Using the iterative solution (\ref{loopiterative}) and
(\ref{2ptfnFourierWb}), we can easily determine the propagator up to
two-loop order in the limit $\Lambda\to\infty$ as
\bea
\tilde G(p)&=&\frac1{p^2+\mu^2}-\frac{g\,\ln(16\pi\,\Lambda^2/\mu^2)}
{16\pi\,(p^2+\mu^2)^2}-\frac{g\,\Lambda^2}{(p^2+\mu^2)^3}\non&&-\,
\frac{g^2}{16\pi\,\mu^2(p^2+\mu^2)^2}\,\left(\frac{\ln(16\pi\,
\Lambda^2/\mu^2)}{16\pi}-\frac{\Lambda^4}{2\mu^4}\right)+\frac{3g^2\,
\Lambda^2\ln(16\pi\,\Lambda^2/\mu^2)}{16\pi\,(p^2+\mu^2)^4}\non&&-\,
\frac{g^2}{16\pi\,(p^2+\mu^2)^3}\,\left(\frac{\ln^2(16\pi\,
\Lambda^2/\mu^2)}{16\pi}+\frac{\Lambda^2}{\mu^2}\right)+\frac{2g^2\,
\Lambda^4}{(p^2+\mu^2)^5}+O\left(g^3\right) \ .
\label{tildeGpert}\eea
The first term in (\ref{tildeGpert}) is the free propagator which as
expected is finite as the momentum cutoff is removed. The second term
recovers the usual, one-loop logarithmic ultraviolet divergence of $\Phi^4$
theory in two dimensions which is generated by the planar bubble
diagram. However, at one-loop order there is an additional quadratic
ultraviolet divergence, which may be attributed to the non-planar
contribution in this case. The divergences are even worse at higher loop
orders. For instance, the usual two-loop ultraviolet divergence
generated by the planar sunset diagram is accompanied by terms in
(\ref{tildeGpert}) whose degree of divergence is different than that
in the usual scalar field theory. These additional divergences in
$\Lambda$ are nothing but divergences in the summations over
degenerate Landau levels in the matrix model, whose degree grows with
the order of perturbation theory.

Another way to try to get rid of the cutoff is to adjust the coupling
constants non-trivially. For this, one
needs to find a critical regime in which there is another scale that
is smaller than $\Lambda$ and the couplings can be thereby adjusted to
achieve a separation of scales, such that non-trivial physics remains
for momentum modes with $|p|\ll\Lambda$. With the exact solution at
hand, it is not difficult to show that this is in fact not
possible. For this, we note that from (\ref{2ptfnFourierWb}) it
follows that the behaviour of the propagator at large distances (much
larger than the inverse momentum cutoff $\Lambda^{-1}$) is determined
by the singularities of the loop amplitude $W(\lambda)$, i.e. its two
complex branch points $\lambda=b_1,b_2$. As a consequence, we can
write down the asymptotic behaviour
\beq
G(x,y)\simeq\e^{-|x-y|/\xi} \ , ~~ |x-y|\gg\Lambda^{-1} \ ,
\label{propasympt}\eeq
where the correlation length $\xi=1/{\rm Im}\,p_0$ is determined by the
condition that the complex number $z=p_0^2/\Lambda^2$ coincides with
one of the two branch points $b_1$ or $b_2$. From
(\ref{bcnonlinconstrNC}) one can show
\beq
({\rm Im}\,z)^2-2\,({\rm Re}\,z)^2=b_1^2\,b_2^2-\frac34\,\Bigl(b_1+b_2
\Bigr)^2>2\tilde g \ .
\label{ImRezineq}\eeq
{}From this inequality one may then show that for any choice of
parameters, the correlation length $\xi$ is bounded as
\beq
\frac1\xi>\left(\frac{\tilde g}3\right)^{1/4}\,\Lambda \ .
\label{xibound}\eeq

{}From (\ref{xibound}) it follows that the only way to keep the
correlation length finite while sending the cutoff to infinity is to
send the coupling constant $\tilde g$ to zero. Careful inspection of
the function (\ref{WxfinalNC}) shows that the only scaling limit which
makes sense is $\tilde g\sim1/\Lambda^4$, which is also suggested by
the inequality (\ref{xibound}). Thus we define
\beq
g=\frac{m^4}{\Lambda^2}
\label{gMLambda}\eeq
with $m$ a dynamically generated mass scale. From the constraint
equations (\ref{bcnonlinconstrNC}) we further find that
$b_1b_2=2\tilde g$ and also $b_1+b_2=O(\tilde g)$, which drops out in
the scaling limit described above. The renormalized two-point function
is thereby given from (\ref{WxfinalNC}) and (\ref{2ptfnFourierWb}) as
\beq
\tilde G(p)=\frac{\sqrt{(p^2+\mu^2)^2+4m^4}-(p^2+\mu^2)}{2m^4} \ ,
\label{2ptfnren}\eeq
so that the additional limit $\Lambda\to\infty$ makes the spectral
density of the external field invisible in the loop equation
(\ref{largeNWxeqn}). The meaning of the scaling limit in
(\ref{2ptfnren}) is easy to understand from perturbation
theory. Comparing the large momentum expansion of (\ref{2ptfnren})
with (\ref{tildeGpert}), it is evident that taking this scaling limit
amounts to a non-perturbative resummation of the leading power
divergences in perturbation theory. As these divergences come from the
Landau level degeneracies and are thereby an artifact of the
particular regularization used, such a resummation procedure does not
lead to a meaningful interacting quantum field theory. This will be
confirmed in the next subsection, wherein we show that all
higher-point Green's functions are trivial.

We conclude that the present model is not a renormalizable,
interacting quantum field theory, because its coupling constant flows
only to the trivial Gaussian fixed point. Within our present
formulation of the model, there are {\it a priori} two scales at
hand. One is the size $1/\sqrt{B}$ of the orbit in the
lowest Landau level, which is the infrared cutoff and which we have
taken to be infinitely large. The other is the size $\sqrt{N/B}$ of
the largest Landau level that we allow for, which is the ultraviolet
cutoff. What we have found above is that there is no intermediate
scale in between the two, no matter what one does with the couplings
of the model. The fields are thereby correlated on the scale of the
cutoff, and the field theory is not renormalizable in the standard
sense. Of course, one needs to analyse the possibility that other
scalings, besides the standard 't~Hooft limit which was described in
Section~\ref{Reg}, may lead to non-trivial fixed points of the
renormalization group flows. In Section~\ref{Scaling} we shall
address this important issue and describe some ways that a non-trivial
interacting quantum field theory may be attained.

Nevertheless, the present model provides an interesting example of a
noncommutative field theory which has a finite cutoff, and which is
exactly solvable. In Section~\ref{Scaling} we will offer an
alternative physical interpretation of the regularization of
Section~\ref{Reg} in which $\Lambda$ is simply the characteristic mass
scale of the theory, and for which the correlation functions at finite
$\Lambda$ can be interpreted as those of the full quantum field
theoretic limit. The Green's functions display some novel features
which are not encountered in ordinary quantum field theory. For
example, because the branch points $b_1$ and $b_2$ are complex-valued,
the propagator exhibits an unusual oscillatory behaviour on top of its
exponential decay. This is similar to the Aharonov-Bohm phases, for
charged particles in the background magnetic field $B=\theta^{-1}$, that
have been observed on top of the usual area law behaviour of Wilson
loops in two-dimensional noncommutative Yang-Mills
theory~\cite{BHN1}. The origin of this behaviour can be understood
from the bare Green's functions (\ref{calGeven}), and the free
propagators (\ref{freepropmassive}) and (\ref{freepropmassless}) for
bosons in a finite external magnetic field, where the Aharonov-Bohm
phase factors $x\cdot By$ are explicit. Similar behaviours for the
Green's functions of scalar field theory on more complicated
noncommutative spacetimes have also been observed in~\cite{BES}.

\subsection{Connected Correlators\label{Connected}}

To substantiate the claim made in the previous subsection about the
triviality of the interacting quantum field theory, in this subsection
we show that the four-point Green's function is trivial, i.e. it
factorizes into products of propagators in the scaling limit described
above, and further argue that this is a generic feature of all higher
order correlators (\ref{Greensfnscont}). For this, we expand the
four-point function analogously to the propagator to formally get
\bea
G^{(4)}(x,y;z,w)&=&\Bigl\langle\Phi^\dag(x)\,\Phi(y)\,\Phi^\dag(z)\,
\Phi(w)\Bigr\rangle\non&=&(4\pi\theta)^2\,\sum_{\ell_1,m_1,\dots,
  \ell_4,m_4=1}^\infty\,\Bigl\langle M^\dag_{m_1\ell_1}\,
  M^\nd_{\ell_2m_2}\,M^\dag_{m_3\ell_3}\,M^\nd_{\ell_4m_4}\Bigr
  \rangle_{E=\tilde{\cal
      E}}\non&&\times\,\phi_{\ell_1,m_1}(x)\,\phi_{m_2,\ell_2}(y)
  \,\phi_{\ell_3,m_3}(z)\,\phi_{m_4,\ell_4}(w) \ .
\label{G4xyzwdef}\eea
As in the previous subsection, we make sense of (\ref{G4xyzwdef}) by
restricting the sums over $\ell_j$ to $1,\dots,N$ and compute the
matrix model correlators in the 't~Hooft limit. Again, we replace
$M\to U\cdot M$, $M^\dag\to M^\dag\cdot U^\dag$ and integrate over all
$N\times N$ unitary matrices $U$. A straightforward computation using
the unitary matrix integral formula~\cite{DZ}
\bea
\int\limits_{U(N)}[\dd U]~U^\dag_{k_1\ell_1}\,U^\nd_{\ell_2k_2}\,
U^\dag_{k_3\ell_3}\,U^\nd_{\ell_4k_4}&=&\frac1{N^2}\,\Biggl[
\delta_{\ell_1\ell_2}~\delta_{k_1k_2}~\delta_{\ell_3\ell_4}~
\delta_{k_3k_4}+\delta_{\ell_1\ell_4}~\delta_{k_1k_4}~
\delta_{\ell_2\ell_3}~\delta_{k_2k_3}\Biggr.\non&&-\left.
  \frac1N\,\Bigl(\delta_{\ell_1\ell_4}~\delta_{k_1k_2}~
  \delta_{\ell_2\ell_3}~\delta_{k_3k_4}+\delta_{\ell_1\ell_2}~
  \delta_{k_1k_4}~\delta_{\ell_3\ell_4}~\delta_{k_2k_3}\Bigr)
  \right]\non&&
\label{UN4ptid}\eea
yields

\vbox{\bea
G_N^{(4)}(x,y;z,w)&=&\frac1{N^2}\,
\sum_{\ell_1,\dots,\ell_4=1}^N\,\Bigl\langle\left(M^\dag
  M\right)_{\ell_1\ell_2}\,\left(M^\dag M\right)_{\ell_3\ell_4}\Bigr
\rangle_{E=\tilde{\cal E}}\non&&\times\left(\gamma_{\ell_1\ell_2}^{~}(x,y)
  \,\gamma^{~}_{\ell_3\ell_4}(z,w)-\frac1N\,\gamma^{~}_{\ell_1\ell_4}(x,w)
  \,\gamma^{~}_{\ell_3\ell_2}(z,y)\right)~+~
 \Bigl(\longleftrightarrow\Bigr) \ , \non&&
\label{G4xyzwgamma}\eea}
\noindent
where the notation $(\longleftrightarrow)$ indicates to include the
same terms but with the spacetime dependence $(x,y);(z,w)$ replaced by
$(x,w);(z,y)$, and we have introduced the generalized Landau sums
\beq
\gamma_{\ell\ell'}^{~}(x,y)=4\pi\theta\,\sum_{m=1}^\infty\phi_{m,\ell}(x)\,
\phi_{\ell',m}(y) \ .
\label{genLandausums}\eeq
The diagonal elements of the functions (\ref{genLandausums}) of course
coincide with (\ref{gammaellxy}),
$\gamma^{~}_\ell(x,y)=\gamma_{\ell\ell}^{~}(x,y)$.

We can now compute the matrix model correlators appearing in
(\ref{G4xyzwgamma}) in terms of the free energy (\ref{freeenergydef})
as
\bea
\Bigl\langle\left(M^\dag
  M\right)_{\ell_1\ell_2}\,\left(M^\dag M\right)_{\ell_3\ell_4}\Bigr
\rangle_E&=&\frac1{N^4}\,\frac1{Z_N(E)}\,\frac{\partial^2Z_N(E)}{\partial
  E_{\ell_2\ell_1}\,\partial E_{\ell_4\ell_3}}\non&=&N^2\,
\frac{\partial\,\Xi_N(E)}{\partial E_{\ell_2\ell_1}}\,
\frac{\partial\,\Xi_N(E)}{\partial E_{\ell_4\ell_3}}-
  \frac{\partial^2\,\Xi_N(E)}{\partial E_{\ell_2\ell_1}\,
  \partial E_{\ell_4\ell_3}} \ .
\label{matrix4ptfreeen}\eea
Since the function $\Xi_N(E)$ depends only on the eigenvalues
$\lambda_\ell$ of the external field $E$, we can compute its
derivatives by replacing $E$ with $E+v/N$ and using Schr\"odinger
perturbation theory to expand it about $v=0$
(see~(\ref{dEchainrule})). One finds
\bea
\frac{\partial\,\Xi_N(E)}{\partial E_{\ell m}}&=&-\frac1{N}~
\delta_{\ell m}~W(\lambda_\ell) \ , \non\frac{\partial^2\,
  \Xi_N(E)}{\partial E_{\ell m}\,\partial E_{\ell'm'}}&=&
-\delta_{\ell
  m}~\delta_{\ell'm'}~W_2(\lambda_\ell,\lambda_{\ell'})
-\frac1N~\delta_{\ell m'}~\delta_{\ell'm}~\Bigl(1-\delta_{\ell m}
\Bigr)\,\frac{W(\lambda_\ell)-W(\lambda_{\ell'})}{\lambda_\ell-
\lambda_{\ell'}} \ , \non&&
\label{XiNEderivs}\eea
where $W(\lambda_\ell)$ is the loop function (\ref{WxlnZdef}) and we
have defined
\beq
W_2(\lambda_\ell,\lambda_{\ell'})=\frac1{N^2}\,\frac{\partial^2
  \ln Z[\lambda]}{\partial\lambda_\ell\,\partial\lambda_{\ell'}} \ .
\label{W2def}\eeq
The matrix correlator (\ref{matrix4ptfreeen}) is thereby given as
\bea
\Bigl\langle\left(M^\dag
  M\right)_{\ell_1\ell_2}\,\left(M^\dag M\right)_{\ell_3\ell_4}\Bigr
\rangle_E&=&\delta_{\ell_1\ell_2}~\delta_{\ell_3\ell_4}~\Bigl(
W(\lambda_{\ell_1})\,W(\lambda_{\ell_3})-W_2(\lambda_{\ell_1},
\lambda_{\ell_3})\Bigr)\non&&-\,\frac1N~\delta_{\ell_1\ell_4}~
\delta_{\ell_2\ell_3}~\Bigl(1-\delta_{\ell_1\ell_2}\Bigr)\,
\frac{W(\lambda_{\ell_1})-W(\lambda_{\ell_3})}
{\lambda_{\ell_1}-\lambda_{\ell_3}} \ . \non&&
\label{matrix4ptfinal}\eea

The crucial point now is the behaviour of the sums over Landau
eigenfunctions in (\ref{genLandausums}) in the large $N$ limit. In the
previous subsection we saw that for $\ell=\ell'$ this sum is of
order $1$. For $\ell\neq\ell'$, by using the analysis of Appendix~G it
is possible to show that the sum is exponentially small in
$|\ell-\ell'|$ for $\ell-\ell'\simeq N$. Putting everything together,
the Green's function (\ref{G4xyzwgamma}) at $N=\infty$ becomes
\bea
G_N^{(4)}(x,y;z,w)&=&\frac1{N^2}\,\sum_{\ell,m=1}^NW(\lambda_\ell)
\,W(\lambda_m)\,\gamma^{~}_{\ell\ell}(x,y)\,\gamma^{~}_{mm}(z,w)~+~
\Bigl(\longleftrightarrow\Bigr)~+O\left(\frac1N\right)\non&=&
G(x,y)\,G(z,w)+G(x,w)\,G(z,y)+O\left(\frac1N\right)
\label{G4trivial}\eea
with $G(x,y)$ the two-point Green's function
(\ref{2ptfnWgammaell}). The right-hand side of (\ref{G4trivial}) is
just the ``trivial'' connected part of the correlator up to order
$\frac1N$, and this proves that the connected four-point Green's
function of the field theory is trivial in the large $N$ limit. From
the analysis above we believe that this triviality is unavoidable in the
't~Hooft limit, as it is simply a consequence of the well-known large
$N$ factorization in the matrix model, which implies that all
connected matrix correlation functions vanish. In particular, it
should be valid for arbitrary interaction potential~$V$. We believe that this
is a generic feature that will generalize to all higher-point functions
of the model, and so the quantum field theory in the scaling limit
obtained in the previous subsection is Gaussian, as all connected
$2r$-point Green's functions are of order $1/N^{r-1}$.

The computation above shows that the connected part of the four-point
Green's function,
\beq
G^{(4)}_{\rm conn}(x,y;z,w)=G^{(4)}(x,y;z,w)-
G(x,y)\,G(z,w)-G(x,w)\,G(z,y) \ ,
\label{G4conn}\eeq
is of order $\frac1N$ in the 't~Hooft limit. However, it also shows
that one can compute the quantity $N\,G^{(4)}_{\rm conn}(x,y;z,w)$
explicitly in the large $N$ limit, and everything needed to write down
a closed formula for it is given above. The non-vanishing leading
term of this and higher Green's functions may be of interest when
applying this model as a generalized mean field theory of a local
quantum field theory, as we discuss in Appendix~A. It
provides the analog of the random phase approximation in conventional mean
field theory. A more detailed investigation of these higher-order
Green's functions would be interesting but is left for future work.

\newsection{Higher Dimensional Generalizations\label{HigherD}}

A remarkable feature of the present class of field theories, and the
formalism that we have developed to analyse them, is that everything
we have said thus far generalizes to {\it arbitrary} even
dimensionality. Namely, the quantum field theory with action
(\ref{SVstar}) is exactly solvable for any $n$. This is quite unlike
the situation that one would expect in ordinary quantum field
theory. In this section we will describe the higher-dimensional
generalization of the exact solution that we have obtained. As we will
see, the $2n$-dimensional case can be mapped {\it exactly} onto the
two-dimensional model (This is also observed in~\cite{L,GMS}). Since
much of the machinery is the same as in the two-dimensional case, we
will only highlight the essential changes which occur.

Let us first fix some notation. We can rotate to a local coordinate
system on $\real^{2n}$ in which the non-degenerate antisymmetric matrix
$\theta=(\theta^{ij})$ assumes its canonical skew-diagonal form
\beq
\theta=\pmatrix{0&\theta^1& & & \cr-\theta^1&0& & & \cr
 & &\ddots& & \cr & & &0&\theta^n\cr & & &-\theta^n&0\cr}
\label{thetaskewdiag}\eeq
with non-zero skew-eigenvalues $\theta^i$, $i=1,\dots,n$. Similarly,
at the special point $B=(B_{ij})=\theta^{-1}$, which we assume
throughout, the antisymmetric matrix $B$ is in its canonical form
with skew-eigenvalues $B_i=(\theta^i)^{-1}$. Corresponding to the
$i^{\,{\rm th}}$ skew block, we introduce the complex coordinates
\beq
z^i=x^{2i-1}+\ii x^{2i} \ , ~~ \overline{z^i}=x^{2i-1}-\ii x^{2i} \ .
\label{skewz}\eeq
We will also use a similar notation for momentum space variables,
$K_i=(p_{2i-1},p_{2i})$, $i=1,\dots,n$.

\subsection{Regularization}

The Hamiltonian $-{\sf D}^2$ in the action (\ref{SVstartilde}) is the sum of
$n$ two-dimensional Landau Hamiltonians in the magnetic field
variables $B_i$. Its eigenfunctions are therefore given as a product
of two-dimensional Landau wavefunctions,
\beq
\phi_{\vell,\vem}(x)=\prod_{i=1}^n\phi_{\ell_i,m_i}\left(z^i\,,\,
\overline{z^i}\,;\,B_i\right) \ ,
\label{2neigen}\eeq
where $\phi_{\ell_i,m_i}(z^i,\overline{z^i}\,;B_i)$ are the
two-dimensional Landau eigenfunctions of Section~\ref{Matrix} in the
noncommutativity parameter $\theta^i=(B_i)^{-1}$, and the
quantum numbers are
\beq
\vell=(\ell_1,\dots,\ell_n) \ , ~~ \vem=(m_1,\dots,m_n) \ ,
{}~~ \ell_i,m_i=1,2,\dots \ .
\label{vellvemdef}\eeq
The functions (\ref{2neigen}) form a complete orthonormal basis in
$L^2(\real^{2n})$. From the corresponding two-dimensional results
of Section~\ref{Matrix} we obtain immediately the eigenvalue equation
\beq
-{\sf D}^2\phi_{\vell,\vem}=4\,\sum_{i=1}^nB_i\left(\ell_i-\frac12
\right)\,\phi_{\vell,\vem} \ ,
\label{neigeneq}\eeq
and the star-product projector relation
\beq
\phi_{\vell,\vem}\star\phi_{\vell',\vem'}=\frac1{\det
(4\pi\theta)^{1/4}}~\delta_{\vem,\vell'}~\phi_{\vell,\vem'}
\label{nstarproj}\eeq
with $\delta_{\vell,\vem}\equiv\prod_i\delta_{\ell_i,m_i}$. An
analogous expression holds for the Hamiltonian $-\tilde{\sf D}^2$.

We can now diagonalize the action (\ref{SVstartilde}) by the expansions
\beq
\Phi(x)=s\,\sum_{\vell,\vem\in\nat^{\,n}}M_{\vell,\vem}~\overline{
\phi_{\vell,\vem}}(x) \ , ~~
\Phi^\dag(x)=s\,\sum_{\vell,\vem\in\nat^{\,n}}
\overline{M_{\vell,\vem}}~\phi_{\vell,\vem}(x) \ ,
\label{nPhiexp}\eeq
where $s$ is a scaling parameter which we will determine below so as to
get the appropriate scaling required for exact solvability of the
model. We regard $M=(M_{\vell,\vem})$ as a compact operator on the
Hilbert space ${\cal H}={\cal S}(\nat^{\,n})$, with trace $\Tr_{\cal
H}^{~}(M)=\sum_\vem M_{\vem,\vem}$. We will use an obvious matrix
notation with $(M^\dag
M)_{\vell,\vell'}=\sum_{\vem}(M^\dag)_{\vell,\vem}\,M_{\vem,\vell'}$
and $(M^\dag)_{\vell,\vem}=\overline{M^\nd_{\vem,\vell}}$. Then the action
(\ref{SVstartilde}) can be written in the matrix form
\beq
\tilde S_{\rm b}=\Tr^{~}_{\cal H}\Bigl(\sigma\,M\,{\cal E}\,M^\dag+
\tilde\sigma\,M^\dag\,{\cal E}\,M+V_0\left(M^\dag M\right)
\Bigr) \ ,
\label{nSbmatrix}\eeq
where the external field is given by
\beq
{\cal
E}_{\vell,\vem}=s^2\,\left[4\,\sum_{i=1}^nB_i\left(\ell_i-\frac12
\right)+\mu^2\right]~\delta_{\vell,\vem}
\label{ncalEdef}\eeq
and we have defined the renormalized interaction potential
\beq
V_0(w)=\sum_{k\geq2}\frac{s^{2k}}{\det(4\pi\theta)^{(k-1)/2}}\,
\frac{g_k}k\,w^k \ .
\label{V0wdef}\eeq

There are many ways to now regularize the noncommutative field theory
in the form (\ref{nSbmatrix}) by mapping it onto a finite dimensional
matrix model. As we will see later on, these regularizations are all
equivalent in the limit where the size of the matrices becomes
infinite. The simplest one, which we will call the ``naive''
regularization, is to restrict each set of Landau quantum numbers to a
common finite range $L<\infty$,
\beq
\ell_i,m_i=1,\dots,L \ , ~~ i=1,\dots,n \ .
\label{naivereg}\eeq
We can then map the integer vectors (\ref{vellvemdef}) bijectively to
single integers $\ell,m$ as
\beq
\ell=\sum_{i=1}^nL^{i-1}\,(\ell_i-1)+1 \ , ~~
m=\sum_{i=1}^nL^{i-1}\,(m_i-1)+1 \ .
\label{ellmbij}\eeq
The precise form of the mapping (\ref{ellmbij}) is not important here,
as any other bijection on $\nat^{\,n}\to\nat$ would work equally well, for
instance any lexicographic ordering $\vell\mapsto\ell={\rm
lex}(\ell_1,\dots,\ell_n)$ and $\vem\mapsto m={\rm
lex}(m_1,\dots,m_n)$. With this mapping, we can then identify matrix
elements as $M_{\vell,\vem}\equiv M_{\ell m}$, and the regularization
above amounts to restricting $\ell,m=1,\dots,N$ with
\beq
N=L^n \ .
\label{NLn}\eeq
It should be stressed though that this is {\it not} the natural
regularization. The one leading to a nice $N\to\infty$ limit is the
isotropic regularization whereby the one-particle energies are bounded
by a momentum scale $\Lambda'$ as
\beq
\sum_{i=1}^n|B_i|\left(\ell_i-\frac12\right)\leq\Lambda^{\prime\,2} \ .
\label{1partLambda}\eeq
The isotropic limit is essentially obtained by letting the
skew-eigenvalues $B_i$ of the magnetic field all approach the same
value. However, to get the correct scaling it is more convenient to proceed
first via the naive regularization above.

The regularized partition function is then given by the $N\times N$
matrix integral
\beq
Z_N({\cal E})=\int\DD M~\DD M^\dag~\e^{-\Tr\bigl(\sigma\,M\,{\cal E}\,M^\dag+
\tilde\sigma\,M^\dag\,{\cal E}\,M+V_0(M^\dag M)\bigr)} \ ,
\label{ZNcalEn}\eeq
similarly to that of Section~\ref{Reg}. To obtain a good large $N$
limit, we now need an overall factor of $N$ coming out in front of
the trace in (\ref{ZNcalEn}). This requires a relation
$s^2\,\det(B)^{1/2n}\,L=4\pi N=4\pi L^n$, which fixes the scaling parameter
$s$ introduced in (\ref{nPhiexp}) as
\beq
s=L^{(n-1)/2}\,\det(4\pi\theta)^{1/4} \ .
\label{sfixed}\eeq
The partition function (\ref{ZNcalEn}) then assumes the form in
(\ref{ZNEdef}), with the renormalized interaction potential
(\ref{V0wdef}) taking the form (\ref{tildeVdef}). As expected from
dimensional analysis, the rescaled coupling constants in
(\ref{tildeVdef}) are now given by
\beq
\tilde\mu^2=\frac{\mu^2}{\Lambda^2} \ , ~~ \tilde g_k=
\frac{g_k}{\Lambda^{2n-2(n-1)k}}
\label{nrescalecoupl}\eeq
with
\beq
\Lambda^2=\frac L{\det(4\pi\theta)^{1/2n}} \ .
\label{nLambdadef}\eeq
The quantity (\ref{nLambdadef}) is the natural ultraviolet cutoff in
any dimension, as then the one-particle energies are bounded as
$4\sum_iB_i\ell_i\leq4bLn=16\pi n\Lambda^2$ (in the isotropic case
where all skew-eigenvalues $B_i=b$ are the same). The one-particle
energies are now given by $\tilde{\cal
E}_{\vell,\vem}=\lambda_\vell~\delta_{\vell,\vem}$, with
\beq
\lambda_\vell=4\pi\left[4\,\sum_{i=1}^n\frac{b_i}L\,\left(\ell_i-
\frac12\right)+\tilde\mu^2\right]
\label{lambdavelldef}\eeq
and
\beq
b_i=\frac{B_i}{\det(B)^{1/2n}} \ .
\label{bidef}\eeq

After shifting the eigenvalues (\ref{lambdavelldef}) by the mass as
in Section~\ref{ExactVac}, they are scaled correctly so that the
spectral density
\beq
\rho(\lambda)=\frac1N\,\sum_{\vell\in(\zed_L)^n}\delta(\lambda-\lambda_\vell)
\label{rholambdavell}\eeq
is given by a finite Riemann sum in the large $N$ limit,
\beq
\rho(\lambda)=\prod_{i=1}^n\,\int\limits_0^{16\pi}\frac{\dd\lambda_i}
{16\pi}~\delta\left(\lambda-\sum_{i=1}^nb_i\lambda_i\right) \ .
\label{rholambdalargeN}\eeq
{}From (\ref{rholambdalargeN}) it follows that the isotropic
regularization in (\ref{1partLambda}), i.e. $\sum_i\ell_i\leq L$, is
simpler and more natural. In that case, rather than the limits
$\lambda_i\leq16\pi$ in the integrals defining the density of states
(\ref{rholambdalargeN}), we have $\sum_ib_i\lambda_i\leq16\pi$. We can
then transform to spherical coordinates to explicitly compute the
integral in (\ref{rholambdalargeN}) and obtain the large $N$
eigenvalue distribution
\beq
\rho(\lambda)=\frac{\lambda^{n-1}}{\left(16\,\sqrt\pi\,\right)^n\,\Gamma\left(
\frac n2\right)} \ ,
\label{rholambdan}\eeq
with compact support $\lambda\in[0,16\pi]$. Here $\Gamma$ is the Euler
function which arises from integrating over the solid angle in $n$
dimensions. Thus one of the main characterizations of the higher
dimensional generalizations is that the spectral density is no longer
flat.

\subsection{The Two-Point Function}

Let us now consider the two-point function $G(x,y)$ in the case of the
noncommutative field theory (\ref{SVstar}). With the identifications
$\vell\equiv\ell$, $\vem\equiv m$, the matrix model representation
(\ref{nPhiexp},\ref{nSbmatrix}) is formally the same as in the
two-dimensional case, and we thus find
\beq
G_N(x,y)=-\frac1N\,\sum_{\vell\in(\zed_L)^n}W(\lambda_\vell)\,
\gamma^{~}_\vell(x,y)
\label{Gxyn}\eeq
with $W(\lambda_\vell)$ the logarithmic derivative of the partition function
(\ref{ZNcalEn})
in the 't~Hooft scaling limit and
\beq
\gamma^{~}_\vell(x,y)=\det(4\pi\theta)^{1/2n}
\,L^{n-1}\,\sum_{\vem\in(\zed_L)^n}
\phi_{\vell,\vem}(x)\,\phi_{\vem,\vell}(y) \ .
\label{gammavell}\eeq
Since the Landau eigenfunctions in (\ref{gammavell}) are just the products
(\ref{2neigen})
of two-dimensional Landau wavefunctions, the sum in (\ref{gammavell}) is a
product of the
$n$ sums computed in Appendix~G and in the scaling limit whereby $\ell/L$ is
finite as
$L\to\infty$ it gives the result
\bea
\gamma^{~}_\vell(x,y)&=&\det(4\pi\theta)^{1/2n}
\,L^{n-1}\,\prod_{i=1}^n\,\sum_{m_i=1}^\infty\phi_{\ell_i,m_i}
\left(z^i\,,\,\overline{z^i}\,;\,B_i\right)\,\phi_{m_i,\ell_i}
\left(w^i\,,\,\overline{w^i}\,;\,B_i\right)\nn\\&=&
4^n\,\Lambda^{2n-2}\,\prod_{i=1}^nJ_0
\left(\,\sqrt{\lambda_i\,b_i}~\Lambda\,\left|z^i-w^i\right|\,\right) \  ,
\label{gammavellscaling}\eea
with $w^i=y^{2i-1}+\ii y^{2i}$, $\overline{w^i}=y^{2i-1}-\ii y^{2i}$, and
$\lambda_i=16\pi\ell_i/L$. In the limit $L\to\infty$, the Green's function
(\ref{Gxyn}) thereby
becomes
\beq
G(x,y)=-\Lambda^{2n-2}\,\prod_{i=1}^n\,\int\limits_0^{16\pi}
\frac{\dd\lambda_i}{4\pi}~
J_0\left(\,\sqrt{\lambda_i\,b_i}~\Lambda\,\left|z^i-w^i
\right|\,\right)\,W\left(\,\sum_{i=1}^nb_i\lambda_i\right) \ .
\label{Gxynscaling}\eeq
We now change variables $\lambda_i\,b_i\,\Lambda^2=K_i^2$ in
(\ref{Gxynscaling}), use the fact that
$\prod_ib_i=1$, and then apply the angular integral representation of the
Bessel function $J_0$
to get
\beq
G(x,y)=-\frac1{\Lambda^2}\,\prod_{i=1}^n~\int\limits_{K_i^2\leq16\pi
b_i\Lambda^2}
\frac{\dd K_i~K_i}{2\pi}~\int\limits_0^{2\pi}\frac{\dd\tau_i}{2\pi}~\e^{\ii
K_i|z^i-w^i|\cos\tau_i}~
W\left(\,\sum_{i=1}^n\frac{K_i^2}{\Lambda^2}\right) \ .
\label{Gxynangint}\eeq

We can thereby identify $p_{2i-1}=K_i\cos\tau_i$,
$p_{2i}=K_i\sin\tau_i$ as the two-dimensional momenta associated with
the skew-block coordinates $(x^{2i-1},x^{2i})$, and write
\beq
G(x,y)=-\frac1{\Lambda^2}~\int\limits_{\bigl
  \{p_{2i-1}^2+p_{2i}^2\leq16\pi\Lambda^2b_i\bigr\}}\,
\frac{\dd^{2n}p}{(2\pi)^{2n}}~W\left(p^2/\Lambda^2\right)~
\e^{\ii p\cdot(x-y)} \ ,
\label{Gxynfinal}\eeq
which is the $2n$-dimensional generalization of the formula
(\ref{2ptfnniceform}) for the  propagator. Thus the scaling limit used
above always looks rotationally symmetric except at very small
distances. However, this anisotropy is due to the anisotropic
regularization that we have used, which is invariant only under an
abelian $SO(2)^n$ subgroup of $SO(2n)$. Had we used the isotropic
scaling described earlier we would have obtained a rotationally
symmetric cutoff $p^2\leq16\pi\Lambda^2$ in momentum space. No
matter how one chooses the skew-eigenvalues $B_i$, in the large $N$
limit the anisotropy is totally washed out and one gets the same
result. Thus, rather remarkably, we are left with an
$SO(2n)$-invariant Green's function, even though rotational symmetry
is broken by the external fields. While other regularizations are
possible as well, at finite cutoff $\Lambda$ they only make a
difference at small distances $|x-y|\ll\Lambda^{-1}$.

The Fourier transform $W(\lambda)$ of the two-point function
satisfies the same loop equation (\ref{largeNWxeqn}) as in the two-dimensional
case
(before the mass shift),
whose solution is given in (\ref{Wxfinal},\ref{bcnonlinconstr}). The only
change is that the
density of states $\rho(\lambda)$ is no longer flat, and with the isotropic
regularization it is
given by the simple explicit expression (\ref{rholambdan}). In particular, from
the iterative solution
(\ref{loopiterative}) it follows that the perturbative expansion of the
propagator now
contains the anticipated power law divergences of scalar field theory in $2n$
dimensions. Moreover,
the analytic structure of the loop function $W(\lambda)$ is fairly insensitive
to the precise form
of the spectral density, as long as its support lies in $[0,\infty)$. In
particular, the singularities
of $W(\lambda)$ are again cuts which sit in the complex plane, and it is
possible to demonstrate
once again that the field theory in {\it any} dimension is not renormalizable.
In the limit $\Lambda\to\infty$, the rotational symmetry breaking $SO(2n)\to
SO(2)^n$ is no longer washed away.

\newsection{Scaling Limits\label{Scaling}}

In the previous sections we have found that, in the scaling limit
whereby the quantity (\ref{LambdaNtheta}) plays the role of an
ultraviolet cutoff, the renormalized quantum field theory is trivial
in the sense explained before. The purpose of this section is to point
out that there exist other scaling limits of the noncommutative field
theory which, while mathematically sharing the same properties as that
studied already, have drastically different physical interpretations
and have the potential of providing non-trivial connected Green's
functions. At the same time, we will give nice physical
characterizations of the large $N$ limit of the matrix model from a
field theoretic point of view. We shall also describe extensions of
our analysis that remove the Landau level degeneracies which spoilt the
renormalizability of the interacting quantum field theory.

\subsection{Scale Transformations\label{ScaleTransf}}

The crux of the existence of different scaling limits of the
noncommutative quantum field theory resides
in the special scaling property of the Landau wavefunctions,
\beq
\phi_{\ell,m}(x;B)=\lambda^n~\phi_{\ell,m}\left(\lambda\,x\,;
  \,\lambda^{-2}\,B\right) \ ,
  \label{Landauscaling}\eeq
which easily follows from the generating function
(\ref{genfnexpl}). The physical interpretation of
(\ref{Landauscaling}) is that a typical skew-eigenvalue $b$ of $B$ defines
the magnetic length $l_{\rm mag}=1/\sqrt{b}$. If the length scale is
changed as $l_{\rm mag}\to\lambda\,l_{\rm mag}$, then the physics is
unchanged provided we rescale the spacetime coordinates also as
$x\to\lambda\,x$. The factor of $\lambda^n$ in (\ref{Landauscaling})
then ensures the correct normalization of the Landau eigenfunctions.

Let us now consider the Green's functions in the regulated field
theory at {\it finite} $N$,
\beq
G_N(x_1,\dots,x_r;y_1,\dots,y_r)=\zeta^{2r}\,\Bigl\langle\Phi^\dag(x_1)\,
\Phi(y_1)\cdots\Phi^\dag(x_r)\,\Phi(y_r)\Bigr\rangle \ ,
\label{GreensfnfiniteN}\eeq
where the parameter $\zeta=\zeta_N>0$ is a multiplicative wavefunction
renormalization. Then the regularized partition function
(\ref{ZNcalEn}) and all Green's functions (\ref{GreensfnfiniteN}) at
finite $N$ are invariant under the scale transformations
\bea
x&\longmapsto&\lambda\,x \ , \nn\\B&\longmapsto&\lambda^{-2}\,B \ ,
\nn\\\theta&\longmapsto&\lambda^2\,\theta \ , \nn\\\mu^2&\longmapsto&
\lambda^{-2}\,\mu^2 \ , \nn\\g_k&\longmapsto&
\lambda^{2k(n-1)-2n}\,g_k \ , \nn\\\zeta&\longmapsto&\lambda^{n-1}\,
\zeta \ ,
\label{scaletransfs}\eeq
where we have redefined coordinates using (\ref{Landauscaling}) so that the
scaling parameter $s=1$ in (\ref{nPhiexp})--(\ref{V0wdef}). This
property is elementary to check by using (\ref{Landauscaling}) and
compensating the transformations (\ref{scaletransfs}) by the change of
matrix integration variables $M\to\lambda\,M$,
$M^\dag\to\lambda\,M^\dag$. The Jacobian $\lambda^{2N^2}$ of this
latter transformation is not important here, as it cancels out in the
Green's functions and leads to a finite, irrelevant shift of the free
energy.

This scaling property is important for the physical interpretation of
the large $N$ limit of the matrix model. It implies that we are free
to make the transformation (\ref{scaletransfs}) with $\lambda=N^\nu$ for
arbitrary real $\nu$, before taking the limit $N\to\infty$. The result
is independent of the exponent $\nu$. However, even though the
mathematical expressions for the Green's functions are unchanged, the
physical interpretation of the large $N$ limit does depend on
$\nu$. Choosing the scaling $x\to N^\nu\,x$ and sending $N\to\infty$
for $\nu>0$ means that we blow up small distances and ``zoom'' into
space. On the other hand, for $\nu<0$ the opposite phenomenon occurs,
we ``zoom'' out so that short distance structures become
invisible. The first scaling limit is an infrared limit while the
second one is an ultraviolet limit. To be more precise, let us put the
system on a hypercubic lattice of side $R$ and spacing $a$, so
that there are $N=(R/a)^{2n}$ lattice points. In conventional quantum
field theory, one usually performs the ultraviolet limit, which means
sending $a\to0$ keeping $R$ fixed (and adjusting all parameters so that
the limit makes mathematical sense). In the infrared (or
thermodynamic) limit, one sends $R\to\infty$ keeping $a$ fixed (so
that the finiteness of the spatial volume no longer matters).

This suggests that the parameter $\Lambda$ introduced in
(\ref{LambdaNtheta}) has different interpretations dependent on the
choice of exponent $\nu$. In the two cases described above we may
identify $\Lambda\propto1/R$ with $a\to0$ or $\Lambda\propto1/a$ with
$R\to\infty$, respectively. There can also be many other
interpolating identifications as well. The definition
(\ref{LambdaNtheta}) is simply the statement that there is a length scale in
the field theory, the renormalized magnetic length. Any field
theoretic interpretation of this length is possible, and this point of
view is consistent with the duality described in
Section~\ref{UVIR}. This is one of the various unconventional
properties possessed by the present class of quantum field
theories. In the ensuing subsections we shall make these statements
precise and explicit.

For this, we will introduce the two basic cutoffs above more
precisely. The infrared cutoff $R$ is like the diameter of space and
is defined by
\beq
R^2=4\pi N'\,\theta \ ,
\label{diameterdef}\eeq
with $N'\neq N$. The significance of this length scale is well-known
from the physics of the quantum Hall effect. To understand its
interpretation better, consider the $\ell=1$ Landau eigenfunctions
which from (\ref{genfnexpl}) and (\ref{Landauextract}) are given
explicitly by
\beq
\phi_{1,m}(x)=\frac{\overline{z}^{\,m-1}}{\sqrt{\pi\theta\,(m-1)!}}~
\e^{-|z|^2/2\theta} \ .
\label{phi0mx}\eeq
The one-particle ``location'' $|\phi_{1,m}(x)|^2$ thereby has a sharp
peak at $|z|\propto\sqrt{(m-1)\,\theta}$, and so restricting the quantum
numbers $m<N'$ is equivalent to the bound $|z|<R$ above. This
argument also works for $\ell>1$. In other words, the infrared
cutoff (\ref{diameterdef}) amounts to constraining the system quantum
numbers to lie below a finite number $N'$ within each Landau level, as
in the quantum Hall effect. This cutoff was ignored previously and
implicitly sent to infinity.

The other cutoff is our standard ultraviolet cutoff which is the
lattice spacing $a$ defined by
\beq
a^2=\frac1{\Lambda^2}=\frac{4\pi\theta}N \ .
\label{latticespacing}\eeq
For clarity, in the remainder of this section we shall work mostly in two
dimensions $n=1$ and only with the finite $N$ two-point function
which is given by
\beq
G_N(x,y)=-\sum_{\ell=1}^{N}\,\sum_{m=1}^\infty w_N(e_\ell)\,
\phi_{m,\ell}(x)\,\phi_{\ell,m}(y) \ .
\label{GNxydef}\eeq
Here $e_\ell=4B(\ell-\frac12)+\mu^2$ and $w_N(\xi)$ is a function
which converges to the rescaled loop function
$\frac1{\Lambda^2}\,W(\xi/\Lambda^2)$ of the matrix model in the limit
$N\to\infty$. According to (\ref{scaletransfs}) we may set $\zeta=1$
in two dimensions. Note that $e_\ell\to p^2+\mu^2$ at large $N$.

\subsection{Infrared Limit}

The particular scaling we studied earlier, whereby a finite
ultraviolet cutoff is kept, will be refered to as the ``infrared limit''
of the noncommutative field theory. Explicitly, it is defined by first
taking $R\to\infty$, then $N\to\infty$ such that $a$ is
finite. Looking at the non-interacting Green's function
(\ref{freepropmassless}), the asymptotic behaviour $K_0(z)\simeq-\ln z$ for
$z\to0$ implies that this limit corresponds to ``zooming'' into
space and the breaking of translation invariance by the magnetic field
becomes invisible. In terms of the scale transformations of the
previous subsection, it corresponds to setting $\nu=-1$ and rescaling
the couplings according to (\ref{scaletransfs}) with
$\lambda=1/\sqrt{4\pi\theta}$. The quantity
(\ref{LambdaNtheta}) is thereby interpreted as an ultraviolet cutoff
and the Green's function is computed in the limit
\beq
G_{\rm ir}(x,y)=\left.\lim_{N\to\infty}\,G_N(x,y)
\right|_{\theta=N/4\pi\Lambda^2}
\label{Guvxylim}\eeq
keeping $\Lambda$ finite. The result is given by
(\ref{2ptfnniceform}), with
$\frac{16\pi\ell}N\to\frac{p^2}{\Lambda^2}$ in the limit. Everything
is known about this limit. The quantum field theory is quasi-free,
i.e. everything can be calculated solely from the explicit knowledge
of the two-point Green's function. There is no new length scale
introduced between $a\to0$ and $\infty$.

\subsection{Ultraviolet Limit}

Combining the result above with the duality described in
Section~\ref{UVIR} immediately implies the existence of another
scaling in the noncommutative field theory in which the physics is
completely different. We will refer to this dual scaling limit as the
``ultraviolet limit''. In fact, it is straightforward to establish the
duality property of the finite $N$ Green's function
(\ref{GNxydef}) for fixed parameters $g$ and $\theta$. This follows
directly from the generating functional (\ref{genfnexpl}), which can
be easily checked to possess the symmetries ${\cal P}_{s,t}(-x)={\cal
  P}_{-s,-t}(x)$ and $\tilde{\cal P}_{s,t}(p)=2\pi\theta\,{\cal P}_{\ii
  s,\ii t}(\theta\,p)$. It follows that the Landau eigenfunctions obey
\bea
\tilde\phi_{\ell,m}(p)&=&-2\pi\ii^{\ell+m}\,\theta\,\phi_{m,\ell}(\theta\,p)
 \ , \nn\\\phi_{\ell,m}(-x)&=&(-1)^{\ell+m}\,\phi_{\ell,m}(x) \ ,
 \label{Landausyms}\eea
and hence that the regulated propagator has the duality
\beq
\tilde G_N(p,q)=(2\pi\theta)^2\,G_N(\theta\,p\,,\,\theta\,q)
\label{tildeGNpq}\eeq
under Fourier transformation.

This property implies that the limit
\beq
\tilde G_{\rm uv}(p,q)=\left.\lim_{N\to\infty}\,\frac1{N^2}\,
\tilde G_N\left(\frac pN\,,\,\frac qN\right)\right|_{\theta=N/
4\pi\Lambda^2}
\label{tildeGirpqdef}\eeq
keeping $\Lambda$ finite is also well-defined. By using
(\ref{2ptfnniceform}) and substituting $x=-p/4\pi\Lambda^2$ it can be
written as
\bea
\tilde G_{\rm uv}(p,q)&=&\frac1{4\Lambda^4}\,G\left(\frac
  p{4\pi\Lambda^2}\,,\,\frac
  q{4\pi\Lambda^2}\right)\nn\\&&{~~~~}^{~~}_{~~}\nn\\
&=&-\frac1{\Lambda^2}~\int\limits_{|x|
  \leq1/\sqrt\pi\,\Lambda}\dd^2x~W_{\rm b}\Bigl(
(4\pi\Lambda\,x)^2\Bigr)~\e^{-\ii x\cdot(p-q)} \ ,
\label{Girniceform}\eea
and it thereby gives an ultra-local Green's function in the ultraviolet
scaling limit,\footnote{\baselineskip=12pt Note that neither the
  ultraviolet nor the infrared limits respect magnetic
  translations. The reason was explained at the beginning of
  Section~\ref{Symms}. Magnetic translation invariance implies
  translation invariance in both position and momentum space (up to
  phase factors). However, the limits $\theta\to\infty$ and
  $\theta\to0$ are singular. In the latter case the translational symmetry
  is broken in position space (and the propagator is ultra-local
  there), while in the former limit it is broken in momentum
  space. The breaking of translational invariance also occurs to all
  orders in perturbation theory of standard noncommutative
  $\Phi^4$-theory with a finite cutoff~\cite{GW}.}
\beq
G_{\rm uv}(x,y)=-\frac1{\Lambda^2}\,W_{\rm b}\Bigl((4\pi\Lambda\,x)^2
\Bigr)~\delta^{(2)}(x-y)
\label{Girultra}\eeq
with $|x|\leq1/\sqrt\pi\,\Lambda$. In particular, in the free case it
reduces to
\beq
G_{\rm uv}^{(0)}(x,y)=\frac1{(4\pi\Lambda^2\,x)^2+\mu^2}~
\delta^{(2)}(x-y) \ .
\label{Gir0}\eeq
It is interesting to note that this scaling limit could also have been
defined directly as
\beq
G_{\rm uv}(x,y)=\left.\lim_{N\to\infty}\,N^2\,G_N(Nx,Ny)\right|_{
  \theta=N/4\pi\Lambda^2} \ .
\label{Girxydirect}\eeq
The limit in (\ref{Girxydirect}) is technically rather difficult to
carry out explicitly. But the duality symmetry of the quantum field
theory enables a quick and easy derivation from the corresponding
results of the infrared limit above.

This scaling corresponds to first taking the limit $a\to0$, then
$N'\to\infty$ such that $R$ is finite. In terms of the scale
transformations of Section~\ref{ScaleTransf}, it corresponds to
setting $\nu=1$ and rescaling the couplings according to
(\ref{scaletransfs}) with $\lambda=1/\sqrt{4\pi\theta}$. Note that
this means now $\theta\to0$ and $\Lambda$ is interpreted as a finite infrared
cutoff. From (\ref{freepropmassless}) and the asymptotic behaviour
$K_0(z)\simeq\sqrt{\frac\pi{2z}}~\e^{-z}$ for $z\to\infty$ it follows
that this limit corresponds to ``zooming'' out of space, i.e. probing
large distances of order $N/\Lambda=\sqrt{4\pi N\,\theta}$, and all
that is observable is the spatial dependence due to the magnetic field
whereas the center of mass dependence becomes ultra-local. Because of
the UV/IR duality, this ultraviolet limit is given by the {\it same}
't~Hooft limit of the matrix model, and all that happens is that the
long and short distance cutoffs of the theory are
interchanged. Everything about this limit is also understood, because
its properties are just dual to those of the previous subsection. In
particular, the quantum field theory is again quasi-free, but with
two-point Green's function given by the ultra-local form
(\ref{Girultra}), and there is no new length scale introduced between
$0$ and $R\to\infty$.

While this argument for going from the infrared to the ultraviolet
limit is mathematically very simple, it is somewhat paradoxical. In
the infrared limit we take $\theta\to\infty$ and then, by a simple
rescaling of the spatial arguments leaving the matrix integral
unchanged, we arrive at a limit whereby $\theta\to0$. The resolution
of this paradox comes from using the invariance of the matrix integral
under $M\to N^\nu\,M$, $M^\dag\to N^\nu\,M^\dag$ before the limit
$N\to\infty$ is taken. Thus if we consider just the matrix model, then
from (\ref{scaletransfs}) it follows that the
only relevant parameters are $\theta\,\mu^2$ and $\theta\,g$. This
suggests a different physical interpretation of the infrared limit above
which makes the existence of the ultraviolet limit somewhat less
paradoxical. The crucial computation for the infrared limit was the
sum over Landau levels (\ref{gammaellJ0lambda}) for large $\ell$ (of
order $N$). However, following the derivation of Appendix~G through,
it is clear from (\ref{GNxydef}) that the same result can be obtained
also in some limit keeping the noncommutativity parameter $\theta$
finite and
\beq
G_{\rm ir}(x,y)=\left.\lim_{N\to\infty}\,\frac1N\,G_N\left(\frac x{\sqrt N}
\,,\,\frac y{\sqrt N}\right)\right|_{\theta=1/4\pi\Lambda^2} \ .
\label{Guvaltdef}\eeq
The results obtained are exactly as before, only now $\Lambda$ has a
different interpretation. It is simply the inverse of the length scale
set by the noncommutativity parameter, and it need not be taken to
infinity any longer. In a similar vein, we may write
\beq
G_{\rm uv}(x,y)=\left.\lim_{N\to\infty}\,N\,G_N\left(\sqrt N\,x\,,\,\sqrt N
\,y\right)\right|_{\theta=1/4\pi\Lambda^2} \ .
\label{Giraltdef}\eeq
This makes clear the interpretations above as zooming in versus out of
space for the infrared versus ultraviolet limits. It is natural from
the point of view of the noncommutative duality.

This discussion motivates the search for alternative scaling limits in
between the other two that we have analysed, in which the Green's
functions have a non-trivial structure on both short
and large distance scales, and whereby the magnetic translation
symmetry is preserved. The simplest one uses
both infrared and ultraviolet cutoffs (\ref{diameterdef}) and
(\ref{latticespacing}), and corresponds to a scaling exponent
$\nu=0$ with $\lambda=1$. It comes from sending $a\to0$ and
$R\to\infty$ such that $\theta$ is finite, i.e. $N'=N\to\infty$ with
$\theta$ finite. This can be regarded as a ``true'' quantum field
theory limit, in which both ultraviolet and infrared cutoffs have been
removed. It is defined by
\beq
G_0(x,y)=\left.\lim_{N\to\infty}\,G_N(x,y)
\right|_{\theta=1/4\pi\Lambda^2} \ ,
\label{Gintxydef}\eeq
and it thereby corresponds to keeping all length scales in between the
infrared and ultraviolet regimes (no ``zooming'' in or out). We
believe that such intermediate scaling limits will only lead to an
interacting quantum field theory if some non-planar limit of the
matrix model is used. Such an analysis is beyond the scope of the
present work.

\subsection{Lifting the Degeneracy of Landau Levels}

The alternative scaling limits of the noncommutative quantum field
theory that we have proposed in this section, and in particular the
introduction of the intrinsic infrared cutoff (\ref{diameterdef}),
have the effect of essentially removing
the degeneracies from each Landau level. As is known from the theory
of the quantum Hall effect, an alternative way to accomplish this is
to add a confining electric potential to the Landau Hamiltonian, as we
did in (\ref{SVstartilde}). Thus this model has the potential of also
providing an exactly solvable interacting quantum field theory. There
are several ways that one may proceed in analysing this perturbation
of the original field theory. For instance, one may regard $\tilde\sigma$ in
(\ref{SVstartilde}) as a small parameter and simply examine the
perturbative corrections in $\sigma$ to the results above. This is
tractable, as we have illustrated how to completely solve the $\tilde\sigma=0$
model. Alternatively, one could pursue a saddle-point analysis of the
eigenvalue representation (\ref{ZNEafterIZ}) in the large $N$
limit. The effective eigenvalue action describes interacting particles
in a common self-interaction potential $\tilde V$ but with an electric
field dependent on each individual particle. Particle $\ell$ feels an
electric field $\ell/\theta$. Thus the equilibrium configuration of
the particles occurs when they are well-ordered. Permutation symmetry
is then broken, and rather than $N!$ identical saddle points, there is
just a single one. This is true for very large $\tilde\sigma$,
i.e. for equilibrium positions $h_\ell$ such that
$\sigma\,|h_\ell-h_m|\gg\theta$. For small $\tilde\sigma$ the
eigenvalues $h_\ell$ with $\ell\gg1$ accumulate close to the origin
and the logarithmic Vandermonde type repulsions at short distances
become important. Finally, one can generalize the auxilliary Hermitian
matrix integral representation of Appendix~C to this model, and
analyse it as described there. The solution of this generalized
noncommutative field theory represents an interesting challenge in the
search for non-trivial, exactly solvable interacting models.

Another, more difficult way to lift the degeneracy of the Landau
levels is to perturb the field theory away from the special point
$B=\theta^{-1}$. This of course destroys its $GL(\infty)$ symmetry and
hence its exact solvability. From (\ref{SVinfmatrix}) we see that
perturbing the action to $B\neq\theta^{-1}$ is like adding hopping
terms, giving a model defined on a two-dimensional lattice $\nat^{\,2}$
with somewhat unusual shifting actions. One way to proceed in solving
this lattice model would be to try to map the terms involving the
shift matrix (\ref{Gammadef}) onto the generators of an auxilliary
Heisenberg algebra represented by infinite-dimensional matrices. The
total kinetic energy term would then assume a more natural form. It
may be that these additional hopping terms produce relevant
deformations of the trivial Wilsonian renormalization group fixed
point which we have found. A similar sort of perturbation expansion
has been studied in~\cite{GW,MRW}.

Coupling scalar field theories to fermionic field theories with enough
supersymmetry is known to be a way to rid the noncommutative quantum
field theory of UV/IR mixing~\cite{MST,KT}. It may be that the
fermionic analog of the quantum field theory with action
(\ref{SVstar}) contributes power-like divergences which cancel those
of the bosonic theory, and therefore that appropriate supersymmetric
extensions yield a renormalizable interacting model. Such
supersymmetric field theories are the topic of the next section.

\newsection{Supersymmetric Extension}

In this final section we will formulate a supersymmetric extension of the
scalar quantum field theory defined by the action (\ref{SVstar}). We
will only present its explicit construction in detail, and leave aside
its exact solution for future work. The particularly interesting
aspect of this formulation is that non-trivial supersymmetric
interactions can only be formulated using the noncommutative
star-product. This fact is even apparent, as we will see, at the level
of the free supersymmetric action, whereby the supersymmetry
transformations are intrinsically operator-valued and hence are
parametrized by elements of a noncommutative algebra. We will begin by
writing down the analog of (\ref{SVstartilde}) with $\sigma=1$,
$\tilde\sigma=0$ for relativistic fermions,
and then proceed to describe the supersymmetric combination of the two
field theories. Throughout this section we work in two spacetime
dimensions and at the special point $B=\theta^{-1}$.

\subsection{Fermionic Models}

Consider the noncommutative field theory of a massive two-component Dirac
fermion field $\Psi(x)$ in two dimensions under the influence of a
constant background magnetic field. The action is
\eq
S_{\rm f}=\int\dd^2x~\Bigl[\Psi^\dag(x)\left(\sigma^i\,{\sf
    D}_i+\mu^{~}_{\rm f}\right)\Psi(x)+V_\star\left(\Psi^\dag\star\Psi
\right)(x)\Bigr] \ ,
\label{SVferm}\eqend
where $\sigma^i$, $i=1,2$ are the usual Pauli spin matrices. The
Dirac operator can be diagonalized in the complex coordinates
(\ref{complexcoords}) by expressing it in terms of the ladder operators
(\ref{ladderops}) as
\beq
\sigma^i\,{\sf D}_i=\frac2{\sqrt\theta}\,\pmatrix{0&{\sf a}
\cr-{\sf a}^\dag&0\cr} \ .
\label{Diracladder}\eeq
In terms of the Landau eigenfunctions, the orthonormal solutions of
the Dirac equation are then given by
\beq
\psi_{\ell,m}^\pm(x)=\frac1{\sqrt2}\,\pmatrix{\phi_{\ell-1,m}(x)
\cr\pm\ii\,\phi_{\ell,m}(x)\cr}
\label{Diracortho}\eeq
with
\beq
\sigma^i\,{\sf D}_i\,\psi_{\ell,m}^\pm=\pm\,2\ii\,\sqrt{\frac{\ell-1}
\theta}~\psi_{\ell,m}^\pm \ ,
\label{Diraceq}\eeq
where we have used (\ref{laddersonLandau}). Here $m=1,2,\dots$, while
$\ell=2,3,\dots$.

The case of $\ell=1$ should be treated separately. The $\pm$
eigenfunctions (\ref{Diracortho}) are interchanged under the action
of the two-dimensional chirality operator, which is the
diagonal Pauli matrix $\sigma^3$ in the basis of Dirac matrices that
we have chosen. In contrast, zero modes of the Dirac equation are all
chiral, as only one independent solution for each $m$ exists. They are given by
\beq
\psi_{1,m}(x)=\pmatrix{0\cr\phi_{1,m}(x)\cr}
\label{Diracorthozero}\eeq
with
\beq
\sigma^i\,{\sf D}_i\,\psi_{1,m}=0 \ .
\label{Diraceqzero}\eeq
The existence of chiral zero modes of the Dirac operator
follows from the index theorem,
according to which the number of zero modes of negative chirality
minus the number of zero modes of positive chirality is equal to the
total magnetic flux divided by $2\pi$. In the present case the total flux is
infinite, as is the number of chiral zero modes.

We can now expand the fermion field of (\ref{SVferm}) in the complete
basis of Landau wavefunctions to get
\beq
\Psi(x)=(4\pi\theta)^{1/4}~\sum_{m=1}^{\infty}\br{\,
\sum_{\ell=2}^{\infty}\,\sum_{s=\pm}
F_{s,m\ell}~\psi_{\ell,m}^s(x)+f_m~\psi_{0,m}^{~}(x)} \ ,
\label{PsiLandauexp}\eeq
where $F_\pm=(F_{\pm,m\ell})_{m\geq1\,,\,\ell\geq2}$ are complex
Grassmann-valued matrices, and $\veff=(f_m)_{m\geq1}$ is a
complex Grassmann-valued vector. Note that $F_{\pm}$ are
rectangular matrices, in contrast to the bosonic case. In terms of
them, the action (\ref{SVferm}) can be expressed as an infinite,
complex fermionic matrix model coupled to an infinite, complex
fermionic vector model by using the star-product identities
\bea
\left(\psi^s_{\ell,m}\right)^\dag\star\psi^{s'}_{\ell',m'}&=&
\frac1{\sqrt{4\pi\theta}}\,\delta^{ss'}\,\delta_{\ell\ell'}~
\phi_{m,m'} \ , \nn\\\psi^\dag_{0,m}\star\psi_{0,m'}&=&
\frac1{\sqrt{4\pi\theta}}\,\phi_{m,m'} \ , \nn\\
\left(\psi^s_{\ell,m}\right)^\dag\star\psi_{0,m'}&=&
\psi_{0,m'}^\dag\star\psi^s_{\ell,m}~=~0
\label{fermprojrel}\eeq
which follow easily from (\ref{Landauprojrel}). The action thereby
becomes
\bea
S_{\rm f}&=&\sum_{s=\pm}\Tr_{\cal H}^{~}\left(F_s\,{\cal K}^s\,
F_s^\dag\right)+\sum_{s=\pm}\sqrt{4\pi\theta}\,\mu^{~}_{\rm f}\,
\Tr_{\cal H}^{~}\left(F_s^\dag\,F_s\right)+\sqrt{4\pi\theta}\,
\mu^{~}_{\rm f}\,\veff^\dag\veff\nn\\&&+\,4\pi\theta\,\Tr_{\cal H}^{~}
\,V\left(\frac{\sum_{s=\pm}F_s\,F_s^\dag+\veff\veff^\dag}
{\sqrt{4\pi\theta}}\right) \ ,
\label{Sfermmatrix}\eea
where
\beq
{\cal K}^\pm_{\ell\ell'}=\pm\ii\,\sqrt{16\pi(\ell-1)}~\delta_{\ell\ell'} \
{}.
\label{defK}\eeq

\subsection{Supersymmetry Transformations}

We will now combine the fermionic model of the previous subsection
with the bosonic model of Section~\ref{BosDefs} and define a
supersymmetric field theory on a noncommutative phase space. At a
first glance, it appears impossible to have supersymmetry in such a
field theory, because bosons and fermions interact with an external
magnetic field in a different way. Indeed, the spectrum of the free
bosonic Hamiltonian, $2B(2\ell-1)$, is offset by the quantity
$2B$ with respect to the
spectrum of the square of the free Dirac operator, $4B(\ell-1)$. However,
this offset can be compensated by a shift in the boson mass. To see
how this works, let us compute the partition functions for free scalar
fields of mass $\mu$ and free fermion fields of mass $\mu^{~}_{\rm f}$. Up
to irrelevant constants, they are given by
\bea
\left(Z_{\rm b}^{(0)}\right)^{-1}
&=&\prod_{\ell,m=1}^\infty\br{\mu^2+4\ell B-2B} \ , \non
Z_{\rm f}^{(0)}&=&\prod_{m=1}^\infty\left[\mu^{~}_{\rm f}\,
\prod_{\ell>1}\br{\mu_{\rm f}^2+4\ell B-4B}\right] \ .
\label{Zbffree}\eea
If we now take $\mu^2=\mu_{\rm f}^2-2B$, then the functions in
(\ref{Zbffree}) are almost the same, up to a minor mismatch in their
zero mode contributions.

It is straightforward to see that the free action for the complex scalar
field and the Dirac fermion field, with masses related as above, is
indeed supersymmetric. Using (\ref{Boxab}) at $B\theta=1$ and
(\ref{Diracladder}), the action can be rewritten in terms of ladder
operators as
\beq
S_{\rm susy}^{(0)}=\int\dd^2x~\left[
\Phi^\dag(x)\br{\Aosc^\dag\Aosc+\mu_{\rm f}^2}\Phi(x)+
\Psi^\dag(x)\,\pmatrix{\mu^{~}_{\rm f}&\Aosc\cr
-\Aosc^\dag&\mu^{~}_{\rm f}\cr}\,\Psi(x)\right]
\label{Ssusy0}\eeq
where $\Aosc=2\,\sqrt B\,\aosc$. The action (\ref{Ssusy0})
is invariant under the infinitesimal supersymmetry transformations
\bea
\D_\epsilon\Phi&=&\epsilon^\dag\,\pmatrix{\Aosc^\dag&0\cr
0&\mu^{~}_{\rm f}\cr}\,\Psi \ , \non
\D_\epsilon\Phi^\dag&=&\Psi^\dag\,\pmatrix{\Aosc&0\cr
0&\mu^{~}_{\rm f}\cr}\,\ep \ , \non
\D_\epsilon\Psi&=&\pmatrix{\mu^{~}_{\rm f}&\Aosc\cr
-\Aosc^\dag&\mu^{~}_{\rm f}\cr}
\pmatrix{\Aosc&0\cr0&\mu^{~}_{\rm f}\cr}\,\ep\,\Phi \ , \non
\D_\epsilon\Psi^\dag&=&\Phi^\dag\,\ep^\dag\,
\pmatrix{\Aosc^\dag&0\cr0&\mu^{~}_{\rm f}\cr}
\pmatrix{\mu^{~}_{\rm f}&\Aosc\cr
-\Aosc^\dag&\mu^{~}_{\rm f}\cr} \ ,
\label{ss}\eea
where the parameters $\ep$ and $\ep^\dag$ of the supersymmetry transformations
are arbitrary Grassmann odd functions of the $\bosc$ and $\bosc^\dag$
operators in (\ref{ladderops}). Here we use the convention that the
action of a differential operator from the right is consistent with
integration by parts,~i.e.
\beq
f\,\aosc\equiv\bosc^\dag\,f \ , ~~ f\,\aosc^\dag\equiv\bosc\,f \ .
\label{arightdef}\eeq

It follows that the supersymmetry transformations in this case contain
infinitely many parameters. This resembles local supersymmetry
somewhat, except that here the parameters of the transformation are
arbitrary functions of differential operators rather than arbitrary
functions of the spacetime coordinates. A closer analogy is the type
of supersymmetry that arises in zero-dimensional supersymmetric matrix
models~\cite{Makeenko:1995sa}--\cite{Ambjorn:1996ym}, in which the parameter
of the transformation is an arbitrary matrix. In fact, the parameters
of the supersymmetry transformation in the present case become matrices
after expanding in the basis of Landau eigenfunctions. For this, we note that
the kinetic energy operator for
the boson field in (\ref{Ssusy0}) can be written in matrix form as
${\cal K^+}{\cal K}^-$, where ${\cal K}^\pm$ are the matrices defined
in \rf{defK}. The action (\ref{Ssusy0}) can thereby be written as the
supersymmetric matrix-vector model
\bea
S_{\rm susy}^{(0)}&=&\Tr^{~}_{\cal H}\br{\,\sum_{s=\pm}F_s\left({\cal K}^s+
\sqrt{4\pi\theta}\,\mu^{~}_{\rm f}\right)F^\dag_s+M^\dag\left(
{\cal K}^+{\cal K}^-+4\pi\theta\,\mu_{\rm f}^2\right)M}\non&&+\,
\sqrt{4\pi\theta}\,\mu^{~}_{\rm f}\,\veff^\dag\veff+4\pi\theta\,
\left(\mu_{\rm f}^2-2\theta^{-1}\right)\,\vbeta^\dag\vbeta \ ,
\label{Ssusy0matrix}\eea
where for convenience we have separated out the zero-mode part
$\beta_m\equiv M_{1m}$ of the scalar field, and in
(\ref{Ssusy0matrix}) it is understood that $M\equiv(M_{\ell
  m})_{\ell\geq2\,,\,m\geq1}$ is a rectangular matrix. The
action (\ref{Ssusy0matrix}) is invariant under the infinitesimal
supersymmetry transformations
\bea
\D_\epsilon F_s&=&-\epsilon_s\,M\,\left({\cal K}^{-s}+
\sqrt{4\pi\theta}\,\mu^{~}_{\rm f}\right) \ , \non
\D_\epsilon F_s^\dag&=&-\left({\cal K}^{-s}+\sqrt{4\pi\theta}\,
\mu^{~}_{\rm f}\right)\,M^\dag\,\epsilon_s^\dag \ ,
\non\D_\epsilon M&=&\sum_{s=\pm}\epsilon_s\,F_s^\dag \ , \non
\D_\epsilon M^\dag&=&\sum_{s=\pm}F_s\,\epsilon_s^\dag  \ ,
\label{matrixss}\eea
where $\epsilon_\pm\equiv(\epsilon_{\pm,\ell\ell'})_{\ell,\ell'\geq2}$
is a pair of infinite dimensional matrices. As expected, the
supersymmetry transformations do not involve zero modes.

Since ${\cal K}^+=({\cal K}^-)^\dag$, we have $(\D_\epsilon
F_s)^\dag\neq\D_\epsilon F_s^\dag$, and hence $F_s$ and $F_s\ddd$
should be regarded as independent variables in the functional integral
defining the corresponding quantum field theory. The
supersymmetry transformation (\ref{matrixss}) is then a legitimate
change of variables, and it leads to the supersymmetric Ward identities
in the usual way, as long as the integration measure is superinvariant.
The condition for this is
\eqq{\label{constr}
\Tr^{~}_{\cal H}\left(\epsilon_\pm+\epsilon_\pm^\dag\right)=0 \ .
}
The absence of zero modes in (\ref{matrixss}) and the constraint
\rf{constr} eliminate only a finite number of degrees of freedom.

\subsection{Supersymmetric Interactions}

Thus far we have considered a non-interacting supersymmetric quantum
field theory for which noncommutativity played no role.
Noncommutativity becomes important if we try to answer
the question of whether or not there exist interactions that preserve the
supersymmetry. We believe that the affirmative answer to this question can be
given {\it only} within the framework of noncommutative field theory.
It seems impossible to construct local interactions that are
invariant under the supersymmetry transformations \rf{ss}, though we
cannot prove this fact rigorously. However, ``star-local'' interacting quantum
field theories which possess the infinite-parameter supersymmetry
\rf{ss} do exist, as we now proceed to demonstrate.

The reason why noncommutativity aids the construction of Lagrangians
which are invariant under the supersymmetry transformations
of the previous subsection
is that the ladder operators (\ref{ladderops}) obey remarkable
Leibniz-type rules with respect to the star-product,
\ar{\label{leib}
\aosc\,(f\star f')&=&(\aosc\,f)\star f' \ , \non
\aosc^\dag\,(f\star f')&=&(\aosc^\dag\,f)\star f' \ , \non
\bosc\,(f\star f')&=&f\star(\bosc\,f') \ , \non
\bosc^\dag\,(f\star f')&=&f\star(\bosc^\dag\,f') \ , \non
(\bosc\,f)\star f'&=&f\star(\aosc^\dag\,f') \ , \non
(\bosc^\dag\,f)\star f'&=&f\star(\aosc\,f') \ .
}
These equalities can be proven by expanding the fields
$f$ and $f'$ in the Landau basis
and using (\ref{Landauprojrel}). Alternatively, the star-product
projector relation (\ref{Landauprojrel}) for the Landau wavefunctions
can be derived from the Leibniz rules (\ref{leib}), which can be
independently checked by a direct computation in the Fourier basis by
using (\ref{fstarg}). Formally, the relations (\ref{leib}) simply
reflect the fact the algebra of functions on $\real^2$, equipped with
the star-product, generates a bimodule for the harmonic oscillator
algebras (\ref{harmoscrels}) realized by the commuting $\aosc$ and
$\bosc$ operators. This was already implicit in the definitions
(\ref{arightdef}).

As usual, the construction of supersymmetric Lagrangians is facilitated
by the introduction of auxilliary fields $\cal F$. With them, we postulate
the supersymmetry transformations

\vbox{\bea
\D_\epsilon\Phi&=&\epsilon^\dag\,\pmatrix{\Aosc^\dag&0\cr
0&\zeta\cr}\,\Psi \ , \non
\D_\epsilon\Phi^\dag&=&\Psi^\dag\,\pmatrix{\Aosc&0\cr
0&\zeta\cr}\,\ep \ , \non
\D_\epsilon\Psi&=&\pmatrix{0&\Aosc\cr-\Aosc^\dag&0\cr}
\pmatrix{\Aosc&0\cr0&\zeta\cr}\,\ep\,\Phi-\ii\,
\pmatrix{\Aosc&0\cr0&\zeta\cr}\,\ep\,{\cal F} \ , \non
\D_\epsilon\Psi^\dag&=&\Phi^\dag\,\ep^\dag\,
\pmatrix{\Aosc^\dag&0\cr0&\zeta\cr}
\pmatrix{0&\Aosc\cr-\Aosc^\dag&0\cr}-
\ii\,{\cal F}^\dag\,\epsilon^\dag\,
\pmatrix{\Aosc^\dag&0\cr0&\zeta\cr} \ , \non
\D_\ep{\cal F}&=&\ii\,\ep^\dag\,\pmatrix{\Aosc^\dag&0\cr0&\zeta\cr}
\pmatrix{0&\Aosc\cr-\Aosc^\dag&0\cr}\,
\Psi \ , \non\D_\ep{\cal F}^\dag&=&\ii\,\Psi^\dag\,
\pmatrix{0&\Aosc\cr-\Aosc^\dag&0\cr}
\pmatrix{\Aosc&0\cr0&\zeta\cr}\,\ep \ ,
\label{sscalF}\eea}
\noindent
where $\zeta$ is an arbitrary parameter of mass dimension~1.
It can be set to unity, but then the left and right components
$\ep$ and $\ep^\dag$ will have different scaling dimensions.

There are two star-quadratic invariants of the supersymmetry
transformations (\ref{sscalF}) which contain at most two derivatives
of the fields. They are given by
\bea
{\cal L}_0&=&\Psi^\dag\star\pmatrix{0&\Aosc\cr
-\Aosc^\dag&0\cr}\,\Psi+\Phi^\dag\star\Aosc^\dag
\Aosc\,\Phi+{\cal F}^\dag\star{\cal F} \ , \non
{\cal L}_1&=&\Psi^\dag\star\Psi+\ii\,{\cal F}^\dag\star\Phi+\ii\,
\Phi^\dag\star{\cal F} \ .
\label{quadinvs}\eea
The superinvariance of the expressions (\ref{quadinvs}) depends crucially on
the bimodule properties \rf{leib} of the ladder operators.
We can now proceed to construct supersymmetric Lagrangians by taking
star-products of ${\cal L}_1$ and ${\cal L}_0$. The most general renormalizable
Lagrangian, i.e. the one with no couplings of negative
dimension, is then given by
\beq
{\cal L}={\cal L}_0+\mu_{\rm f}\,{\cal L}_1+\frac g2\,{\cal L}_1
\star{\cal L}_1 \ .
\label{mostgenrenL}\eeq
The auxilliary fields can be eliminated from (\ref{mostgenrenL}) by
using the equations of motion. Then the second term gives masses to
the boson and fermion fields, while the third term induces various
interactions. These interactions are star-local, but non-polynomial in
$\Phi$ and $\Psi$. Explicitly, one finds
\bea
S_{\rm susy}&=&\int\dd^2x~\left[
\Phi^\dag(x)\left(\Aosc^\dag\Aosc+\mu_{\rm f}^2\right)\Phi(x)+
\Psi^\dag(x)\,\pmatrix{\mu^{~}_{\rm f}&\Aosc\cr
-\Aosc^\dag&\mu^{~}_{\rm f}\cr}\,\Psi(x)\right.\non&&
+\,2g\mu_{\rm f}^2\,\Phi^\dag\star\Phi\star
\Phi^\dag\star\Phi(x)+2g\mu_{\rm f}\,\Phi^\dag\star\Phi\star
\Psi^\dag\star\Psi(x)+\frac g2\,\Psi^\dag\star\Psi\star
\Psi^\dag\star\Psi(x)\non&&+\Biggl.\,O\left(g^2\right)\Biggr] \ .
\label{susyints}\eea

The action (\ref{susyints}) possesses the usual star-local $GL(\infty)$
symmetry as before and hence yields a potentially solvable supersymmetric
quantum field theory. It would be interesting to seek a superspace
formulation of this model, which may aid in finding its
exact solution. However, it is not clear what the concept of a
superspace could mean in the present context. The supersymmetry here
acts as a fermionic rotation on the Landau levels, which has nothing
in common with ordinary supersymmetry. In particular, the commutator
of supercharges is not a spacetime
translation~\cite{SS,Ambjorn:1996ym}. The definition of noncommutative
superspaces has been addressed recently within various different contexts
in~\cite{CZ}--\cite{FLM}. In the present case, supersymmetric
rotations are the super-analogs of the bosonic $GL(\infty)$ symmetry,
so it is tempting to speculate that together they generate an infinite
dimensional $GL(\infty|\infty)$ supergroup. Geometrically,
this noncommutative supersymmetry would then correspond to the superspace
generalization of area-preserving diffeomorphisms.

\subsection*{Acknowledgments}

This paper is dedicated to the memory of Ian Kogan, who will be deeply
missed as both a good friend and physicist. We thank M.~Abou-Zeid, G.~Akemann,
A.~Alekseev, J.~Ambj\o rn, H.~Braden, N.~Dorey, V.~Elias, A.~Gorsky,
H.~Grosse, P.~Horv\'athy, D.~Johnston, D.~O'Connor, V.~Kazakov, I.~Kogan,
D.~Mateos, A.~Orlov, J.-H.~Park and T.~Turgut for helpful discussions and
correspondence. E.L. and K.Z. would like to thank the Erwin
Schr\"odinger Institute in Vienna, and E.L. would like to thank the
Feza Gursey Institute in Istanbul, for hospitality during part of this
work. This work was supported in part by the Swedish Science Research
Council~(VR) and the G\"oran Gustafssons Foundation. The work of
R.J.S. was supported in part by an Advanced Fellowship from the
Particle Physics and Astronomy Research Council~(U.K.). The work of
K.Z. was supported in part by RFBR grant 01--01--00549 and grant
00--15--96557 for the promotion of scientific schools.

\setcounter{section}{0}

\appendix{Mean Field Analysis of Commutative Complex $\Phi^4$-Theory}

In this appendix we point out that the models studied in this paper
also have a concrete physical motivation as a novel kind of mean field
theory for conventional (commutative) complex $\Phi^4$-theory. For this, we
consider the action $S_0+S_{\rm int}$ for ordinary two-dimensional
complex $\Phi^4$-theory written in momentum space, with free part
\beq
S_0=\int\dd^2k~\left(k^2+\mu^2\right)\,\tilde\Phi^\dag(k)\,\tilde\Phi(k)
\ ,
\label{freecommPhi4}\eeq
and the interaction
\beq
S_{\rm int}=\frac{g_0}2\,\int\dd^2k~\dd^2k'~\dd^2q~\dd^2q'~
\delta^{(2)}(k'-k+q'-q)\,\tilde\Phi^\dag(k')\,\tilde\Phi(k)\,
\tilde\Phi^\dag(q')\,\tilde\Phi(q)
\label{intcommPhi4}\eeq
describing all possible momentum conserving processes whereby
particles with incoming momenta $k$ and $q$ are scattered to new
momenta $k'$ and $q'$. Standard mean field theory can be applied to
this theory by truncating the interaction to retain only the Hartree
and Fock terms. In the former interactions the individual particle
momenta are conserved, $k=k'$ and $q=q'$, while in the latter terms
they are exchanged, $k=q'$ and $q=k'$. In the present case these two
types of terms are actually the same, and their sum yields the
Hartree-Fock interaction
\beq
S_{\rm HF}=g_0\,\int\dd^2k~\dd^2q~\tilde\Phi^\dag(k)\,\tilde\Phi(k)
\,\tilde\Phi^\dag(q)\,\tilde\Phi(q) \ .
\label{HFcommPhi4}\eeq

The quantum field theory with action $S_0+S_{\rm HF}$ can be solved
exactly, and its exact solution is identical to mean field theory for
commutative complex $\Phi^4$-theory. Up to now our discussion could have been
easily extended to arbitrary spacetime dimensionality, but in two
dimensions there are special mixed interactions which are of Hartree
type in one component of the momenta and of Fock type in the other
component, $k'=(k_1',k_2')=(k_1,q_2)$ and $q'=(q_1',q_2')=(q_1,k_2)$,
or vice versa~\cite{ELmf}. The sum of all of these mixed terms yields
the interaction
\beq
S_{\rm mixed}=g_0\,\int\dd^2k~\dd^2q~\tilde\Phi^\dag(k_1,q_2)\,
\tilde\Phi(k_1,k_2)\,\tilde\Phi^\dag(q_1,k_2)\,\tilde\Phi(q_1,q_2) \ .
\label{mixedcommPhi4}\eeq
We now claim that the model with action $S_0+S_{\rm mixed}$ is
equivalent to the one studied in this paper. For this, we introduce a
regularization by restricting the momenta to
$k=(k_1,k_2)=\frac{2\pi}R\,(\ell,m)$ with
$\ell,m=0,\pm\,1,\pm\,2,\dots,\pm\,R/a\equiv(N-1)/2$, where $R<\infty$
is a large but finite radius for the size of
spacetime serving as an infrared cutoff, and $a>0$ is a small but
finite lattice spacing serving as an ultraviolet cutoff. We can then
identify the Fourier modes of the fields with the elements of an
$N\times N$ matrix $M=(M_{\ell m})$ as
\beq
M_{\ell m}=\left(\frac{2\pi}R\right)^2\,\tilde\Phi\left(\frac{2\pi\ell}R
\,,\,\frac{2\pi m}R\right) \ .
\label{McommPhi4}\eeq
By introducing a diagonal matrix $E$ with elements
\beq
E_{\ell m}=\ell^2~\delta_{\ell m} \ ,
\label{calFdef}\eeq
we may thereby write the regulated action as a complex matrix model
\beq
S_0+S_{\rm mixed}=\Tr\left(\sigma\,M^\dag\,E\,M+
  \tilde\sigma\,M\,E\,M^\dag+\mu^2\,M^\dag M+\frac g2\,
  M^\dag MM^\dag M\right)
\label{summatrixcommPhi4}\eeq
with $\sigma=\tilde\sigma=1$, $g=2g_0/R^2$, and $\Tr$ the usual
$N\times N$ matrix trace as before.

The matrix integral that we evaluate in Section~\ref{Toda} provides an
exact and explicit formula for the free energy of this model. In
Sections~\ref{Exact} and~\ref{CorrFns} we find the complete solution
of the model in the limit $R\to\infty$ for the special maximally
anisotropic case $\tilde\sigma=0$. In section~\ref{Scaling} we examine
other possible ways of removing the cutoffs in the anisotropic
model. The physical interpretation of the model presented here gives a
strong motivation for extending the results of this paper to the more
complicated isotropic case where $\sigma=\tilde\sigma$.

\appendix{UV/IR Duality versus T-Duality}

Combining the translation and symplectomorphism invariances of the
  action (\ref{SVstar}) with
  its Morita-type duality symmetry found in~\cite{LS} lends further
  support to the observation of~\cite{LS} that the quantum field
  theory defined by (\ref{SVstar}) may be regarded
  as a discrete noncommutative $\zeds_2$ gauge theory when $B=\theta^{-1}$. For
  this, we introduce a constant, background metric $H=(H_{\mu\nu})$ on
$\real^{2n}$ and use it to rewrite the action (\ref{SVstar}) in a
covariant form. We regard the action $S_{\rm b}$ as a functional of
the dynamical field $\Phi$, the background fields $H$ and $B$, and the
coupling parameters $g$ and $\theta$. The action then has a duality
under Fourier transformation~\cite{LS},
\beq
S_{\rm b}[\Phi;H,B,g,\theta]=\Bigl|\det(B/2\pi)\Bigr|~
S_{\rm b}\left[\tilde\Phi;
\tilde H,\tilde B,\tilde g,\tilde\theta\right] \ ,
\label{Sbduality}\eeq
where the dual parameters in momentum space are given by

\vbox{\bea
\tilde H&=&B^{-1}\,H\,B^{-1} \ , \nn\\\tilde B&=&B^{-1} \ ,
\nn\\\tilde g&=&\frac{g}{\Bigl|\det(2\pi\theta)\Bigr|} \ , \nn\\
\tilde\theta&=&\theta^{-1} \ .
\label{dualpars}\eea}
\noindent
Note that, in contrast to the transformations of Section~\ref{UVIR}, here
we do not rescale the arguments of the fields.

There is a novel gauge Morita equivalence interpretation of this noncommutative
duality, wherein we may heuristically regard the model (\ref{SVstar})
as a noncommutative gauge theory defined along $2n$
finite discrete directions, i.e. on a two-sheeted manifold
$\real^{2n}\times\zed_2$. Noncommutative Yang-Mills theory defined on a
$2n$-dimensional {\it torus} is manifestly invariant under the standard
$SO(2n,2n,\zed)$ open string T-duality transformations~\cite{sz1}
\bea
\tilde\theta&=&({\sf A}\,\theta+{\sf B})^\top\,(\theta\,{\sf Q}-
{\sf N})^{-1} \ , \nn\\\tilde H&=&(\theta\,{\sf Q}-{\sf N})^\top\,
H\,(\theta\,{\sf Q}-{\sf N}) \ , \nn\\ \tilde g_{\rm YM}^2&=&
g_{\rm YM}^2~\det(\theta\,{\sf Q}-{\sf N}) \ ,
\label{Moritaduality}\eea
where the superscript $\top$ denotes transposition, $g^{~}_{\rm YM}$ is the
Yang-Mills coupling constant, and in this equation $\theta$ is the
dimensionless noncommutativity parameter. The $2n\times2n$ symmetric integral
matrix $\sf N$ is proportional to the rank $N$ of the gauge theory, while the
antisymmetric integral matrix $\sf Q$ is determined by the magnetic fluxes of
the gauge bundle around the various two-cycles of the torus. The symmetric and
antisymmetric integral matrices $\sf A$ and $\sf B$, respectively, are chosen
to solve the generalized Diophantine equation
\beq
{\sf A}\,{\sf N}+{\sf B}\,{\sf Q}=\id_{2n} \ .
\label{Dioeqn}\eeq

Going back to the scalar field theory we may, in the usual Connes-Lott type
interpretation of quartic scalar field theory~\cite{CL}, regard $\Phi$ as the
off-diagonal components of a superconnection, and the field theory
(\ref{SVstar}) as induced by the limit in which the diagonal gauge degrees of
freedom are frozen out in the usual noncommutative Yang-Mills action for the
superconnection. The noncommutative scalar field theory is then formally the
zero rank limit of a noncommutative gauge theory. All formulas above still make
mathematical sense in this limit and it is heuristically the choice that should
be made for a discrete gauge theory. It means that one should set
\beq
{\sf N}={\sf 0} \ .
\label{sfNsf0}\eeq
The equation (\ref{Dioeqn}) is then solved by
\beq
{\sf A}={\sf 0} \ , ~~ {\sf Q}=-{\sf B}=\pmatrix{0&-1\cr1&0\cr}
\otimes\id_n \ .
\label{AQBchoice}\eeq
Then the transformation rules (\ref{Moritaduality}) coincide exactly
with (\ref{dualpars}) at the special point $B=\theta^{-1}$. This holds
with the identification $g_{\rm YM}^2=((2\pi)^n\,g)^{-1}$. For this
``discrete'' truncation of the toroidal compactification, the choice
of flux matrix $\sf Q$ in (\ref{AQBchoice}) is very natural, given the
charges carried by the scalar fields $\Phi$ (see
(\ref{SVlargesymspace})). This gauge theoretic interpretation also
indicates why the duality property is uniquely possessed by a quartic
interaction of the scalar field, which has other special features as
well which are displayed in the main text. It would be interesting to
pursue this gauge theoretic interpretation further using the usual
techniques of noncommutative geometry, as it may help elucidate
further features of the model and its generalizations.

\appendix{Alternative Derivation of the Loop Equations}

In this appendix we will present another derivation of the Schwinger-Dyson
equations
(\ref{largeNWxeqn}) for the matrix model (\ref{ZNEdef}) with potential
(\ref{Phi4pot}). This
derivation has the advantage of being somewhat more amenable to generalization
to
other noncommutative complex $\Phi^4$ field theories, such as that associated
with the
matrix model
(\ref{ZNEtildeE}). It also gives another physical interpretation to the loop
function (\ref{WxlnZdef}) in terms
of the spectral distribution of an auxilliary Hermitian one-matrix model.

We introduce an auxilliary $N\times N$ Hermitian matrix $X=(X_{\ell m})$
through the
Hubbard-Stratonovich transformation
\beq
\e^{-\frac g2\,\Tr(MM^\dag)^2}=\int\DD X~\e^{-\Tr(\frac1{2g}\,X^2+\ii
XMM^\dag)} \ ,
\label{HStransf}\eeq
where $\DD X=\prod_\ell\dd X_{\ell\ell}~\prod_{m<\ell}\dd\,{\rm Re}\,X_{\ell
m}~\dd\,{\rm Im}\,X_{\ell m}/2\pi$. After a rescaling of $X$ and $M$ by the
matrix dimension $N$, the partition function (\ref{ZNEdef}) may
then be written in the form
\beq
Z_N(E)=\int\DD M~\DD M^\dag~\int
\DD X~\e^{-\frac N{2\tilde g}\,\Tr\,X^2-\vec M^\dag(\id\otimes E^\top+
\ii X\otimes\id)\vec M} \ ,
\label{ZNEHS}\eeq
where for convenience we have changed variables from $N\times N$ complex
matrices
$M=(M_{\ell m})$ to vectors $\vec M\in\complex^{N^2}$ through the rule
\beq
\vec M=(M_{11},M_{12},\dots,M_{1N},M_{21},\dots,M_{NN})^\top \ .
\label{vecMdef}\eeq
The Gaussian integration over $M$ and $M^\dag$
can now be carried out in this manner, giving the effective Hermitian
one-matrix integral with a Penner type
interaction potential,
\beq
Z_N(E)=\int\DD X~\e^{-\frac N{2\tilde g}\,\Tr\,X^2-\Tr_{\rm
Ad}^{~}\,\ln(\id\otimes E^\top+\ii
X\otimes\id)} \ ,
\label{ZNEX}\eeq
where $\Tr_{\rm Ad}^{~}$ is the matrix trace in the adjoint representation.

The matrix integral (\ref{ZNEX}) is invariant under unitary transformations of
both $X$
and $E$, and can thus be written as an eigenvalue model. For this, we
diagonalize
$X=U\,{\rm diag}(\mu_1,\dots,\mu_N)\,U^\dag$,
with $\mu_\ell\in\real$ and $U\in U(N)$, compute the Jacobian of the
integration measure to get
$\DD X=[\dd U]~\prod_\ell\dd\mu_\ell~\Delta_N[\mu]^2$, and thereby write
(\ref{ZNEX}) as
\beq
Z_N(E)=\prod_{\ell=1}^N~\int\limits_{-\infty}^\infty\dd\mu_\ell~\e^{-
\frac N{2\tilde g}\,\mu_\ell^2}~\prod_{m=1}^N\frac1{\lambda_m+\ii\mu_\ell}~
\Delta_N[\mu]^2 \ .
\label{ZNEmu}\eeq
The saddle point equation for the large $N$ limit of the eigenvalue integral
(\ref{ZNEmu})  is
\beq
\frac{\mu_\ell}{\tilde g}-\frac1N\,\sum_{m\neq\ell}\frac1{\mu_\ell-\mu_m}+
\frac1N\,\sum_{m=1}^N\frac1{\mu_\ell-\ii\lambda_m}=0 \ .
\label{saddleptX}\eeq
We now multiply the expression (\ref{saddleptX}) through by
$\frac1N\,(\mu_\ell-z)^{-1}$ for
$z\in\complex$, sum over all $\ell=1,\dots,N$,
and expand the terms to rewrite them in terms of the resolvent function
\beq
\Sigma(z)=\frac1N\,\sum_{\ell=1}^N
\frac1{\mu_\ell-z} \ .
\label{Sigmaz}\eeq
By rewriting the sums over the eigenvalues $\lambda_m$ of the external field
$E$ as integrals
over the spectral density (\ref{rhoE}), and dropping the term $\Sigma'(z)/N$ at
$N=\infty$,
we may in this way write (\ref{saddleptX}) as the large $N$ loop equation
\beq
\frac1{\tilde
g}\,\Bigl(1+z\,\Sigma(z)\Bigr)+\Sigma^2(z)+\int\limits_{a_1}^{a_2}\dd\lambda~
\frac{\rho(\lambda)}{z+\ii\lambda}\,\Bigl(\Sigma(z)-\Sigma(-\ii\lambda)\Bigr)=0
\ .
\label{loopeqnX}\eeq
The loop equation (\ref{loopeqnX}) is essentially the same as
(\ref{largeNWxeqn}).
Since the resolvent function (\ref{Sigmaz}) has the asymptotic behaviour
$\Sigma(z)\simeq-\frac1z+
O(\frac1{z^2})$ for $z\to\infty$, it then follows that it may be identified
with the function (\ref{WxlnZdef})
through
\beq
\Sigma(z)=-\ii W(-\ii z) \ .
\label{SigmaWrel}\eeq

This coincidence is not surprising, as both functions $W$ and $\Sigma$
are the generating functions of correlators of derivatives of $M^\dag
M$ acting in the complex matrix model. Indeed, a straightforward
calculation using the Hubbard-Stratonovich transformation above
shows
\beq
\delta_{\ell\ell'}~W(\lambda_\ell)\equiv\frac1N\,\Bigl\langle
\left(M^\dag M\right)_{\ell\ell'}\Bigr\rangle_E=\delta_{\ell\ell'}~
\left\langle\,\frac1N\,\sum_{m=1}^N\frac1{\lambda_\ell+\ii\mu_m}\,
\right\rangle_{\rm HS} \ ,
\label{HScorr}\eeq
where the second expectation value is taken with respect to the
Hubbard-Stratonovich representation (\ref{ZNEX}). The usefulness of
this approach, however, is that the eigenvalue model (\ref{ZNEmu}) may
be straightforwardly generalized to include also the external field
$\tilde E$ that appears in the matrix model (\ref{ZNEtildeE}), and its
loop equations may be analysed by the technique presented here. The
exact solvability of the matrix model (\ref{ZNEtildeE}) may thereby be
attributed to the $U(N)$ symmetry of the $N\times N$ matrix model of
the Hubbard-Stratonovich field $X$. This
then has the potential of presenting an exact, non-perturbative
solution of the generalized noncommutative quantum field theory with
action (\ref{SVstartilde}) and interaction potential given by
(\ref{Phi4pot}). Furthermore, this technique has the advantage of
straightforwardly supplying an explicit formula for the generating
functional (\ref{ZbJgen}) of the connected Green's functions of the model.

\appendix{Explicit Solution of the Master Equation at Large $N$}

To solve the non-linear integral equation (\ref{largeNWxeqn}), we will
reduce it to a standard Riemann-Hilbert problem of the type that
appears for ordinary Hermitian one-matrix models. For this, we
introduce the resolvent function of the external field $E$ which is
defined as the Hilbert transform
\beq
\omega(z)=\int\limits_{a_1}^{a_2}\dd\lambda~\frac{\rho(\lambda)}
{z-\lambda}
\label{omegadef}\eeq
of the spectral density (\ref{rhoE}), and the function
\beq
\Omega(z)=\int\limits_{a_1}^{a_2}\dd\lambda~\frac{\rho(\lambda)
\,W(\lambda)}{z-\lambda} \ .
\label{Omegadef}\eeq
Both of these functions are analytic everywhere in the complex
$z$-plane, except on the support $[a_1,a_2]$ of the spectral
distribution where they each have a branch cut. Let
\beq
\omega_\pm(\lambda)=\frac12\,\Bigl(\omega(\lambda+\ii0)\pm
\omega(\lambda-\ii0)\Bigr) \ , ~~ \lambda\in[a_1,a_2]
\label{omegapmdef}\eeq
be the continuous and singular parts of the function $\omega(z)$
across its branch cut. We similarly define $\Omega_\pm(\lambda)$ for
$\lambda\in[a_1,a_2]$. Note that
$\omega_-(\lambda)=\ii\pi\,\rho(\lambda)$ and
$\Omega_-(\lambda)=\ii\pi\,W(\lambda)\,\rho(\lambda)$.

{}From the definitions (\ref{omegadef}) and (\ref{Omegadef}) it follows
that the singular parts of these functions are related through
\beq
\Omega_-(\lambda)=W(\lambda)\,\omega_-(\lambda) \ , ~~ \forall
\lambda\in\real \ .
\label{Omegaomegasingrel}\eeq
Furthermore, from the large $N$ Schwinger-Dyson equations
(\ref{largeNWxeqn}) it follows that the continuous part of
(\ref{Omegadef}) obeys
\beq
\tilde g\,\Omega_+(\lambda)=\tilde g\,W^2(\lambda)-\Bigl(\lambda-
\tilde g\,\omega_+(\lambda)\Bigr)\,W(\lambda)-1 \ , ~~
\lambda\in[a_1,a_2] \ .
\label{Omegaconteqn}\eeq
Assuming $\tilde g\neq0$, this suggests the analytic ansatz
\beq
\Omega(z)=W^2(z)-\left(\,\frac z{\tilde g}-\omega(z)\right)\,
W(z)-\frac1{\tilde g} \ , ~~ \forall z\in\complex \ .
\label{analyticansatz}\eeq
Substituting (\ref{analyticansatz}) into (\ref{Omegaomegasingrel}) and
(\ref{Omegaconteqn}) then implies the respective restrictions
\bea
W_-(\lambda)\,\left(\,\frac\lambda{\tilde g}-\omega_+(\lambda)-2\,
W_+(\lambda)\right)&=&0 \ , ~~ \forall\lambda\in\real \ , \label{Wminusreal}\\
W_-(\lambda)\,\Bigl(W_-(\lambda)+\omega_-(\lambda)\Bigr)&=&0 \ ,
{}~~ \lambda\in[a_1,a_2]
\label{Wminusa}\eea
on the real-valued analytic function $W(z)$ in the complex
$z$-plane. In deriving (\ref{Wminusreal}) from (\ref{analyticansatz})
we have assumed that $W(z)$ has no poles.

In addition to (\ref{Wminusreal}) and (\ref{Wminusa}), a third
restriction on the function $W(z)$ is imposed by the boundary
conditions required to solve the Schwinger-Dyson equations. From
(\ref{largeNWxeqn}) it follows that there are two branches of
solution, one with the asymptotic behaviour $W(z)\simeq\frac z{\tilde
  g}$ for $z\to\infty$, and the other with $W(z)\simeq-\frac1z$ for
$z\to\infty$. We will take the latter boundary condition, as it is the
one which matches that of the perturbative solution
(\ref{loopiterative}) of the complex matrix model. This branch reduces
continuously to the Gaussian solution
$W^{(0)}(\lambda)=-\frac1\lambda$ at $\tilde
g=0$.\footnote{\baselineskip=12pt The former asymptotic behaviour is
  the pertinent boundary condition to use in solving the corresponding
  Hermitian matrix model in an external field, obtained by the formal
  substitution $X=M^\dag M$ in the matrix integral (\ref{ZNEdef}),
  along with the appropriate change of integration domain. This
  Gaussian Hermitian matrix integral is proportional to
  $\e^{N\,\Tr\,E^2/2\tilde g}$, and it thereby produces a solution
  which is singular at $\tilde g=0$. More precisely, the loop equation
  (\ref{largeNWxeqn}) is very similar to that of the Kontsevich-Penner
  model~\cite{CM1}--\cite{ZC} whose solution is given by the singular
  branch. While their equations of motion are the same, the complex
  and Hermitian matrix models differ in the choice of boundary
  conditions. The situation here is in marked contrast to that in the
  absence of the external field, whereby the difference between
  partition functions calculated in the large $N$ limit with different
  integration domains is exponentially small~\cite{AJM}, and hence does not
  affect the solution at leading order $N=\infty$.} Thus we will in
addition require
\beq
W(z)\simeq-\frac1z+O\left(\frac1{z^2}\right)~~~~{\rm as}~~z\to\infty \
{}.
\label{Wasympt}\eeq

We now solve (\ref{Wminusreal})--(\ref{Wasympt}) by making a one-cut
ansatz, which will lead to a rational parametrization of the
solution. For this, we assume that $W_-(\lambda)$ is non-vanishing on
a single connected interval $[b_1,b_2]$ in the complex
$\lambda$-plane. To satisfy (\ref{Wminusa}) we must then have
\beq
[a_1,a_2]\cap[b_1,b_2]=\emptyset \ .
\label{abintempty}\eeq
Since the resolvent $\omega(z)$ is analytic everywhere away from its
branch cut on $[a_1,a_2]$, from (\ref{abintempty}) it follows that
$\omega_-(\lambda)=0$ for $\lambda\in[b_1,b_2]$. From
(\ref{Wminusreal}) we then have
\beq
W_+(\lambda)=\frac12\,\omega(\lambda)-\frac\lambda{2\tilde g} \ , ~~
\lambda\in[b_1,b_2] \ .
\label{Wconteq}\eeq
We can now turn (\ref{Wconteq}) into a standard Riemann-Hilbert equation
\beq
\left(\frac{W(\lambda)}{\sqrt{(\lambda-b_1)(\lambda-b_2)}}
\right)_-=\frac{\frac\lambda{\tilde g}-\omega(\lambda)}
{2\ii\,\sqrt{(b_2-\lambda)(\lambda-b_1)}} \ , ~~ \lambda\in[b_1,b_2] \
  .
\label{Wdisconteq}\eeq

It follows that the function $W(z)$ is given everywhere in the complex
$z$-plane by the contour integral
\beq
W(z)=\oint\limits_{\stackrel{\scriptstyle[b_1,b_2]}{\scriptstyle
w\neq z}}\frac{\dd w}{4\pi\ii}~\frac{\omega(w)-
\frac w{\tilde g}}{w-z}~\sqrt{\frac{(z-b_1)(z-b_2)}
{(w-b_1)(w-b_2)}} \ .
\label{Wzeverywhere}\eeq
The discontinuity equation (\ref{Wdisconteq}) determines the solution
(\ref{Wzeverywhere}) uniquely up to terms which are regular at $z=0$,
and the large $z$ behaviour (\ref{Wasympt}) implies that the regular
terms vanish. Substituting (\ref{omegadef}) into (\ref{Wzeverywhere})
and integrating along the cut $[b_1,b_2]$ then yields
\bea
W(z)&=&\frac z{2\tilde g}-\frac{\sqrt{(z-b_1)(z-b_2)}}{2\tilde g}
\nn\\&&-\,\frac12\,
\int\limits_{a_1}^{a_2}\dd\lambda~\frac{\rho(\lambda)}{z-\lambda}~
\frac{\sqrt{(z-b_1)(z-b_2)}-\sqrt{(\lambda-b_1)(\lambda-b_2)}}
{\sqrt{(\lambda-b_1)(\lambda-b_2)}} \ .
\label{Wzfinal}\eea
The large $z$ behaviour (\ref{Wasympt}) also generates two boundary
conditions which unambiguously determine the branch points $b_1$ and
$b_2$ of the function $W(z)$. Expanding the right-hand side of
(\ref{Wzfinal}) for $z\to\infty$ we encounter a constant term, which
must vanish, and a term proportional to $\frac1z$, with known residue
$-1$. This supplements (\ref{Wzfinal}) with the respective constraints
\bea
b_1+b_2&=&-2\tilde g\,\int\limits_{a_1}^{a_2}\dd\lambda~
\frac{\rho(\lambda)}{\sqrt{(\lambda-b_1)(\lambda-b_2)}} \ , \nn\\
3\left(b_1^2+b_2^2\right)+2\,b_1b_2+8\tilde g&=&-8\tilde g\,
\int\limits_{a_1}^{a_2}\dd\lambda~\frac{\lambda\,\rho(\lambda)}
{\sqrt{(\lambda-b_1)(\lambda-b_2)}} \ .
\label{b1b2eqns}\eeq

\appendix{Calculation of the Convolution Kernels\label{ConvKerApp}}

The trace formula (\ref{Landautraceformula}) for the integration
kernel ${\cal G}_L$ can be written in terms of
the generating function (\ref{genfnexpl}) for the Landau wavefunctions as
\bea
{\cal G}_L(x_1,\dots,x_L)&=&\left(\frac1{\sqrt{4\pi\theta}}
\right)^{L-2}~\prod_{I=1}^L\,\int\frac{\dd u_I~
\dd\overline{u_I}}{\ii\pi}~\e^{-|u_I|^2}\nn\\&&\times\,
{\cal P}_{\overline{u_1},u_2}(x_L)\,{\cal
P}_{\overline{u_2},u_3}(x_{L-1})\cdots
{\cal P}_{\overline{u_L},u_1}(x_1) \ .
\label{calGgenfn}\eea
This representation can be proven by substituting the definitions
(\ref{Landaugenfn}) into (\ref{calGgenfn}) and repeatedly applying the
integral identity
\beq
\int\frac{\dd u~\dd\overline{u}}{\ii\pi}~\e^{-|u|^2}~u^\ell~
\overline{u}^{\,m}=\ell!~\delta_{\ell m}
\label{duidentity}\eeq
to reduce (\ref{calGgenfn}) to (\ref{Landautraceformula}). Inserting
the explicit forms of the generating functions (\ref{genfnexpl}) into
(\ref{calGgenfn}) then leaves a set of $L$ coupled Gaussian
integrals. The result of these integrations is most efficiently
presented by introducing an $L$ dimensional vector notation involving
the $L\times L$ shift matrix
\beq
\mS_L=\pmatrix{0&0&\dots&0&1\cr1&0&\dots&0&0\cr0&1&\dots&0&0\cr
\vdots&\vdots&\ddots&\vdots&\vdots\cr0&0&\dots&1&0\cr} \ ,
\label{shiftmatrix}\eeq
and the complex $L$-vectors
\beq
\vxi^\dag=(\,\overline{\xi_1}\,,\,\dots\,,\,\overline{\xi_L}\,) \ , ~~
\vxi=\pmatrix{\xi_1\cr\vdots\cr\xi_L\cr}
\label{xivectors}\eeq
where
\beq
\xi=\sqrt B\,z \ , ~~ \overline{\xi}=\sqrt B\,\overline{z}
\label{zrescaled}\eeq
are the dimensionless rescalings of the complex coordinates
(\ref{complexcoords}) on $\real^2$ by the magnetic length.

After some careful inspection, it is easy to see that the integrated
form can then be written as
\beq
{\cal G}_L(x_1,\dots,x_L)=2\left(\frac1{2\pi\theta}\right)^{L-1}~
\det(\id+\mS_L)^{-1}~\e^{\vxi^\dag\,\mS_L(\id+\mS_L)^{-1}\,\vxi-
\frac12\,\vxi^\dag\cdot\vxi} \ .
\label{calGshiftxi}\eeq
This same result could also have been obtained directly from
(\ref{calGdeltas}) by using the Fourier representation of the
delta-function. The expression (\ref{calGshiftxi}) as it stands is
formal, because while for $L$ odd all eigenvalues of $\id+\mS_L$ are
non-zero, for $L$ even there is exactly one zero eigenvalue and the
matrix $\id+\mS_L$ is singular. We will therefore regulate the expression
(\ref{calGshiftxi}) by replacing the shift matrix $\mS_L$ with
$\alpha\,\mS_L$ and at the end take the limit $\alpha\to1$. Thus we
will instead compute
\beq
{\cal G}_L(x_1,\dots,x_L)=2\left(\frac1{2\pi\theta}\right)^{L-1}~
\lim_{\alpha\to1}~\det(\id+\alpha\,\mS_L)^{-1}~\exp\left(-\frac12\,
\vxi^\dag\,\frac{\id-\alpha\,\mS_L}{\id+\alpha\,\mS_L}\,\vxi\right) \
{}.
\label{calGreg}\eeq

The determinant in (\ref{calGreg}) can be computed by expanding in
minors along the first row to produce two triangular determinants and
hence get
\beq
\det(\id+\alpha\,\mS_L)=\left|\matrix{1&0&\dots&0&\alpha\cr
\alpha&1&\dots&0&0\cr0&\alpha&\dots&0&0\cr\vdots&\vdots&
\ddots&\vdots&\vdots\cr0&0&\dots&\alpha&1\cr}\right|=1-(-\alpha)^L \ .
\label{shiftdetalpha}\eeq
The calculation of the inverse matrix in (\ref{calGreg}) is also
straightforward owing to the identity $\mS_L^L=\id$, which yields
\beq
(\id+\alpha\,\mS_L)^{-1}=\sum_{k=0}^\infty(-\alpha)^k\,\mS_L^k=
\frac1{1-(-\alpha)^L}\,\sum_{p=0}^{L-1}(-\alpha)^p\,\mS_L^p \ .
\label{shiftinvalpha}\eeq
Since the $p^{\rm th}$ power of the shift matrix generates a cyclic
permutation of order $p$ when acting on $L$-vectors, i.e.
\beq
(\mS_L^p\,\vxi)_I=\xi_{(I+p)_{{\rm mod}\,L}} \ ,
\label{shiftpaction}\eeq
after a reshuffling of indices the quadratic form in (\ref{calGreg})
can be written as
\bea
\vxi^\dag\,\frac{\id-\alpha\,\mS_L}{\id+\alpha\,\mS_L}\,\vxi&=&
\frac1{1-(-\alpha)^L}\,\left[\Bigl(1+(-\alpha)^L\Bigr)\,
\sum_{I=1}^L\overline{\xi_I}\,\xi_I\right.\nn\\&&+\left.2\,\sum_{I<J}
(-\alpha)^{J-I}\,\Bigl(\,\overline{\xi_I}\,\xi_J+(-\alpha)^L\,
\overline{\xi_J}\,\xi_I\Bigr)\right] \ .
\label{quadformreg}\eea
To get the final result for the kernel (\ref{calGreg}), the cases of
$L$ odd and of $L$ even should now be treated separately.

When $L=2r+1$ is odd, the limit $\alpha\to1$ is non-singular, and
hence we can simply set $\alpha=1$ in (\ref{shiftdetalpha}) and
(\ref{quadformreg}) to compute (\ref{calGreg}) as
\beq
{\cal G}_{2r+1}(x_1,\dots,x_{2r+1})=\left(\frac1{2\pi\theta}
\right)^{2r}\,\exp\left(-\ii\,\sum_{I<J}(-1)^{J-I}\,x_I\cdot Bx_J
\right) \ .
\label{calGodd}\eeq
When $L=2r$ is even, the regulator $\alpha$ must be removed
carefully. For this, we set $\alpha=1-\varepsilon$, and expand
(\ref{shiftdetalpha}) and (\ref{quadformreg}) to second order in
$\varepsilon\downarrow0$ to get
\bea
\det(\id+\alpha\,\mS_{2r})&=&2r\varepsilon+O\left(\varepsilon^2\right) \ ,
\label{detepsilon}\\\vxi^\dag\,\frac{\id-\alpha\,\mS_{2r}}
{\id+\alpha\,\mS_{2r}}\,\vxi&=&\frac B{r\varepsilon}\,
\left(\,\sum_{I=1}^{2r}(-1)^I\,x_I\right)^2-B\left(\,\sum_{I=1}^{2r}(-1)^I
\,x_I\right)^2\nn\\&&-\,\frac{2\ii}r\,\sum_{I<J}(-1)^{J-I}\,(J-I-r)
\,x_I\cdot Bx_J+O(\varepsilon) \ .
\label{quadepsilon}\eea
The first term in (\ref{quadepsilon}) combines with (\ref{detepsilon})
to produce a delta-function in the limit $\varepsilon\downarrow0$ owing to
the identity
\beq
\lim_{\varepsilon\downarrow0}\,\frac B{2\pi\varepsilon}~\e^{-\frac B
{2\varepsilon}\,|x|^2}=\delta^{(2)}(x) \ .
\label{deltalim}\eeq
This leads to the expression (\ref{calGeven}) quoted in the main
text.

\appendix{Calculation of the Free Propagator\label{FreePropApp}}

In this appendix we shall compute the propagator
(\ref{freepropdef}). We will start by evaluating the Landau heat kernel
\eq
P_t(x,y)\equiv\langle x|\e^{-t\,\hat{\sf D}^2}|y\rangle
=\sum_{\ell,m=1}^\infty\e^{-4B(\ell-\frac12)t}~
\phi_{m,\ell}(x)\,\phi_{\ell,m}(y)
\label{Ptxy}\eqend
for $0\leq t<\infty$. For this, we will find it convenient to compute the
function
\eq
\chi^{~}_t(k,x)=\e^{t\,{\sf D}^2}~\e^{\ii k\cdot x}~\phi_0(x)=\int\dd^2y~
P_t(x,y)~\e^{\ii k\cdot y}~\phi_0(y) \ ,
\label{chi}\eqend
where $\phi_0$ is the ground state wavefunction
(\ref{groundstate}). By a simple calculation we find
\beq
\chi_t^{~}=\e^{t\,{\sf D}^2}~\e^{\ii({\sf c}+{\sf c}^\dag)}~\phi_0
\label{chitsimple}\eeq
where
\beq
{\sf c}=\overline{\kappa}\,{\sf a}+\kappa\,{\sf b} \ , ~~
\kappa=\frac1{\sqrt{4B}}\,\Bigl(k_1+\ii k_2\Bigr) \ ,
\label{Aixii}\eeq
with $\sf a$ and $\sf b$ the ladder operators (\ref{ladderops}). By using the
Baker-Campbell-Hausdorff formula with $[{\sf c},{\sf c}^\dag]=2|\kappa|^2$ and
${\sf c}\,\phi_0=0$ we obtain
\beq
\chi_t^{~}=\exp\left(\ii~\e^{t\,{\sf D}^2}~{\sf c}^\dag~
\e^{-t\,{\sf D}^2}\right)~
\e^{-|\kappa|^2}{}~\e^{t\,{\sf D}^2}~\phi_0 \ .
\eeq
{}From the oscillator representation (\ref{Boxab}) of the Landau
Hamiltonian at $B\theta=1$ and (\ref{Didiag}) we get
\beq
\chi_t^{~}=\e^{-2Bt}~\e^{-|\kappa|^2}~\exp\ii\left(\kappa~
\e^{-4Bt}~{\sf a}^\dag+\overline{\kappa}\,{\sf b}^\dag\right)~\phi_0 \ .
\label{chitexplicit}\eeq
Inserting
$\phi_0=\exp\ii(\kappa~\e^{-4Bt}~{\sf b}+\overline{\kappa}\,{\sf
  a})\,\phi_0$ into (\ref{chitexplicit}) and using the
Baker-Campbell-Hausdorff formula again yields
\beq
\chi^{~}_t(k,x)=\e^{-2Bt}~\e^{-|\kappa|^2}~\e^{|\kappa|^2\exp(-4Bt)}
{}~\exp\ii\left(\kappa~\e^{-4Bt}~\overline{\xi}+\overline{\kappa}\,\xi
\right)~\phi_0(x) \ ,
\label{chitfinal}\eeq
where we have used (\ref{ladderops}) and the rescaling
(\ref{zrescaled}).

Along with (\ref{chi}), the expression (\ref{chitfinal}) allows us to compute
the heat kernel (\ref{Ptxy}) via the Fourier transform
\bea
P_t(x,y)&=&\frac{\phi_0(x)}{\phi_0(y)}\,
\int\frac{\dd^2 k}{(2\pi)^2}~\e^{-\ii k\cdot y}~\e^{-2Bt}~
\e^{-|\kappa|^2\bigl(1-\exp(-4Bt)\bigr)}\nn\\&&\times\,
\exp\ii\left(\kappa~\e^{-4Bt}~\overline{\xi}+\overline\kappa\,\xi
\right) \ .
\label{ptiFourier}\eeq
The integral in (\ref{ptiFourier}) is Gaussian, and after a straightforward
computation we arrive at the final result
\eq
P_t(x,y)=\frac B{2\pi\sinh(2Bt)}~\e^{-\frac B2\,\coth(2Bt)\,
|x-y|^2}~\e^{-\ii x\cdot By}\ .
\label{Ptxyfinal}\eqend
Note that from (\ref{deltalim}) it follows that
$P_t(x,y)\to\delta^{(2)}(x-y)$ for $B\downarrow0$, as it
should. Finally, the propagator (\ref{freepropdef}) is given by
\bea
C_\mu(x,y)&=&\int\limits_0^\infty\dd t~\e^{-t\mu^2}\,\langle x|
\e^{-t\,\hat{\sf D}^2}|y\rangle\nn\\&=&\frac B{2\pi}~\e^{-\ii x\cdot By}\,
\int\limits_0^\infty\dd t~\e^{-t\mu^2}~
\frac{\e^{-\frac B2\,\coth(2Bt)\,|x-y|^2}}{\sinh(2Bt)} \ .
\label{Cmxygen}\eea
The change of variables
\beq
u=\frac{B\,|x-y|^2}2\,\Bigl(\coth(2Bt)-1\Bigr)
\label{tuvarchange}\eeq
then brings the propagator (\ref{Cmxygen}) into the form
(\ref{freepropmassive}) given in the main text.

\appendix{Scaling Limit of the Function $\gamma^{~}_\ell(x,y)$}

To compute the sum over Landau eigenfunctions (\ref{gammaellxy}), it is
convenient to introduce the generating function
\beq
\gamma(x,y;t)=\sum_{\ell=1}^\infty\frac{t^{2(\ell-1)}}{(\ell-1)!}
{}~\gamma^{~}_\ell(x,y) \ .
\label{gammaxyt}\eeq
Proceeding as with the representation (\ref{calGgenfn}), by using
(\ref{gammaellxy}) we may then express (\ref{gammaxyt}) in terms of the
generating functions (\ref{Landaugenfn}) as
\beq
\gamma(x,y;t)=4\pi\theta\,\int\frac{\dd u~\dd\overline{u}}{\ii\pi}~
\e^{-|u|^2}\,\int\limits_0^{2\pi}\frac{\dd\tau}{2\pi}~
{\cal P}_{\overline{u}
\,,\,t\,\e^{-\ii\tau}}(x)\,{\cal P}_{t\,\e^{\ii\tau}\,,\,u}(y) \ .
\label{gammaxytcalP}\eeq
Upon substituting in the explicit form (\ref{genfnexpl}), we observe that the
integral over $u$ is Gaussian, and that the resulting $\tau$ integration yields
the Bessel function $J_0$ of the first kind of order~0, giving
\beq
\gamma(x,y;t)=4~\e^{-|x-y|^2/2\theta+\ii x\cdot By+t^2}~
J_0\Bigl(2t\,|x-y|\,/\,\sqrt\theta\,\Bigr) \ .
\label{gammaxytJ0}\eeq
By substituting the power series expansions of the exponential and
Bessel functions into (\ref{gammaxytJ0}) and comparing with
(\ref{gammaxyt}) we find the functions (\ref{gammaellxy}) for finite
$\ell$ and $\theta$ in the form
\beq
\gamma^{~}_\ell(x,y)=4(\ell-1)!~\e^{-|x-y|^2/2\theta+\ii x\cdot By}~
\sum_{k=0}^{\ell-1}\frac{(-1)^k}{(\ell-k-1)!\,(k!)^2}\,\left(
\frac{|x-y|^2}\theta\right)^k \ .
\label{gammaellxyfinite}\eeq

We are interested in the function (\ref{gammaellxyfinite}) in the scaling limit
$\theta\to\infty$, $\ell\to\infty$ with $\frac\ell\theta$ fixed. In this limit,
the exponential factor in (\ref{gammaxytJ0}) reduces to $\e^{t^2}$. The
extraction of large orders in $\ell$ in the Taylor series (\ref{gammaxyt}) is
accomplished by the standard method of contour integration to write
\beq
\gamma^{~}_\ell(x,y)=(\ell-1)!~\oint\limits_{t=0}\frac{\dd t}{2\pi\ii
t^{2\ell-1}}~\gamma(x,y;t) \ .
\label{gammaellxycontint}\eeq
By rescaling the integrand of (\ref{gammaellxycontint}) as
$t=\sqrt{\ell-1}\,w$ and using (\ref{gammaxytJ0}) we get
\beq
\gamma^{~}_\ell(x,y)=4\,(\ell-1)!~\e^{\ell-1}~\oint\limits_{w=0}
\frac{\dd w}{2\pi\ii w}~J_0\Bigl(2w\,|x-y|\,\sqrt{(\ell-1)/\theta}\,
\Bigr)~\e^{(\ell-1)\,(w^2-2\ln w)} \ .
\label{gammaellxyrescale}\eeq
In the large $\ell$ limit, the integral (\ref{gammaellxyrescale}) can be
evaluated by using the saddle-point approximation. There are two saddle points,
at $w=\pm\,1$. By expanding (\ref{gammaellxyrescale}) near these saddle points
and using the Stirling approximation
$\ell!\simeq\sqrt{2\pi\ell}~\e^{-\ell}~\ell^{\,\ell}$ for $\ell\to\infty$, we
finally find
\beq
\gamma_\ell^{~}(x,y)=4\,J_0\Bigl(2\,|x-y|\,\sqrt{\ell/\theta}\,\Bigr) \ ,
\label{gammaellJ0finalapp}\eeq
which on trading the integers $\ell$ for the eigenvalues $\lambda_\ell$ defined
in (\ref{NClambda}) yields the result (\ref{gammaellJ0lambda}) in the main
text.

\end{document}